\documentclass[%
superscriptaddress,
 reprint,
 amsmath,amssymb,
 aps,
]{revtex4-2}

\usepackage{graphicx} 
\usepackage[final]{hyperref} 
\usepackage{enumerate}
\usepackage{amsfonts}
\usepackage{braket}
\usepackage{tikz}
\usepackage{empheq}
\usepackage{bbm}
\usepackage{multirow}
\usepackage{mathtools}

\usepackage[mathscr]{euscript}	
\usepackage{enumitem}
\usepackage{manfnt}
\usepackage{titletoc}
\usepackage[matrix,arrow,curve,arc]{xy}

\setlist[itemize]{itemsep=1pt, topsep=2pt}
\allowdisplaybreaks		

\def\EndDoc{\end{document}\begin{document}}
\newcommand{\xyhole}{{\txt{\phantom{$\cdot$}}}}	
\newcommand{\phdot}{{\phantom{\mkern1.3mu\cdot\mkern1.3mu}}}
\newcommand{\scr}{\scriptstyle}
\newcommand{\sscr}{\scriptscriptstyle}
\usepackage{color}
\xyoption{color}
\xyoption{line}
\UseCrayolaColors

\newcommand{\bigket}[1]{ {\big|{#1}\big\rangle} }
\newcommand{\Bigket}[1]{ {\Big|{#1}\Big\rangle} }
\newcommand{\mcC}{ {\mathcal{C}} }
\newcommand{\mcA}{ {\mathcal{A}} }
\newcommand{\tqd}{ {\mathcal{D}} }
\newcommand{\qTr}{\operatorname{\widetilde{Tr}}}
\newcommand{\llll}{\mathfrak{l}}
\newcommand{\mmm}{\mathfrak{m}}
\newcommand{\sss}{\mathfrak{s}}
\newcommand{\hhh}{\mathfrak{h}}
\newcommand{\vvv}{\mathfrak{v}}
\newcommand{\T}{\mathcal{T}}
\newcommand{\uuu}{\mathfrak{u}}
\newcommand{\p}{\mathfrak{p}}
\newcommand{\ttt}{\mathfrak{t}}
\newcommand{\Dehn}{\mathbf{D}}
\newcommand{\V}{\mathrm{V}}
\newcommand{\M}{\mathrm{M}}
\newcommand{\K}{\mathrm{K}}
\newcommand{\R}{\mathcal{R}}
\newcommand{\defeq}{\overset{\mathrm{def}}{=}}

\definecolor{purple}{rgb}{0.5,0,0.5}
\newcommand{\zhuan}[1]{{\color{orange}\ifmmode\text{\footnotesize(ZL) #1}\else\footnotesize{(ZL) #1}\fi}}
\newcommand{\roger}[1]{{\color{purple}\ifmmode\text{\footnotesize(RM) #1}\else\footnotesize{(RM) #1}\fi}}
\newcommand{\michael}[1]{{\color{blue}\ifmmode\text{\footnotesize(ML) #1}\else\footnotesize{(ML) #1}\fi}}

\newcommand*\widefbox[1]{\fbox{\hspace{2em}#1\hspace{2em}}}

\begin{document}

\preprint{APS/123-QED}

\title{Detecting topological order from modular transformations of ground states on the torus
}%

\author{Zhuan Li}
 \affiliation{
 Department of Physics and Astronomy, University of Pittsburgh, Pittsburgh, Pennsylvania 15260, USA
}%


\author{Roger S. K. Mong}
\affiliation{
 Department of Physics and Astronomy, University of Pittsburgh, Pittsburgh, Pennsylvania 15260, USA
}%

\date{\today}
\begin{abstract}
    The ground states encode the information of the topological phases of a 2-dimensional system, which makes them crucial in determining the associated topological quantum field theory (TQFT).
    Most numerical methods for detecting the TQFT relied on the use of minimum entanglement states (MESs), extracting the anyon mutual statistics and self statistics via overlaps and/or the entanglement spectra.
    The MESs are the eigenstates of the Wilson loop operators, and are labeled by the anyons corresponding to their eigenvalues.
    Here we revisit the definition of the Wilson loop operators and MESs. We derive the modular transformation of the ground states purely from the Wilson loop algebra, and as a result, the modular $S$- and $T$-matrices naturally show up in the overlap of MESs.
    Importantly, we show that due to the phase degree of freedom of the Wilson loop operators, the MES-anyon assignment is not unique. This ambiguity obstructs our attempt to detect the topological order, that is, there exist different TQFTs that cannot be distinguished solely by the overlap of MESs.
    In this paper, we provide the upper limit of the information one may obtain from the overlap of MESs without other additional structure.
    Finally, we show that if the phase is enriched by rotational symmetry, there may be additional TQFT information that can be extracted from overlap of MESs.
\end{abstract}

\maketitle

\section{Introduction}

After the discovery of the Integer Quantum Hall effect and the Fractional Quantum Hall effect, more and more bizarre phases of matter cannot be described by the  Landau-Ginzburg symmetry
breaking paradigm were found. Instead of normal order parameters, these phases are depicted by non-local topological orders \cite{wen1990TopologicalOrder}.
Other than fermion and boson, the point excitations of these phases bears the non-trivial braiding statistics, i.e, the exchange of positions of two anyon results in a phase factor or, for non-Abelian anyons, a transformation of the wave function. These excitations are called anyons, and can only exist in 2-dimensional space. 
In a topological phase, their classes of excitation, fusion relations, and braiding statistics are robust to small deformations to the Hamiltonian, which makes non-Abelian anyon a potential candidate for the fault-tolerant quantum computation \cite{ToricCode,QuantumComputation,QuantumComputationOneMobileAnyon}.

While there are multiple exact solvable microscopic Hamiltonians that realize the non-trivial topological phase \cite{ToricCode,Wen'sPlaquettesModel,kitaev2006anyons,Levin-Wen,Shetengel04,Santos15Rokhsar-KivelsonModels}, the inverse problem---i.e, determining the topological phase from the microscopic Hamiltonian---is extremely challenging.
It is believed that the ground state alone encodes all the information regarding the topological order, i.e., it is possible to determine properties of the excitation spectrum such as anyon content, braiding statistics from the lowest eigenstate of the system%
    ~\footnote{Ref.~\onlinecite{SingleEigenstate} argues that it is sufficient to determine the properties of a many body Hamiltonian from a simple eigenstate. Thus, the ground state, as a special case, encodes all the topological data of the Hamiltonian.}%
    .
A well known example is that: on a torus, the ground state degeneracy equals to the number of anyons of the system. Recently, more progress have been made in determining the topological data from the ground states. 
For example, numerical methods such as exact diagonalization and the density-matrix renormalization group (DMRG) gives the ground state, from which we can extract the quantum dimensions~\cite{TEEkitaev,TEELevinWen}, edge excitation spectrum~\cite{EntanglementSpectrumLiHaldane}, fractional charges, chiral central charge~\cite{CentralChargeFromWaveFunction}, topological spin~\cite{MomentumPolarization,MomentumPolarizationRoger}, etc.

A crucial characterization of a topological order are the $T$- and $S$-matrices, which encode the topological spin and mutual-statistics (braiding phase) respectively.
Remarkably, the modular transformations of ground states on a torus generate these matrices~\cite{wen1990TopologicalOrder}.
Although the $S$- and $T$-matrices do not constitute a complete description of a topological order in general, there are many simple anyon models that are uniquely determined by the $S$- and $T$-matrices~\footnote{The twisted Drinfeld doubles of ﬁnite groups up to order 31 are uniquely determined by $S$ and $T$~\cite{MCareNotDeterminedByS&T}.}.
(In fact, all models with no more than 5 anyons are uniquely determined by their $S$- and $T$-matrices~\cite{ClassificationOfAnyon,classificationofAnyonrank5}.)
The $S$- and $T$-matrices appear when taking the overlap of ground states in the Minimum Entanglement States (MESs) bases with the proper labeling~\cite{YiZhang2012GroundStateEntanglement,zhang2015overlap}.
The MESs---natural byproducts of tensor-network numerical algorithms such as DMRG---are the ground states with the local minimum bipartite entanglement entropy along a certain cut~\cite{YiZhangoverlapSmatrix,SemionAntisemionOverlap}.
However, it may not be possible a priori to determine the proper labeling of the MES---i.e., the correspondence between anyons and MESs---which in turn means that the MES overlaps alone are not enough to get the modular data~\cite{zhang2015overlap}. 
In this paper, we carefully reexamine these ideas. We ask, why the $S$- and $T$-matrices appear in modular transformations of MESs?  To what extent the modular matrices can be extracted from the grounds states alone?

The aim of this paper is twofold: (1) to revisit the fundamental relations between the Wilson loop operators and the MESs on the torus, and (2) to explore the information we may obtain from the overlap of MESs.
We first re-examine the definition of the Wilson loop operators and the MESs.
In contrast from the previous works, we demonstrate how the $S$- and $T$-matrices naturally appear in the modular transformation of the MESs without reliance on a any conformal field theory (CFT) structure \cite{MooreSeibergCFT} or connection in moduli space \cite{wen1990TopologicalOrder}.
Our approach is to define the Dehn twist operators, then compute the transformations of the Wilson loop operators and the ground states under the Dehn twist operators.

Given the MESs along different cuts, we show that---absent of symmetry or entanglement spectra---their overlaps can only provide the fusion rule $N_{ab}^c$ and the triplet spins $\theta_a\theta_b/\theta_c$ for $N_{ab}^c>0$.
This means that there exist distinct TQFT models that cannot be distinguished by the overlap of MESs solely.

We also consider the system with the global rotational symmetries. The introduction of global symmetry reduces the complexity of the algorithm, which makes the MESs easier to be found. Moreover, the symmetry enriches the topological phase, i.e., in the presence of global symmetry, two equivalent topological phases without symmetry would be distinct. One of the signatures of the symmetry enriched topological phases is that their excitations may change the species under the rotation~\cite{SymmetryEnrichedReview,spatialsymmetries}.
We find that this feature is detectable from the overlap of MESs, and if the non-trivial permutation of anyon exists, it can be used in determining the previous equivalent phases.

This paper is organized as follows. In Section \ref{defW} and \ref{sec:basisW}, we construct the microscopic definition of the Wilson loop operators and come up with the Wilson loop algebra from the Dehn twist operators. Then, the MESs are defined as the eigenstates of the Wilson loop operators in section \ref{sec:defMES}. The modular transformations of MESs are obtained in section \ref{sec:MT}, derived from the Wilson loop algebra.

In Section \ref{sec:indistinguishablemodel} and \ref{sec:algorithm}, we first define the indistinguishable models, and then show how to determine the non-indistinguishable models by using the algorithm from \cite{YiZhangoverlapSmatrix}. We also list several equivalent conditions that determine the indistinguishable models in section \ref{sec:MaxInfo} and \ref{sec:discussion}. {From these conditions we get the maximum information we can obtain from the overlap of MESs, i.e., the fusion rule and triplet spins.} 

In Section \ref{sec:RotationalSymmetry}, we consider the system with 3-, 4- and 6-fold rotational symmetries. For the 3- and 6-fold rotations, we can not only obtain the $S$-matrix and the square of $T$-matrix from the overlap, but also the permutation of anyons under the rotation. While for the 4-fold rotation, we can only get the $S$-matrix and the permutation with the help of the algorithm from \cite{YiZhangoverlapSmatrix}. We show that the permutation of anyon in these case are useful in determining the topological phase.

\section{Brief review of tensor category and torus}\label{sec:notations}

\begin{table*}[bt]
\caption{\label{tab:table1} Here we list some notations that we will use in this paper.}
\begin{tabular}{|c|l|}
\hline
{\textbf{Notation} }&
{ \textbf{Definition} } \\
\hline
$\mcC$ & \parbox{155mm}{\raggedright A unitary modular tensor category (TQFT model), contains a set of anyons, the fusion and braiding rules. }\\
\hline
$\mcA$ & The Abelian subcategory of $\mcC$.\\
\hline
$a,b,c,\cdots$ & The anyons in $\mcC$, they obey the commutative, associative fusion rule. We use $0$ to denote the trivial anyon. \\
\hline
$N_{ab}^c$ & The fusion symbols, indicating the number of ways $a$ and $b$ fuse to anyon $c$.\\
\hline 
{$d_a$} & The quantum dimension of the anyon $a$, $d_a = {\xy (-5,0)="a";  "a";"a"+(0,1),**\dir{}, "a",{\ellipse(5){-}}; {\ar@{>} (-0.05,0.85);(-0.05,0.9)}; (2,0)*{ a} \endxy} $. \\
\hline
$\tqd$ & Total quantum dimension of the TQFT model, $ \tqd \defeq \sqrt{ \sum_{a\in \mcC} {d_a^2}}$.\\
\hline
{ $\theta_a$} & The topological spin of anyon $a$.  \\
\hline
$T$ & The modular $T$ matrix: a diagonal matrix with elements $T_{a,b} \defeq \delta_{a,b}\theta_{a}$, where $\theta_a = {\xy (0,0)*{\phdot}="o"; (-3,3)="tl"; (3,3)="tr"; (-3,-3)="bl"; (3,-3)="br"; (3,5)*{a}; 
			"o";"tl"**\dir{-}; "bl",{\ellipse^{}}; "tr"**\dir{-}?(1)*\dir{>}; "br",{\ellipse_{}}; "o"**\dir{-}; \endxy}$ \\
\hline
$S$ & The modular $S$ matrix: a symmetric matrix with elements   
 $\displaystyle S_{a,b} \defeq  \frac{1}{\tqd} \sum_e N_{ab}^e d_c \frac{\theta_a\theta_b}{\theta_c}=\frac{1}{\tqd}{\xy (-9,0)*{}="Bl"; (-4,0)="a"; (2,0)="b"; (1,0.7)="aa",*+!L{a}; {\ar@{>} "aa"+(0,-.01);"aa"}; (7,0.7)="bb",*+!L{ b}; {\ar@{>} "bb"+(0,-.01);"bb"};
		"a";"b",**\dir{}, "a",{\ellipse(5):a(154),=:a(-20){-}}; "b";"a",**\dir{}, "b",{\ellipse(5):a(154),=:a(-20){-}}; \endxy}.$\\
\hline 
{$\Theta$} & \parbox{150mm}{\raggedright
$\Theta$ is a phase related to the central charge.
$\Theta \defeq \frac{1}{\tqd} \sum_{a\in\mcC} d_a^2 \theta_a$. } \\
\hline
$\llll,\mmm,\dots$ & \parbox{155mm}{\raggedright Simple loops on the torus i.e. the loops without self-intersection. We use the winding number to classify the loops on the torus such as $\llll=(p_1,q_1), p_1,q_1 \in \mathbb{Z} $.}   \\
\hline
$\llll\times\mmm$& The intersection number between loop $\llll$ and $\mmm$. 
\\
\hline
$(\llll,\mmm),\cdots$ & \parbox{155mm}{\raggedright A coordinate basis for the torus. Two simple loops form a basis if and only if they intersect once, i.e. $\llll\times \mmm = -1.$}\\
\hline
$\tau_\llll$ & The Dehn twist about the simple loop $\llll$ on the torus.\\
\hline
\multirow{2}{*}{ $W_\llll(a)$} & The Wilson loop operator of anyon $a$ along the simple loop $\llll$. \\
&The detailed definition for the Wilson loop operator can be find in section \ref{defW}.\\
\hline
\multirow{2}{*}{ $\Dehn_\llll$} & The Dehn twist operator about the simple loop $\llll$. \\
& $\Dehn_\llll \defeq \frac{1}{\tqd} \sum_{a\in\mcC} {d_a \theta_a^*}W_\llll(a)$. \\
\hline
$\ket{a^\llll_\mmm}$ & The standard basis of the ground states in the coordinate system of $(\llll,\mmm)$ with the label $a$.\\
\hline
{ $\lambda(a)$ }& \parbox{155mm}{\raggedright The phase such that $\lambda(a)\lambda(b) = \lambda(c)$ for $N_{ab}^c>0$.} \\

\hline
\end{tabular}
\end{table*}

\subsection{Modular tensor category}
In this section we briefly review some definitions of modular tensor category. For more details we refer the reader to Appendix~\ref{TQFT}, or Refs.~\onlinecite{kitaev2006anyons,SymmetryEnrichedReview}

A TQFT model---also known as a modular tensor category---$\mcC$ consists of a set of anyon $\{a,b,c,\dots\}$, fusion rules among the anyons, and other data associated with braiding and manipulations of the anyons. 

The fusion rules for anyons are described by
\begin{align}
    a\times b = \bigoplus_{c\in\mcC} N_{ab}^c c,
\end{align}
where $N_{ab}^c$ are non-negative integers, which called the fusion symbols.
Several important modular data for a TQFT model, such as quantum dimensions $d_a$, topological spin $\theta_a$, total quantum dimension $\tqd$, the modular matrices $S$ and $T$, are all defined in Table~\ref{tab:table1}. 

The $S$-matrix gives all the 1-dimensional representation of the fusion algebra, i.e., for the ${\pi}(a) \in \mathbb{C}$ such that
\begin{align}
   \pi(a)\pi(b)=\sum_c N_{ab}^c\pi(c),
\end{align}
then there is an anyon $e \in \mcC $ so that $\pi(a) = S_{e,a}/S_{e,0}$. 
These 1-dimensional representation are also called fusion character.
In particular, the quantum dimension $d_a$ is the fusion character that corresponds to the trivial anyon $d_a = S_{0,a}/S_{0,0}$. 

We call $\lambda(a) \in \mathrm{U}(1)$ a fusion phase if it satisfies 
\begin{align}
    \lambda(a){\lambda}(b)= {\lambda}(c), \text{ if } N_{ab}^c >0.
\end{align}
Then, for an arbitrary fusion character $\pi(a)$ 
we can get another fusion character $\pi(a)'$ by attaching the fusion phase, $\pi(a)' = \lambda(a)\pi(a)$.
We show that all the fusion phases are given by the Abelian subcategory $\mcA \subseteq \mcC$,
\begin{align}
    \lambda(a) = \frac{S_{r,a}}{S_{0,a}}, \text{ where } r \in \mcA.
\end{align}
This fact can be easily checked because $\lambda(a)d_a$ is a fusion character, while all the fusion characters are given by $S_{e,a}/S_{0,a}$. Because for the non-Abelian anyon $e$, there exists the anyon $a\in\mcC$ such that $d_aS_{e,a}/S_{0,a} \not\in \mathrm{U}(1)$. Therefore, the fusion phase can only be written as $ \lambda(a) = {S_{r,a}}/{S_{0,a}},$ for the $ r \in \mcA.$

\subsection{Topology of the torus}
Let's start with a general oriented closed 2-dimensional manifold $M$. The mapping class group of $M$ is defined to be the group of self-homeomorphisms of $M$ modulo homotopy. The Dehn-Lickorish theorem claims that the mapping class group of a general surface can be generated by a finite number of Dehn twists.

\begin{figure}[b]
    \centering
    \includegraphics[width=\linewidth]{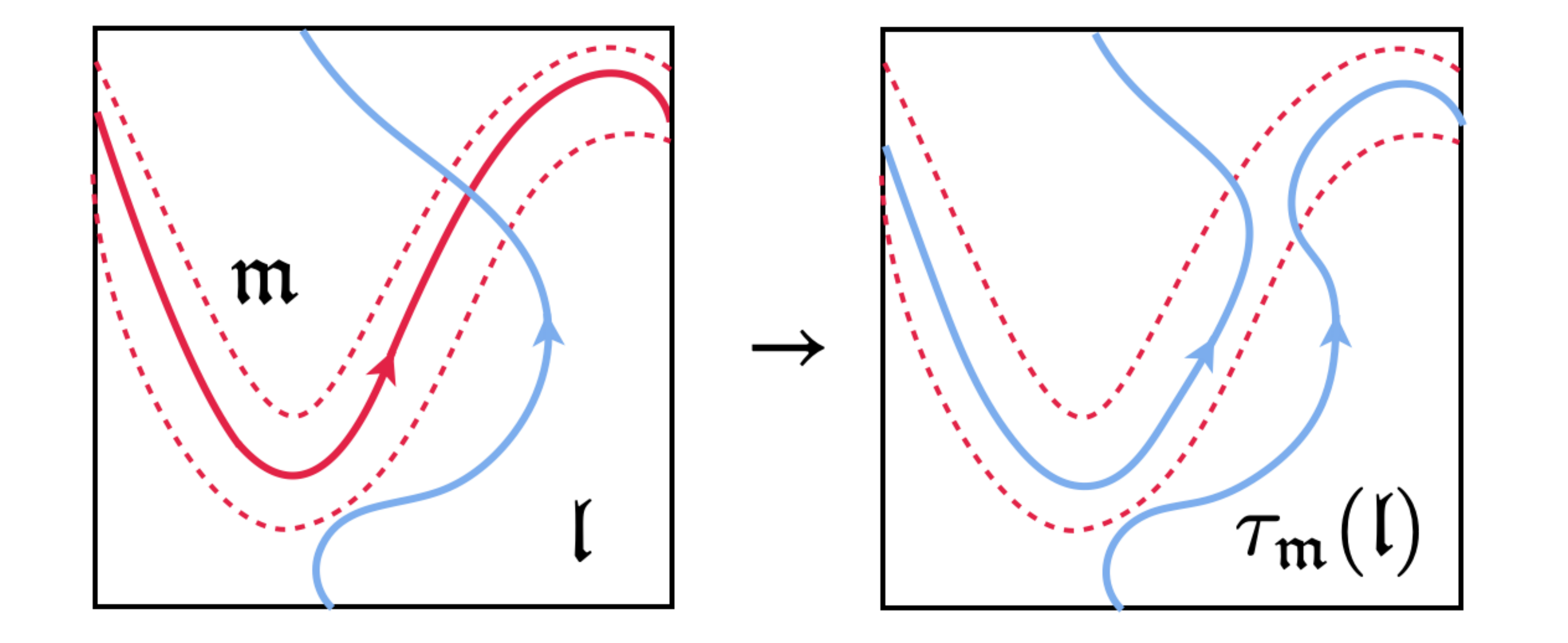}
    \caption{$\llll$ and $\mmm$ are two simple loops on torus, $\tau_{\mmm}$ is the Dehn twist about loop $\mmm$. The orientation of the surface is defined to be $\llll\times \mmm = -1$.}
    \label{fig:DehnTwist}
\end{figure}

Let $\mmm$ be a simple loop, i.e., the loop without the self-intersection.
A Dehn twist $\tau_\mmm$ is a homeomorphism of the surface given by the construction (see Fig.~\ref{fig:DehnTwist}):
first we cut the surface along loop $\mmm$, twist one side of the incision $360^\circ$ and glue them back.
Rigorously, the Dehn twist is defined as follows.
Let $A = S^1 \times [0,1]$, there is a homeomorphism $\phi$ from a neighborhood $N$ of $\mmm$ (region enclosed by the dash line in figure \ref{fig:DehnTwist}) to $A$. The map $\tau_{\mmm}$ is given by
\begin{align}
            \tau_{\mmm}(x) = \begin{cases}
            \phi\circ\tau\circ\phi^{-1}(x), \ \ \ x\in N \\
            x, \ \ \ x\not\in N
            \end{cases}
\end{align}
where $\tau$ is the twist map of the annulus
\begin{align} \begin{aligned}
    \tau &: A \to A , \\
    &\;\;\; (\theta,t) \mapsto (\theta+2\pi t,t) .
\end{aligned} \end{align}

In the case of torus, the mapping class group is the special linear group $\mathrm{SL}(2,\mathbb{Z})$, which is also called the modular group~\footnote{Technically, the modular group is $\mathrm{PSL}(2,\mathbb{Z})$, doubled-covered by $\mathrm{SL}(2,\mathbb{Z})$.}.
The modular group can be generated by two matrices
\begin{align}
    &u = \begin{pmatrix}
    1&0\\
    -1&1
    \end{pmatrix},& &t = \begin{pmatrix}
    1&1\\
    0&1
    \end{pmatrix}.
\end{align}

We can analyse the Dehn twist transformation by its action on the homopty classes of the torus. The loops on the torus are classified by their winding numbers $(p,q)$. 
A homotopy class $[\llll]=(p,q)$ is called simple if the loops in this class are simple, i.e., the greatest common divisor of $p$ and $q$ is $1$. 
Let $\mmm$ and $\llll$ be two arbitrary loops with homotopy class $[\mmm]=(p_1,q_1),[\llll] = (p_2,q_2)$, the intersection number $\mmm\times\llll$ between them is given by
\begin{align}
    \mmm\times\llll = \det\begin{vmatrix}
    p_1 & q_1\\
    p_2 & q_2
    \end{vmatrix}.
\end{align}
The Dehn twist about a simple loop define a nature way to generate a new loop $[\tau_\mmm(\llll)]$ from two existing loops $[\mmm]=(p_1,q_1)$ and $[\llll]= (p_2,q_2)$, (Fig.\ref{fig:DehnTwist}): the homotopy class of Dehn twist loop $\tau_\mmm(\llll)$ is given by 
\begin{align}
    [\tau_\mmm(\llll)] =  [\mmm] \det\begin{vmatrix}
    p_1 & q_1\\
    p_2 &q_2
    \end{vmatrix} + [\llll].
\end{align}
Due to the linearity of the determinant, the Dehn twist $\tau_\mmm$ is linear in the argument: $[\tau_\mmm(\llll_1+\llll_2)] = [\tau_\mmm(\llll_1)]+[\tau_\mmm(\llll_2)]$.

If $\llll\times\mmm = -1$, then we denote $(\llll,\mmm)$ as the basis of the torus.
In particular, if $\llll$ and $\mmm$ only intersect once, we have $[\tau_\mmm(\llll) ] = (\pm p_1 +p_2,\pm q_1+ q_2)$. 
 The actions of the Dehn twists on these basis are given by 
\begin{align}
   & \tau_\llll : \left\{ \begin{aligned}
       &[\llll] \mapsto [\llll] \\
       &[\mmm] \mapsto [-\llll+\mmm]
    \end{aligned} \right. ,&& \tau_\mmm: \left\{ \begin{aligned}
      & [\llll] \mapsto [\llll+\mmm] \\
       &[\mmm] \mapsto [\mmm]
    \end{aligned} \right. .&
\end{align}
We noticed that these two transformations are exactly the generators of the modular group. 
Therefore, if $\llll$ and $\mmm$ only intersect once, all the simple loop classes can be generated from two classes $[\llll]$ and $[\mmm]$ by using the Dehn twist transformations $\tau_\llll$ and $\tau_\mmm$.

\section{Wilson loop operators and the ground states}\label{sec:WilsonloopAlgebra&MESs}
In this section, we give a microscopic definition of the Wilson loop operators, we derive the Wilson loop algebra and define the Minimum Entanglement States (MES) from the Wilson loops.
Here we assume that we know all the TQFT data, and derive relations between Wilson loops and ground states.

Here is a brief summary of this section.

We define the Wilson loop operators $W_\llll(a)$ from the process of creating a particle-antiparticle pair ($a$,$\bar{a}$), moving the particle $a$ along a loop $\llll$, and then annihilating the particles back into the vacuum.
Generically such processes are affected by various geometric phases arising from microscopic details of the model.  We show that there exist a choice of phase and normalization, such that the Wilson loop operators satisfy the relations
 \begin{align}
    W_\llll(a)W_\llll(b) &= \sum_c N_{ab}^cW_\llll(c) \label{fusionW},\\W_{\llll}^\dagger(a) &= W_{\llll}(\bar{a}), \label{conjureW}
     \\
    W_{-\llll}(a) &= W_{\llll}(\bar{a}) \label{antiW},
\end{align}
for each loop $\llll$.

There are an infinite number of simple loops on a torus characterized by their winding number $(p,q)$, and associated with each loop is a set of Wilson loop operators. We show that the Wilson loop operators along an arbitrary loop on a torus can be constructed from the Wilson loop operators along a coordinate basis ($\llll$,$\mmm$): $W_{\llll}(a)$ and $W_\mmm(a)$.
We accomplish this by introducing the Dehn twist operators $\Dehn_\mmm$, given by
\begin{align}
  \Dehn_{\mmm} = \frac{1}{\tqd}\sum_x \theta^*_x d_x W_\mmm(x) .
\end{align}
The operator transform Wilson loops along $\llll$ to those along $\tau_{\llll}(\mmm)$:
\begin{align}
    W_{\tau_{\mmm}(\llll)}(a) = \Dehn_{\mmm} W_{\llll}(a) \Dehn_{\mmm}^\dagger .
    \label{Dehn}
\end{align}    
Figure \ref{fig:DehnTwist} also demonstrates how the Dehn twist operator $\Dehn_{\mmm}$ act on the Wilson loop operators.
As a result, we only need to define the Wilson loop operators along a single coordinate basis on the torus---e.g.\ ($\llll$,$\mmm$)---to fully define all Wilson loop operators.


For every coordinate basis $(\llll,\mmm)$,
we define the basis of the ground states $\ket{b^\llll_\mmm}$ from the Wilson loop operators, which we referred to as the ``standard basis''.
States within the standard basis are labelled by anyons in $a\in\mcC$, with the property that $\ket{b^\llll_\mmm}$ are each MES along the $\mmm$-cut, and
\begin{align}
    W_\mmm(a)\ket{b^\llll_\mmm} &= \frac{S^*_{ab}}{S_{0b}}\ket{b^\llll_\mmm} ,\label{defMm}
\\   W_\llll(a)\ket{b^\llll_\mmm} &= \sum_cN_{ab}^c\ket{c^\llll_\mmm} .\label{defMl}
\end{align}
Here we define an MES to have a \emph{local} minimum in its bipartite entanglement entropy (as opposed to the original definition in Ref.~\onlinecite{YiZhang2012GroundStateEntanglement} which stipulated a global minimum of entropy.)
Eq.~\eqref{defMm} says that the MES are simultaneous eigenstates of the Wilson loop operators; this is always possible as the Wilson loop operators along the same path commute with each other.
Eq.~\eqref{defMl} relates the different eigenstates to one another; doing so fixes the relative phrases between the different MESs.
For each coordinate basis, the standard basis is uniquely defined (up to an overall phase) from these relations. The eigenstates of the Wilson loop operators are also called the minimum entanglement states (MES), because they have the local minimum bipartite entanglement entropy along a certain direction. Here our definition of the basis is slightly different from the previous definition of MESs: we use another direction ($\llll$ in our case) to fix the relative phases of among the MESs.

Our definition of the standard basis is such that the overlaps between the different bases give us the modular matrices,
\begin{align}
    \ket{a^{-\mmm}_\llll} &= \kappa_1 \sum_b S_{ab}\ket{b^\llll_\mmm}, \\
     \ket{a^{\llll+\mmm}_\mmm} & =\kappa_2 \theta_a\ket{a^\llll_\mmm}.
\end{align}
where $\kappa_1$ and $\kappa_2$ are two overall phases only depend on $\llll$ and $\mmm$.
This means the Hilbert space spanned by the ground states and the transformation of coordinate bases form a projective representation of the modular group.

The Wilson loop operators obeying Eqs.~\eqref{fusionW}--\eqref{conjureW} are not unique.
In particular, we can apply a gauge transformation $W_\llll(a)' = \lambda(a) W_\llll(a)$, where $\lambda(a)$ is a fusion phase.
We prove that the number of fusion phase is equal to the number of the Abelian anyons $|\mcA|$.
Because there are two independent directions, there are $|\mcA|^2$ gauge transformations in total which are compatitible with the Wilson loop algebra.
Once we fix the gauge of the basis, we can determine the phase for any other directions by using equation \eqref{Dehn}. The detailed discussion of this gauge transformation of the Wilson loop operators can be find in Appendix \ref{sec:gaugetransformation}.

A gauge transformation of Wilson loop operators will have two effects on the standard basis.
First, it may induce a rearrangement of the states within the standard basis, i.e., permute the anyon labelling of the ground states $\ket{b_\mmm^\llll}$.
Second, it can change relative phase among states within the basis.
In total, there are $|\mathcal{A}|^2$ rearrangements/phases of MESs which is same with the number of gauge transformation of Wilson loop operators. The explicit formula of the rearrangement of the basis can be find in Appendix \ref{sec:rearrange}.

While we only focus on the torus in this work, the Wilson loop algebra and many conclusions here can be easily generalized to the higher genus manifolds.

\subsection{Microscopic construction of the Wilson loop operators}\label{defW}

\begin{figure}[b]
    \centering
    \includegraphics[width=\linewidth]{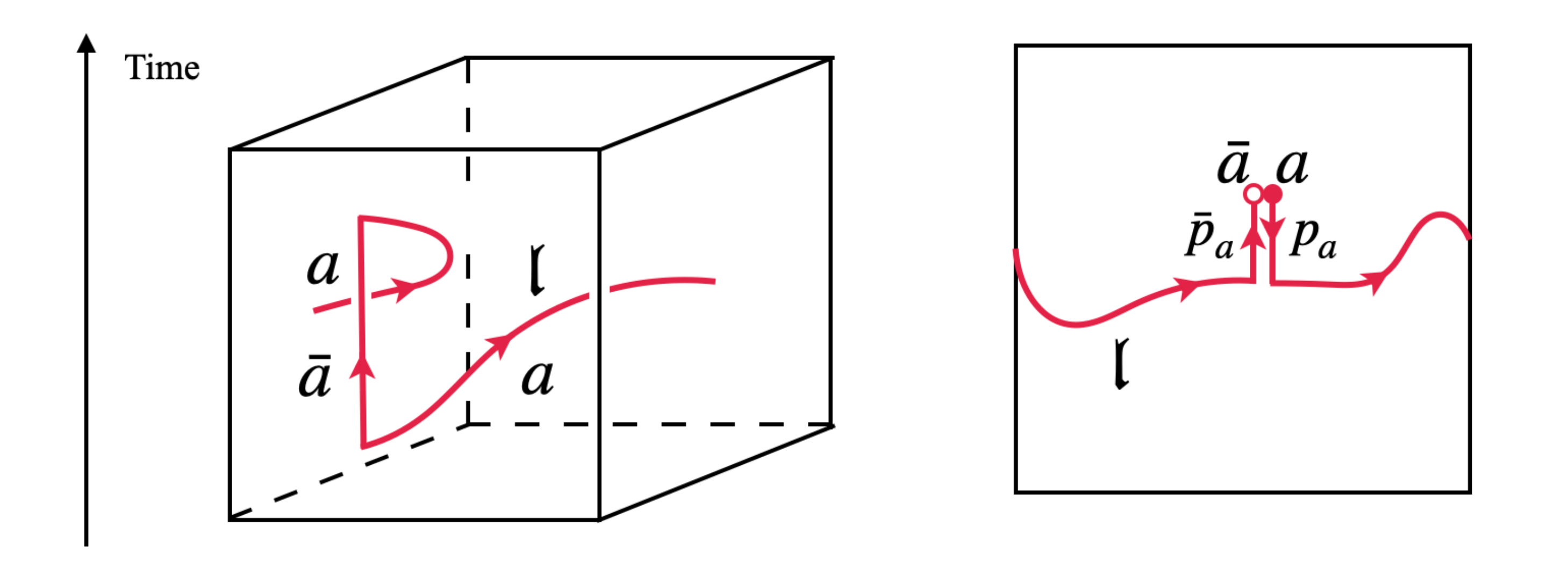}
    \caption{Left: Definition of Wilson loop operator along an arbitrary path $\llll$ in a world line diagram. Right: The definition of the paths in equation \eqref{eq:defW}. The geometric phases of $p_a$ and $\bar{p}_a$ cancel each other, therefore, the total geometric phase is independent from the position where we create the particle anti-particle pair but only depend on the loop $\llll$.}
    \label{fig:defW}
\end{figure}

The Wilson loop operators are the non-local topological invariant line operators. 
After creating a particle-antiparticle pair $(a,\bar{a})$ from a ground state, dragging $a$ alone a closed path $\llll$ adiabatically, and then annihilating with $\bar{a}$, we get another ground state (Fig.~\ref{fig:defW}).
In general, any microscopic process that involve moving excitations may have dynamical (time/energy dependent) and geometric (path dependent) components involved.
Here we want to define the Wilson loop operators independent of the dynamic and geometric phase of this process, such that only the topological phase remains.

Follow Ref.~\onlinecite{kawagoe2020microscopicDefinitionOfFandR}, we decompose the Wilson loop operator into three pieces: creation, movement and annihilation 
\begin{align}
    W_\llll(a) = \alpha_\llll(a) \, \Phi^0_{a\bar{a}} \, {\M}_{p_a}(a) \, {\M}_{\llll}(a) \, {\M}_{\bar{p}_a}(a) \, \Phi_0^{a\bar{a}},
    \label{eq:defW}
\end{align}
where $\Phi_0^{a\bar{a}}$ and $\Phi^0_{a\bar{a}}= {\Phi_0^{a\bar{a}}}^\dagger$ are the creation and annihilation operators respectively, and here we choose the isotopy normalization for the creation and annihilation operators $\Phi^0_{a\bar{a}}\Phi_0^{a\bar{a}} = d_a\mathbbm{1}_{0}$. ${\M}_\llll(a)$ is the movement operator which move the anyon $a$ along the path $\llll$.
Path $\llll $ and $p_a$ are defined in figure \ref{fig:defW}.

Here we deliberately create the pair not along the path $\llll$ so that the phases of $\M_{p_a}^a$ and $\M_{\bar{p}_a}^a$ cancel each other, as do the phases of $\Phi_0^{a\bar{a}}$ and $\Phi^0_{a\bar{a}}$.
Therefore, the phase $\alpha_\llll(a)$ is designed to only fix the phase of $\M_\llll(a)$. It also indicates that the Wilson loop operators do not depend on the location where we create the particle pair and the point where the particle enter the loop. 

There exists a consistent way to choose $\alpha_\llll(a)$ for every simple loop $\llll$ and every anyon $a$, such that the Wilson loop operators defined in this way satisfy the algebra,
\begin{empheq}[box=\widefbox]{align}
    W_\llll(a)W_\llll(b) &= \sum_c N_{ab}^cW_\llll(c), \label{eq:fusionW1}
    \\ W_{\llll}^\dag(a) &= W_{\bar{\llll}}({a}) \label{eq:conjureW1},
    \\ W_{\llll}^\dag(a) &= W_{\llll}(\bar{a}) \label{eq:antiW1}, 
\end{empheq}
where $a,b,c$ are anyons.
Notice that the existence of such phase assignment is not completely trivial, since Eq.~\eqref{eq:fusionW1} consists of $\mathcal{O}(|\mcC|^2)$ equations but there are only $|\mcC|-1$ phases.

Here we outline the argument for why $\alpha_\llll(a)$ exists.
To simplify the notation, we denote $\widetilde{\M}_{\llll}(a) =  {\M}_{p_a}(a){\M}_{\llll}(a){\M}_{\bar{p}_a}(a)$. 
Then the Wilson loop operator becomes 
\begin{align}
    W_\llll(a) = \alpha_\llll(a) \Phi_{a\bar{a}}^0\widetilde{\M}_\llll(a)\Phi_{0}^{a\bar{a}}.
\end{align}
The key point is to show the ``fusion rules" of the movement operators:
\begin{align}
    \alpha_\llll(b)\widetilde{\M}_{\llll}(b)
    &\alpha_\llll(a)\widetilde{\M}_{\llll}(a)\nonumber\\
    &{=} \sum_{c,\mu}\sqrt{\frac{d_c}{d_ad_b}}\Phi_{c,\mu}^{ab}\alpha_\llll(c)\widetilde{\M}_{{\llll}}(c)
    \Phi_{ab}^{c,\mu},
    \label{eq:fusionM}
\end{align}
where $\Phi^{ab}_{c,\mu}$ and ${\Phi_{ab}^{c,\mu}}=\big(\Phi^{ab}_{c,\mu}\big)^\dagger$ are the fusion and splitting operators respectively.
This identity shows that the following two processes are equivalent: (1) moving two anyons $a$ and $b$ around the handle of a torus separately. (2) fusing $a$ and $b$, and moving the resulting anyon around the handle of a torus and then splitting the anyon back into $a$ and $b$.
The Wilson loop algebra \eqref{eq:fusionW1}--\eqref{eq:antiW1} follows from the fusion rules of the movement operators~\eqref{eq:fusionM} with further manipulation of the anyons.
The technical details can be found in Appendix~\ref{sec:Mfusion}.
Here, we briefly describe how to prove the fusion rule~\eqref{eq:fusionM}. We first cut an annular neighborhood of $\llll$, then close one of the punctures to get a disk with an anyon $d$ at the center. By doing this, the movement operators around the handle of the torus become the movement operators around the anyon $d$.
We argue that moving an anyon around $d$ is equivalent to moving $d$ around that anyon, up to quantifiable geometric phases. By cancelling these phases, we can get a set of solution to equation \eqref{eq:fusionM}.
This then implies that there exists a phase assignment such that the Wilson loop operators defined in \eqref{eq:defW} satisfy the Wilson loop algebra \eqref{eq:fusionW1}--\eqref{eq:antiW1}.

Closing the puncture of the annulus into an anyon $d$ is a key step in our approach for solving for $\alpha_\llll$.  However, the identification of $d$ is not unique, but is ambiguous by fusion with an Abelian anyon.
As such, the choice of phases for the Wilson loop operators are also not unique.
In Appendix \ref{sec:gaugetransformation} we show that the number of different sets of $\alpha_\llll(a)$ equals to the number of Abelian anyon in the system $|\mcA|$, and different choices of $\alpha_\llll(a)$ correspond to the gauge transformations of the Wilson loop operators.

\subsection{Dehn twists and Wilson loop algebra}\label{sec:basisW}

Recall that the Dehn twists generate all the simple loop classes on a torus from two classes $[\llll]$ and $[\mmm]$. 
Let's consider the Dehn twist acts on a close loop.
Suppose $\llll$ and $\mmm$ be two simple loops that only intersect once at $p$ (such that $\llll\times\mmm = -1$), the Dehn twist loop $\tau_\mmm(\llll)$ is given by: starting at $p$, traverse $\llll$ once returning to $p$, and then traverse $\mmm$.
With slight abuse of notation, we denote this combination as $\llll + \mmm$.
Thus the movement operators for $\tau_\mmm(\llll)$ comprise of the movement operators for $\llll$ and $\mmm$.
Perhaps unsurprisingly, the Wilson loop operators along $\tau_\mmm(\llll)$ can also be constructed from those along $\llll$ and $\mmm$---although this is a non-trivial task (at least for non-Abelian theories).
Here we only consider the case where $\llll\times\mmm = \pm1$; if $\llll,\mmm$ intersect more than once, the Dehn twist path becomes more complex but remains a combination of $\llll$ and $\mmm$.  For the purposes of this paper, it is sufficient to only consider the single intersection loops.

\begin{figure}
    \centering
    \includegraphics[width = \linewidth]{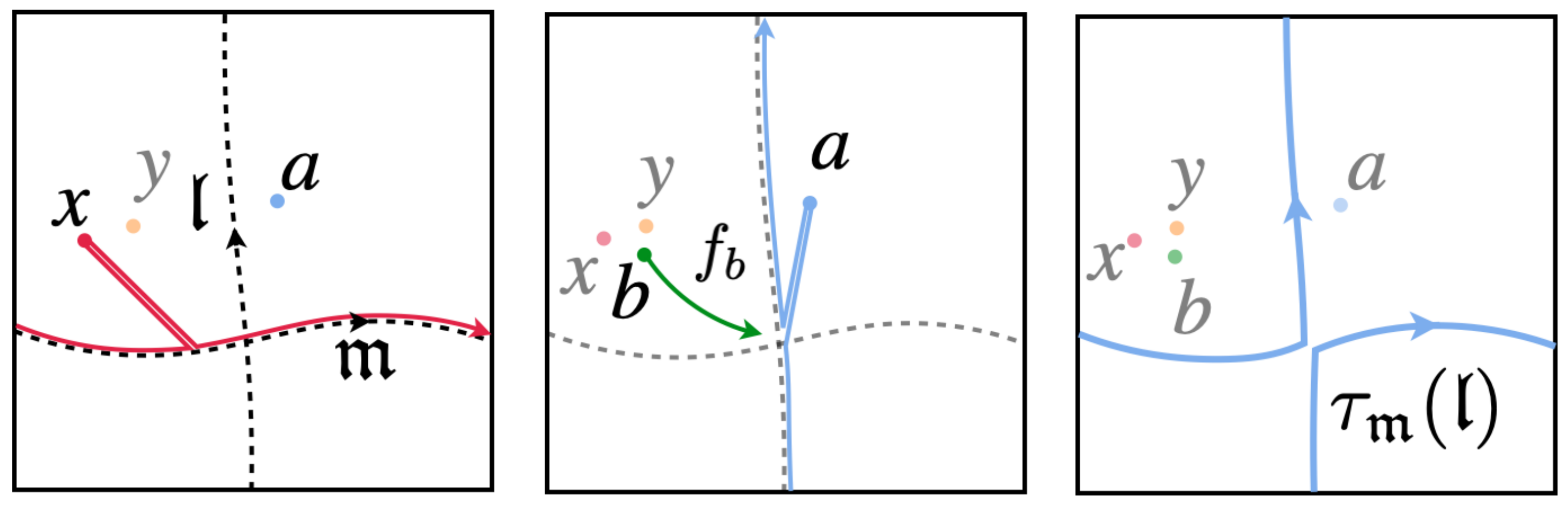}
    \caption{$\mmm$ and $\llll$ are two simple loops intersect at point $p$. $x$, $y$ and $a$ are anyons on the torus. The Wilson loop operator $W_\mmm(x)$ and $W_\llll(a)$ are showed in the first and second diagram respectively. $b$ is the anyon such that $N_{xy}^b >0$, and $f_b$ is the path that connects $b$ with the point $p$. The Dehn twist path $\tau_\mmm(\llll)$ is showed in the last diagram. }
    \label{fig:dehntwistsloop}
\end{figure}

In this subsection, we will show that this Dehn twist operator $\Dehn_{\mmm}$ defined by
\begin{empheq}[box=\widefbox]{align}
    \Dehn_{\mmm} &= \frac{1}{\tqd} \sum_x d_x\theta_x^*W_{\mmm}(x) ,\label{eq:DehnTwistOperator}
\end{empheq}
satisfies the relation
\begin{empheq}[box=\widefbox]{align}
    W_{\tau_{\mmm}(\llll)}(a) \label{eq:DehnTwist} &= \Dehn_{\mmm} W_{\llll}(a)\Dehn_{\mmm}^\dag.
\end{empheq}
(Recall that $\tqd = \sqrt{\sum_xd_x^2}$ is the total quantum dimension and  $\theta_x$ is the topological spin of anyon $x$.)
Formula \eqref{eq:DehnTwist} applies for any anyon.  If $a$ is Abelian, then the equation simplifies to
\begin{align}
    W_{\llll+\mmm}(a) &= \theta_a W_{\llll}(a) W_{\mmm}(a).
    \qquad \text{(for $a \in \mcA$)}
\end{align}
Figure \ref{fig:dehntwistsloop} shows the path taken by the anyons in the Wilson loops operators $W_\mmm$, $W_\llll$, and $W_{\tau_\mmm(\llll)}$.
Importantly, in our construction, the Dehn twist loop $\tau_\mmm(\llll)$ is the combination of the loop $\mmm$ and $\llll$, thus, we do not need to consider the geometric phase caused by combining the two loops.

The form of Dehn twist operator $\Dehn_\llll$ is motivated from the diagram
($\Theta$ is defined in Table \ref{tab:table1})
\begin{align}
    	\frac{1}{\tqd} \sum_a d_a\theta_a \;
		{\xy (0,-6);(0,-4)**\dir{-}; (0,-2);(0,6)**\dir{-}; (1.3,5.5)*{\scr x}; (1,0);(0,0)*{}="o",**\dir{},{\ellipse(6,3):a(5),=:a(-10){-}}; (4,-2.23)="ar"; {\ar@{>} "ar"+(-0.01,-.003);"ar"}; (5.5,-2.8)*{\scr a}; \endxy}
	\;&=\; \Theta
		{\xy (0,6)="t"; (0,-6)="b"; (1,0)="c"; "c"+(0,0)*{\phdot}="ch"; (5,0)="m"; "b"+(-1.3,1)*{\scr x};
			"b";"ch"**\crv{(0,-4) & (0,-1.5)}; "m"**\crv{"c"+(0.8,1.2) & (3.5,2.2) & "m"+(0,1)}; "c"**\crv{"m"+(0,-1) & (3.5,-2.2) & "c"+(0.8,-1.2)}; "t"**\crv{(0,1.5) & (0,4)}; \endxy} \;.
			\label{eq:Dehn}
\end{align}
This identity gives us a way to twist the world line and also proves that the Dehn twist operators we define are unitary.

\begin{figure}[tb]
    \centering
    \includegraphics[width=\linewidth]{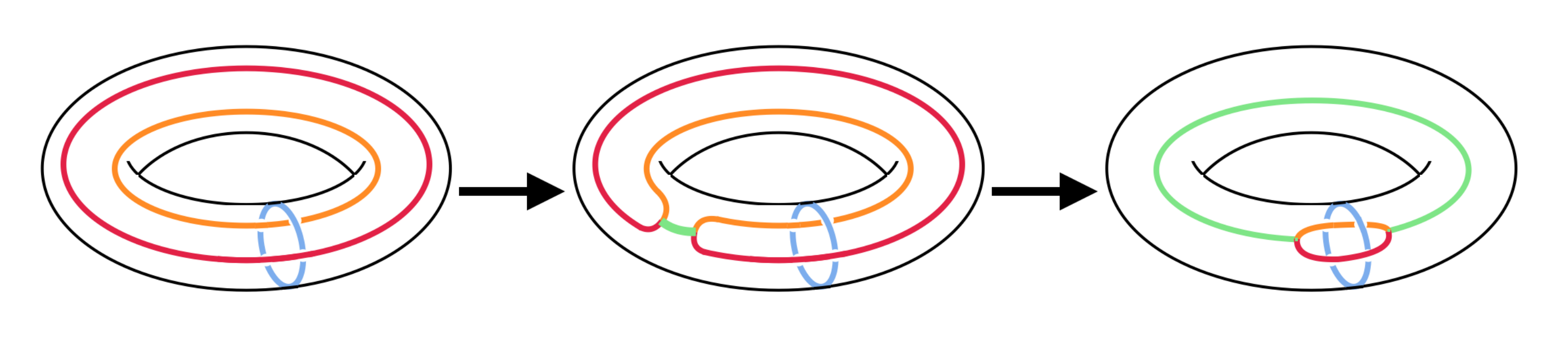}
    \caption{Since $\llll$ and $\mmm$ only intersect once, we can merge two $W_{\llll}$ and move the junction around the torus. The orange loop represent the anyon $y$, the red loop is anyon $x$, the blue loop is anyon $a$ and the green loop line is anyon $b$.}
    \label{fig:threading}
\end{figure}

The R.H.S. of equation \eqref{eq:DehnTwist} contains a product of 3-Wilson loop terms $W_{\mmm}(x)W_{\llll}(a)W_{\mmm}(y)$.
Figure~\ref{fig:threading} gives a summary of how we manipulate the 3-loop process. 
We first fuse $x$ and $y$ along $\mmm$ into a small segment $b$, we then move the fusion vertex around the torus along $\mmm$ to give the threading diagram on the right. We will show a more rigorous way to get this threading diagram based on the microscopic definition of the Wilson loop operator in Appendix \ref{sec:threading}.

\begin{figure}[b]
    \centering
    \includegraphics[width=\linewidth]{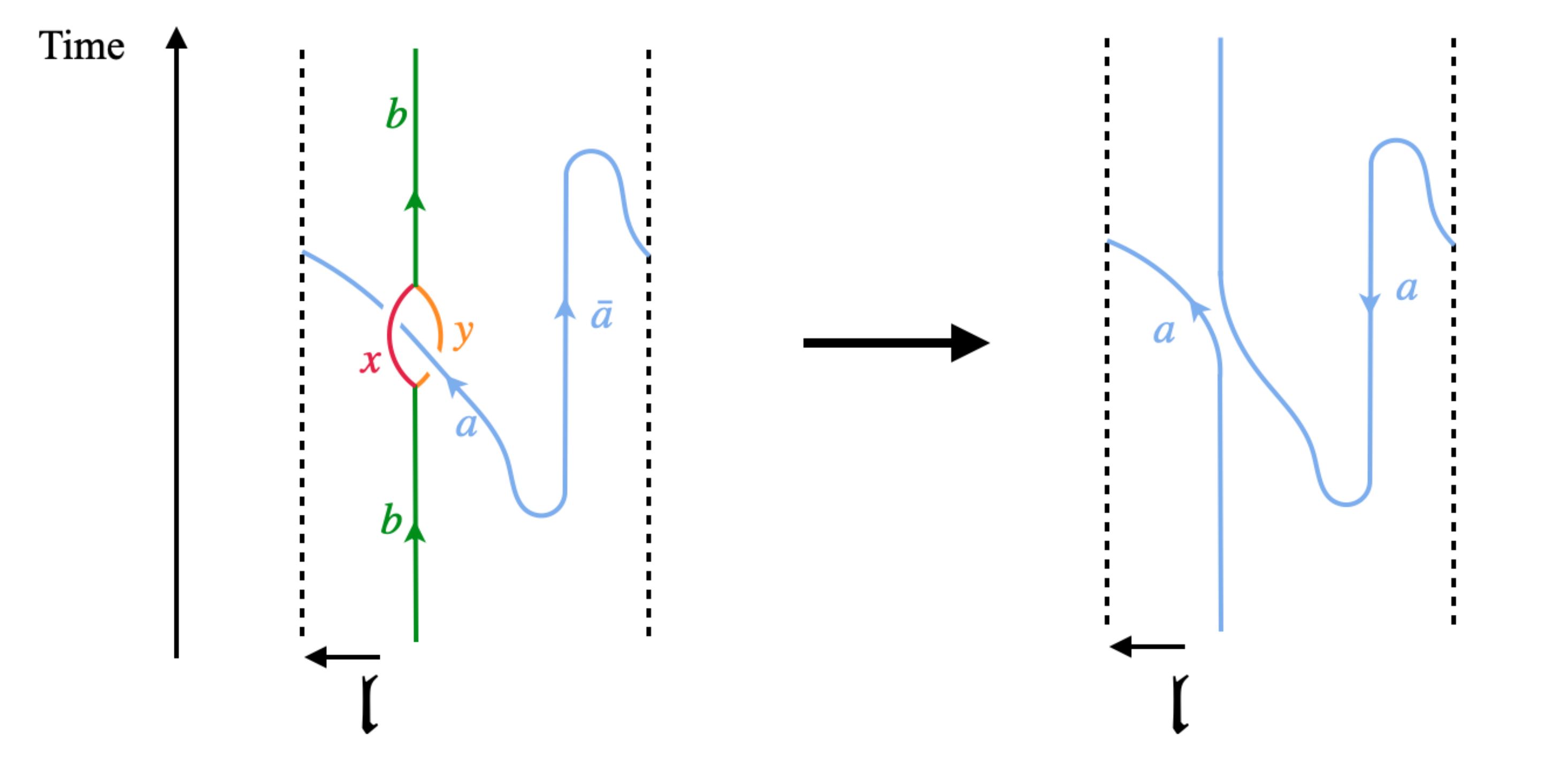}
    \caption{Here we draw the world line diagram of the result of equation \ref{l+m}. The dash line indicate the periodic boundary condition in $\llll$ direction. We can see that the threading diagram transform the Wilson loop operator $W_\llll(a)$ into a movement operator $\M_\llll(a)$. It seems we get an extra Schur indicator, but this indicator is cancelled by the phase attached to the Wilson loop operator.  }
    \label{fig:MovementU}
\end{figure}

Then, back to the equation \eqref{eq:DehnTwist}, on the R.H.S. we need sum over all possible  $x,y\in \mcC$ with the coefficients $d_xd_y\theta_x^* \theta_y/\tqd^2$. In terms of the interlocked diagram we have get the following diagrams
\begin{align}
	&\frac{1}{\tqd^2} \!\sum_{\substack{ x,y ,\mu}}\!\! \sqrt{ d_x d_y d_b} \, {\theta_y}{\theta^*_x}
	{\xy (0,-8)="b1"; (0,-4)="b2"; (0,4)="b3"; (0,8)="b4"; (6,-8)="a1"; (1.5,-2)="a2"; (-1.5,2)="a3"; (-6,8)="a4";
		"a2"*{\phdot}="a2h"; "a1"; "a3"*{\phdot}="a3h"**[blue]\dir{-}; "a4"**[blue]\dir{-};
		"b1";"b2"**[PineGreen]\dir{-}; "a3"**[red]\crv{(-3,0)}; "b3"**[red]\crv{(0,4)};   "b2"; "a2h"**[orange]\crv{(0,-4)}; "b3"**[orange]\crv{(3,0)}; "b4"**[PineGreen]\dir{-};
		"a1"+(1,1)*{\scr a}; "b1"+(-1,1)*{\scr b};  "b2"+(-3,3)*{\scr x}; "b3"+(3,-3)*{\scr y}; "b2"+(+1,-1.1)*{\sscr\mu}; "b3"+(1,1)*{\sscr\mu}; \endxy} \nonumber
		\\& =
		\frac{1}{\tqd^2} \!\sum_{\substack{ x,y \\ \mu\in  V_b^{xy} }}\!\! \sqrt{ d_x d_y d_b} \, {\theta_y}{\theta^*_x} \varkappa_a
		{\xy  (10,-10)="a1";(8,4)="a2";(2,4)="a3";(2,2)="a4";(2,-2)="a5";(2,-4)="a6";(-2,-7)="a7";(-10,10)="a8"; 
		(-2,-10)="b1";(-2,-4)="b2";(-2,-2)="b3";(-2,2)="b4";(-2,4)="b5";(-2,10)="b6";
		"a1";"a3"**[blue]\crv{(4,15)&(1.7,7)};"a4"**[blue]\dir{-};"a5"*{\phdot}="a5h"**[blue]\dir{-};"a6"**[blue]\dir{-};"a7"*{\phdot}="a7h"**[blue]\crv{(2,-7)};"a8"**[blue]\crv{(-6,-6.5)};
		"b1";"b2"**[PineGreen]\dir{-};"b3"**[PineGreen]\dir{-};"b4"**[orange]\dir{-};"b5"**[PineGreen]\dir{-};"b6"**[PineGreen]\dir{-};
		"b3";"a5"**[red]\crv{(0,-3)};"a4"*{\phdot}="a4h"**[red]\crv{(5,0)};"b4"**[red]\crv{(0,3)};
		"b1"+(1,0)*{\scr b};"b6"+(1,0)*{\scr b};
		"a1"+(-2,1.5)*{\scr a};"a8"+(2,-1.5)*{\scr a};
		"b3"+(-1,2)*{\scr x}; "b3"+(-1,0)*{\sscr \mu};"b4"+(-1,0)*{\sscr \mu};
		"a5"+(2.7,2)*{\scr y};
		\endxy} \nonumber
		\\
		&=
		\frac{\varkappa_a}{\tqd^2} \!\sum_{\substack{ x,y \\ \mu\in V_b^{xy} }}\!\! \sqrt{ d_x d_y d_b} \, \frac{\theta_y}{\theta_x} \left(R^{\bar{y}b}_x\right)_{\mu\nu}\left(R^{b\bar{y}}_x\right)_{\mu\nu}
		{\xy  (10,-10)="a1";(8,4)="a2";(2,4)="a3";(2,2)="a4";(2,-2)="a5";(2,-4)="a6";(-2,-7)="a7";(-10,10)="a8"; 
		(-2,-10)="b1";(-2,-4)="b2";(-2,-2)="b3";(-2,2)="b4";(-2,4)="b5";(-2,10)="b6";
		"a1";"a3"**[blue]\crv{(4,15)&(1.7,7)};"a4"**[blue]\dir{-};"a5"**[blue]\dir{-};"a6"*{\phdot}="a6h"**[blue]\dir{-};"a7"*{\phdot}="a7h"**[blue]\crv{(2,-7)};"a8"**[blue]\crv{(-6,-6.5)};
		"b1";"b2"*{\phdot}="b2h"**[PineGreen]\dir{-};"b3"**[PineGreen]\dir{-};"b4"**[orange]\dir{-};"b5"**[PineGreen]\dir{-};"b6"**[PineGreen]\dir{-};
		"b3";"b2"**[red]\crv{(-6,-3)};"a6"**[red]\crv{(0,-4.6)};"a3"*{\phdot}="a3h"**[red]\crv{(6,-3)&(6,3)};"b5"*{\phdot}="b5h"**[red]\crv{(0,4.6)};"b4"**[red]\crv{(-6,3)};
		"b1"+(-1,0)*{\scr b};"b6"+(-1,0)*{\scr b};
		"a1"+(-2,1.5)*{\scr a};"a8"+(2,-1.5)*{\scr a};
		"b3"+(-1,2)*{\scr x}; 
		"b3"+(1,0)*{\sscr \nu};"b4"+(1,0)*{\sscr \nu};
		"a5"+(1.7,2)*{\scr y};
		\endxy} \nonumber
		\\
		&=
		\frac{1}{\tqd^2} \! \sum_{\substack{ x,y \\ \mu,\nu\in  V_b^{xy} }}\!\! \sqrt{ d_x d_y d_b} \frac{\theta_y}{\theta_x} \frac{\theta_x}{\theta_y\theta_b}\delta_{\mu\nu}\varkappa_a
		{\xy  (10,-10)="a1";(8,4)="a2";(2,4)="a3";(2,2)="a4";(2,-2)="a5";(2,-4)="a6";(-2,-7)="a7";(-10,10)="a8"; 
		(-2,-10)="b1";(-2,-4)="b2";(-2,-2)="b3";(-2,2)="b4";(-2,4)="b5";(-2,10)="b6";
		"a1";"a3"**[blue]\crv{(4,15)&(1.7,7)};"a4"**[blue]\dir{-};"a5"**[blue]\dir{-};"a6"*{\phdot}="a6h"**[blue]\dir{-};"a7"*{\phdot}="a7h"**[blue]\crv{(2,-7)};"a8"**[blue]\crv{(-6,-6.5)};
		"b1";"b2"*{\phdot}="b2h"**[PineGreen]\dir{-};"b3"**[PineGreen]\dir{-};"b4"**[orange]\dir{-};"b5"**[PineGreen]\dir{-};"b6"**[PineGreen]\dir{-};
		"b3";"b2"**[red]\crv{(-6,-3)};"a6"**[red]\crv{(0,-4.6)};"a3"*{\phdot}="a3h"**[red]\crv{(6,-3)&(6,3)};"b5"*{\phdot}="b5h"**[red]\crv{(0,4.6)};"b4"**[red]\crv{(-6,3)};
		"b1"+(1,1.5)*{\scr b};"b6"+(1,-1.5)*{\scr b};
		"a1"+(-2,1.5)*{\scr a};"a8"+(2,-1.5)*{\scr a};
		"b3"+(-1,2)*{\scr x}; 
		"b3"+(1,0)*{\sscr \nu};"b4"+(1,0)*{\sscr \nu};
		"a5"+(1.7,2)*{\scr y};
		\endxy} \nonumber
		\\
		&=
		\frac{1}{\tqd^2} \ \sum_{y}\ d_y d_b \, \theta^*_b\varkappa_a
		{\xy  (10,-10)="a1";(8,4)="a2";(2,4)="a3";(2,2)="a4";(2,-2)="a5";(2,-4)="a6";(-2,-7)="a7";(-10,10)="a8"; 
		(-2,-10)="b1";(-2,-4)="b2";(-2,-2)="b3";(-2,2)="b4";(-2,4)="b5";(-2,10)="b6";
		"a1";"a3"**[blue]\crv{(4,15)&(1.7,7)};"a4"**[blue]\dir{-};"a5"*{\phdot}="a5h"**[blue]\dir{-};"a6"**[blue]\dir{-};"a7"*{\phdot}="a7h"**[blue]\crv{(2,-7)};"a8"**[blue]\crv{(-6,-6.5)};
		"b1";"b2"**[PineGreen]\dir{-};"b3"*{\phdot}="b3h"**[PineGreen]\dir{-};"b4"**[PineGreen]\dir{-};"b5"**[PineGreen]\dir{-};"b6"**[PineGreen]\dir{-};
		"b3";"a5"**[red]\crv{(0,-2.5)};"a4"*{\phdot}="a4h"**[red]\crv{(7,0)};"b4"*{\phdot}="b4h"**[red]\crv{(0,2.5)};"b3"**[red]\crv{(-7,0)};
		"b1"+(1,0)*{\scr b};"b6"+(1,0)*{\scr b};
		"a1"+(-2,1.5)*{\scr a};"a8"+(2,-1.5)*{\scr a};
		"a5"+(3,-0.5)*{\scr y};
		\endxy} \nonumber
		\\
		&= \theta^*_b \delta_{ab}\varkappa_a
		{\xy  (10,-10)="a1";(8,4)="a2";(2,4)="a3";(2,2)="a4";(2,-2)="a5";(2,-4)="a6";(-2,-7)="a7";(-10,10)="a8"; 
		(-2,-10)="b1";(-2,-4)="b2";(-2,-2)="b3";(-2,2)="b4";(-2,4)="b5";(-2,10)="b6";
		"a1";"a3"**[blue]\crv{(6,10)&(2,7)};"b5"**[blue]\crv{(2,0)&(-2,0)};
		"a6";"a7"*{\phdot}="a7h"**[blue]\crv{(2,-7)};"a8"**[blue]\crv{(-6,-6.5)};
		"b1";"b2"**[blue]\dir{-};"a6"**[blue]\crv{(-2,0)&(2,0)};
		"b5";"b6"**[blue]\dir{-};
		"b1"+(-1,1.5)*{\scr b};"b6"+(-1,-1.5)*{\scr b};
		"a1"+(-2,1.5)*{\scr a};"a8"+(2,-1.5)*{\scr a};
		\endxy} \nonumber
		\\
		&=
		{\delta_{ab}} 
		{\xy  (8,-10)="a1";(-8,10)="a8"; 
		(0,-10)="b1";(0,10)="b6";
		"a1";"b6"**[blue]\crv{(-1,0)};
		"b1";"a8"**[blue]\crv{(1,0)};
		(-3,0)*{\scr a};(3,0)*{\scr a};
		\endxy}
		\label{l+m}
\end{align}
Figure \ref{fig:MovementU} demonstrates that if we put this result back to the torus (Fig.~\ref{fig:threading} right),
the movement operator along the $\llll$ direction becomes the movement operator along the $\llll+\mmm$ direction.
Thus we obtain the Wilson loop operators along $\llll+\mmm$ direction, which proves equation \eqref{eq:DehnTwist}.
[Equations \eqref{eq:DehnTwistOperator}, \eqref{eq:DehnTwist} are also applicable to higher genus manifolds.]

More generally, for an arbitrary loops $\sss$, $\uuu$ that may have any number of intersections,
\begin{align}
    	\Dehn_{\sss} W_{\uuu}(a)\Dehn_{\sss }^\dagger =  W_{(\sss\times\uuu)\sss+\uuu}(a)=W_{\tau_{\sss }(\uuu)}(a).
    	\label{eq:DehnTwist1}
\end{align}
although we will not show this explicitly.
Equation~\eqref{eq:DehnTwist1} demonstrates that the operator $\Dehn_{\mmm }$ is indeed the Dehn twist operator on the torus.

So far, we established the Dehn twist actions on the Wilson loop operators.
Since the Wilson loop operators only depend on the homotopy class of the loop, we have thus shown that every Wilson loop operator on a torus can be generated from 2 sets of Wilson loop operators $W_\llll(a)$ and $W_\mmm(a)$ for a coordinate basis $(\llll,\mmm)$, along with the knowledge of the topological spins $\theta_a$.

\subsection{Basis of Ground States}\label{sec:defMES}
In this subsection, we will define the ``standard basis'' of ground states $\ket{b^\llll_\mmm}$.
For an arbitrary coordinate basis ($\llll$,$\mmm$), we will explicitly construct these wavefunctions
such that they obey the relations
\begin{empheq}[box=\widefbox]{align}
\label{eq:MESm}
   W_\mmm(a)\ket{b^\llll_\mmm} & = \frac{S^*_{a,b}}{S_{0,b}}\ket{b^\llll_\mmm},  \\
    W_\llll(a)\ket{b^\llll_\mmm} &= \sum_cN_{ab}^c\ket{c^\llll_\mmm}. \label{eq:MESl}
   \end{empheq}
The first equation shows that the basis are defined to be the eigenstates of the Wilson loop operators $W_\mmm(a)$. These eigenstates are also called the minimum entanglement state (MES), because their bipartite entanglement for a pair of cuts parallel to $\mmm$ is a local minimum. The second equation fixed the relative phase between the MESs. We see that the $\llll$ direction is important in fixing the phase.  For example, the set $\ket{a^\llll_\mmm}$ is different from the set $\ket{a^{\mmm+\llll}_\mmm}$, although both are MESs along an $\mmm$-cut.

Now we will show that there exists such states, and these states are unique up to an overall phase.
Because the Wilson loop operators along the same direction commute with each other, they can be diagonalized at the same time.
Let $\ket{X}$ be a common eigenstate of $\{W_\mmm(a)\}$. Then we have
\begin{align}
      &W_\mmm(a)W_\mmm(b)\ket{X} = \varphi(a)\varphi(b)\ket{X} \nonumber \\
      = &\sum_c N_{ab}^{c} W_\mmm(c)\ket{X} = \sum_c \varphi(c) N_{ab}^{c} \ket{X},
\end{align}
where $a,b,c$ are anyons and $\varphi(a),\varphi(b),\varphi(c)$ are the eigenvalues for the corresponding Wilson loop operators. So the eigenvalues form a fusion character
\begin{align}
    \varphi(a)\varphi(b)= \sum_c N_{ab}^{c} \varphi(c) .
\end{align}
Therefore, the eigenvalues of state $\ket{X}$ can be obtained from the modular $S$-matrix
\begin{align}
    \varphi(a) = \frac{S_{a,x}^*}{S_{0,x}},
\end{align}
where $x$ is an anyon. 
Let $\ket{0_\mmm}$ be the special state whose eigenvalues for all the Wilson loop operators are positive
\begin{align}
       W_\mmm(a)\ket{0_\mmm} = d_a\ket{0_\mmm} .
\end{align}
For a given set of Wilson loop operators, the state $\ket{0_\mmm}$ is unique up to a phase. 
We then define
\begin{align}
    \ket{b^\llll_\mmm}= W_\llll(b)\ket{0_\mmm} .
\end{align}  
We will now show that $\ket{b^\llll_\mmm}$ is also an eigenstate of $W_{\mmm}(a)$ satisfies Eqs.~\eqref{eq:MESm} and \eqref{eq:MESl}.

To show the first point, we need to compute the term $W_{\mmm}(a)\ket{b^\llll_\mmm} = W_{\mmm}(a) W_{\llll}(b) \ket{0_\mmm}$.  In order to use use the same trick in figure \ref{fig:threading}, we rewrite $\ket{0_\mmm} = \frac{1}{\tqd^2} \sum_c W_\mmm(c) d_c \ket{0_\mmm}$. Then the term with three Wilson loop operators can be evaluated by the following threading equation
\begin{align}
&    	\sum_{c,\mu} \sqrt{ \frac{d_d }{d_ad_c}} d_c
{\xy (0,-8)="b1"; (0,-4)="b2"; (0,4)="b3"; (0,8)="b4"; (6,-8)="a1"; (1.5,-2)="a2"; (-1.5,2)="a3"; (-6,8)="a4";
		"a2"*{\phdot}="a2h"; "a1"; "a3"*{\phdot}="a3h"**[blue]\dir{-}; "a4"**[blue]\dir{-};
		"b1";"b2"**[PineGreen]\dir{-}; "a3"**[red]\crv{(-3,0)}; "b3"**[red]\crv{(0,4)};   "b2"; "a2h"**[orange]\crv{(0,-4)}; "b3"**[orange]\crv{(3,0)}; "b4"**[PineGreen]\dir{-};
		"a1"+(1,1)*{\scr b}; "b1"+(-1,1)*{\scr d};  "b2"+(-3,3)*{\scr a}; "b3"+(3,-3)*{\scr c}; "b2"+(+1,-1.1)*{\sscr\mu}; "b3"+(1,1)*{\sscr\mu}; \endxy} 
\; = \sum_{c,\mu} \sqrt{ \frac{ d_c}{d_dd_a}}d_d 
         {\xy 
        (-2,-8)="d1";(-2,-2)="d2";(-2,2)="d3";(-2,6)="d4";(-2,8)="d5"; (6,-8)="b1";(2,-4)="b2";(2,-2)="b3";(2,2)="b4";(-6,8)="b5";
        "d1";"d2"**[PineGreen]\dir{-};"d3"**[orange]\dir{-};"d4"*{\phdot}**[PineGreen]\dir{-};"d5"**[PineGreen]\dir{-};
        "b1";"b2"**[blue]\crv{(2,-6)};"b3"*{\phdot}**[blue]\dir{-};"b4"**[blue]\dir{-};"b5"**[blue]\crv{(2,4.5)};
        "d2";"b3"**[red]\crv{(0,-3)};"b4"*{\phdot}**[red]\crv{(5,0)};"d3"**[red]\crv{(0,3)};
        (5.5,0)*{\scr a};(-1,0)*{\scr c};(-3.5,-6)*{\scr d};(-3,2)*{\scr \mu};(-3,-2)*{\scr \mu};(6,-6)*{\scr b};
        \endxy}
		\:  \;\;\nonumber \\
	&\quad= d_d  \ {\xy 
        (-2,-8)="d1";(-2,-2)="d2";(-2,2)="d3";(-2,6)="d4";(-2,8)="d5"; (6,-8)="b1";(2,-4)="b2";(2,-2)="b3";(2,2)="b4";(-6,8)="b5";
        "d1";"d2"**[PineGreen]\dir{-};"d4"*{\phdot}**[PineGreen]\dir{-};"d5"**[PineGreen]\dir{-};
        "b1";"b2"**[blue]\crv{(2,-6)};"b3"*{\phdot}**[blue]\dir{-};"b4"**[blue]\dir{-};"b5"**[blue]\crv{(2,4.5)};
        "b4"*{\phdot};"b3"**[red]\crv{(-2,2)&(-2,-2)};"b4"*{\phdot}**[red]\crv{(6,-2)&(6,2)};
        (6.5,0)*{\scr a};(-3.5,-6)*{\scr d};(6,-6)*{\scr b};
        \endxy}\:  \;\;  = d_d \frac{S^*_{a,b}}{S_{0,b}}
	{\xy (0,-8)="d1"; (0,-1)="d2u"; (0,0)="d2";(0,1)="d3"; (0,8)="d7";
	(6,-8)="b1";(-6,8)="b7";
	"d1";"d2u"**[PineGreen]\dir{-};
	"d3";"d7"**[PineGreen]\dir{-};
	"b1";"d2"**[blue]\dir{-};"b7"**[blue]\dir{-};	
	"d1"+(-2,1)*{\scr d}; "b1"+(1,1)*{\scr b};
	\endxy}\:  \;\;  .
	\label{lm}
\end{align}
Plug this result back into figure \ref{fig:threading} we get
\begin{align}
\begin{aligned}
    \sum_cW_\mmm(a)W_\llll(b)&W_\mmm(c)d_c \\
    &= \frac{S^*_{a,b}}{S_{0,b}}\sum_dW_\llll(b)W_\mmm(d) \, d_d \label{eq:www}.
\end{aligned}
\end{align}
Now, we can apply both sides on the state $\ket{0_\mmm}$. The LHS equals to 
\begin{align}
\begin{aligned}
    \sum_cW_\mmm(a)W_\llll(b)W_\mmm(c)&d_c \ket{0^\llll_\mmm} \\
    &= W_{\mmm}(a)W_\llll(b) \ket{0_\mmm}.
\end{aligned}
\end{align}
While the RHS becomes
\begin{align}
     \frac{S^*_{a,b}}{S_{0,b}} \sum_dW_\llll(b)W_{\mmm}(d)d_d\ket{0^\llll_\mmm} = \frac{S^*_{a,b}}{S_{0,b}} W_\llll(b) \ket{0_\mmm}.
\end{align}
So $W_\llll(b)\ket{0^\llll_\mmm}=\ket{b^\llll_\mmm}$ is an eigenstate of $W_\mmm(a)$ with eigenvalue $S^*_{ab}/S_{0b}$. And from the Wilson loop algebra we have,
\begin{align}
\begin{aligned}
    W_\llll(a) \ket{b^\llll_\mmm} &= W_\llll(a)W_\llll(b)\ket{0^\llll_\mmm}\\
    &= \sum_cW_\llll(c)\ket{0^\llll_\mmm} = \sum_c\ket{c^\llll_\mmm}.
\end{aligned}
\end{align}
Therefore, $\ket{b^\llll_\mmm}$ form a basis of ground states and satisfy equation \eqref{eq:MESm} and \eqref{eq:MESl}.

\subsection{Modular transformation}\label{sec:MT}
Now we will consider the relationship between the basis in different coordinates.
It is believed that overlaps between the different standard bases give us the modular $S$- and $T$-matrix, which together form a projective representation of the modular group~\cite{wen1990TopologicalOrder,YiZhang2012GroundStateEntanglement}.
Here we proof this with the Wilson loop algebra established earlier.

There are two contexts for which the modular group appears.
The first point of view (i.e., the ``passive transformation'' approach) is to consider the possible transformations of the coordinate basis $(\llll$,$\mmm) \to (\llll',\mmm')$, and the relations between the ground state bases.
Alternately, the operator algebra between Dehn twists in various directions results in a projective representation of the modular group.  For $\llll \times \mmm = -1$, their Dehn twists $\Dehn_\llll$ and $\Dehn_\mmm$ corresponds to $u = sts^{-1}$ and $t$ of the modular group; their actions on (in some specific) ground state basis will give us the modular matrices.


We first take the first point of view; studying passives transformation on the ground state space.
Consider the possible transformation of the coordinate basis.
Recall that a pair of loops $(\llll,\mmm)$ forms a basis if $\llll \times \mmm = -1$, i.e., they only intersect once.
Therefore, an admissible transformation of the coordinate basis $(\llll,\mmm)$ to $(\llll',\mmm')$ can be characterized by a $2\times 2$ matrices with integer coefficients and determinant 1. (The determinant needs to be positive because we only consider the transformation preserve the orientation.)
Hence they form the special linear group $\mathrm{SL}(2,\mathbb{Z})$, isomorphic to the modular group.

More precisely, let $P$ be an arbitrary element in the modular group represented by the matrix in $\mathrm{SL}(2,\mathbb{Z})$.  Its action on a coordinate basis is
\begin{align}
    P = \begin{pmatrix} a & b \\
    c & d
    \end{pmatrix}: \ \ (\llll,\mmm) \mapsto (a\llll+b\mmm, c\llll+d\mmm),
\end{align}
with $ad-bc=1$. We noticed that the passive transformation of the ground states is directly related to the modular transformation in conformal field theory: 
\begin{align}
    P = \begin{pmatrix} a & b \\
    c & d
    \end{pmatrix}:\ \ \tau \mapsto \frac{a\tau+b}{c\tau+d}.
\end{align}

The overlap matrix for the transformation is defined as $ \braket{ x^\llll_\mmm | y^{a\llll+b\mmm}_{c\llll+d\mmm} }$.
As the modular group is generated by two elements, it suffices to compute the overlap matrices corresponding to the $s$- and $t$- transformations:
\begin{align} \begin{aligned}
    s &= \begin{pmatrix} 0 & -1 \\ 1 & 0 \end{pmatrix} &&:& (\llll,\mmm) &\mapsto (-\mmm,\llll),
\\   t &= \begin{pmatrix} 1 & 1 \\ 0 & 1 \end{pmatrix} &&:& (\llll,\mmm) &\mapsto (\llll+\mmm,\mmm).
\end{aligned} \end{align}
In the remainder of this section, we will show that their overlap matrices are indeed the same as the modular matrices defined from TQFT:
\begin{empheq}[box=\widefbox]{align}
   &\braket{a^\llll_\mmm|b^{-\mmm}_\llll} \propto  S_{a,b},
  &\braket{a^\llll_\mmm|b^{\llll+\mmm}_\mmm }\propto T_{a,b} = \delta_{ab}\theta_a.&
   \end{empheq}

The states $\ket{b^{-\mmm}_\llll}$ are the eigenstates of $W_\llll(a)$, and their relative phases are fixed by $W_{-\mmm}(a)$. These states can be obtained by repeating the calculation in section \ref{sec:defMES}. We modify equation \eqref{lm} by choosing the blue line to be the $-\mmm$ direction and green line to be the $\llll$ direction (the orientation of the torus does not change). Then we have
\begin{align}
\begin{aligned}
        \sum_cW_\llll(a)W_{-\mmm}(b)&W_\llll(c)d_c \\
        &= \frac{S^*_{ab}}{S_{0b}}\sum_cW_{-\mmm}(b)W_\llll(c)d_c.
\end{aligned}
\end{align}
If we apply both sides on the state $\ket{0^\llll_\mmm}$, then the LHS becomes
\begin{align}
\begin{aligned}
    \sum_cW_\llll(a)W_{-\mmm}(b)&W_\llll(c)d_c\ket{0_\mmm^\llll} 
    \\&= W_\llll(a)\sum_c d_cW_{-\mmm}(b) \ket{c^\llll_\mmm}. 
\end{aligned}
\end{align}
And the RHS is 
\begin{align}
\begin{aligned}
    \frac{S^*_{ab}}{S_{0b}}\sum_cW_{-\mmm}(b)&W_\llll(c)d_c \ket{0_\mmm^\llll} \\
    &=  \frac{S^*_{ab}}{S_{0b}}\sum_cd_cW_{-\mmm}(b)\ket{c^\llll_\mmm}.
\end{aligned}
\end{align}
So, $\sum_cd_cW_{-\mmm}(b)\ket{c^\llll_\mmm}$ is the eigenstate of $W_\llll(a)$ with eigenvalue $S^*_{ab}/S_{0b}$. It is easy to check that the relative phases between these states are fixed by $W_{-\mmm}(a)$:
\begin{align}
    W_{-\mmm}(a)\sum_c \frac{d_c}{\tqd} W_{-\mmm}(b)\ket{c^\llll_\mmm}
    &= \sum_{c,e}N_{ab}^e\frac{d_c}{\tqd}W_{-\mmm}(e)\ket{c^\llll_\mmm}\nonumber\\
    &=\sum_{e}N_{ab}^e\ket{e^{-\mmm}_\llll}.
\end{align}
Thus, we can write
\begin{align}
    \boxed{
    \ket{b^{-\mmm}_\llll} =\kappa_1 \sum_c\frac{d_c}{\tqd}W_{-\mmm}(b)\ket{c^\llll_\mmm}= \kappa_1 \sum_cS_{b,c}\ket{c^\llll_\mmm}
    ,}
\end{align}
where $\kappa_1$ is an overall phase.
Therefore, the generator $s$ give us the modular $S$-matrix $\braket{a^\llll_\mmm|b^{-\mmm}_\llll} = \kappa_1 S_{ab}$.

For the generator $t$, we need to define the state $\ket{b^{\llll+\mmm}_\mmm}$. Notice that this state is also the eigenstate of $W_{\mmm}(a)$, so it differs from the state $\ket{b^{\llll}_\mmm}$ by a phase factor $\ket{b^{\llll+\mmm}_\mmm} = e^{i\phi(b)} \ket{b^{\llll}_\mmm}$.
The phase factor here can be determined by comparing the Wilson loop operators $W_{\llll+\mmm}(a)$ to $W_{\llll}(a)$.
From section \ref{sec:basisW}, we have $W_{\llll+\mmm}(a) = \Dehn_\mmm W_\llll(a)\Dehn_\mmm^{-1}$, and $W_\mmm(a) = \Dehn_\mmm W_\mmm(a)\Dehn_\mmm^{-1}$,
    that is, the set of operators $(W_{\llll+\mmm},W_\mmm)$ can be written as a similarity transformation of $(W_{\llll},W_\mmm)$.
Hence the set of states $\Dehn_\mmm\ket{b^\llll_\mmm}$ form a basis for the coordinate system $(\llll,\mmm)$.
Therefore, we can denote $\ket{b^{\llll+\mmm}_\mmm} = \kappa_2 \Dehn_\mmm\ket{b^{\llll}_\mmm}$ for some phase $\kappa_2$.
Finally we have
\begin{align}
    \ket{b^{\llll+\mmm}_\mmm} &=  \kappa_2\Dehn_\mmm\ket{b^{\llll}_\mmm} \nonumber \\ &= \kappa_2\sum_x\frac{d_x\theta_x^*}{\tqd}W_\mmm(x)\ket{b^{\llll}_\mmm} \nonumber\\
    &=  \kappa_2\sum_x\frac{d_x\theta_x^*}{\tqd}\frac{S^*_{x,b}}{S_{0,b}}\ket{b^{\llll}_\mmm} \nonumber\\
    & =  \kappa_2\Theta^* \theta_b\ket{b_\mmm^\llll}
\end{align}
where the last step is from the identity \eqref{eq:Dehn}.
So the generator $T$ give us the modular $T$-matrix $\braket{a^{\llll}_\mmm|b^{\llll+\mmm}_\mmm} = \delta_{ab}\kappa_2\theta_a$.

In the second viewpoint, we consider the Dehn twist operators as active transformations in the ground state space.
In contrast to the first viewpoint, we analysis the Dehn twist operators as an algebra. In section \ref{sec:basisW} we have constructed two generators of Dehn twist operators,
\begin{align}
    &\Dehn_{\llll} = \sum_x \frac{\theta_x^*d_x}{\tqd} W_\llll(x),&\Dehn_{\mmm} =  \sum_x \frac{\theta_x^*d_x}{\tqd} W_\mmm(x)&.
\end{align}
Their actions on the coordinate basis are given by, 
\begin{align} \begin{aligned}
    \Dehn_{\llll} :\; & \llll \mapsto \llll ,
&&&   \Dehn_{\mmm} :\; & \llll \mapsto \llll+\mmm ,
\\ & \mmm \mapsto \mmm-\llll ,
&&& & \mmm \mapsto \mmm .
\end{aligned} \end{align}
This is a linear representation with
\begin{align}
    \Dehn_\llll \mapsto u &= \begin{pmatrix} 1 & 0 \\ -1 & 1 \end{pmatrix} ,
&  \Dehn_\mmm \mapsto t &= \begin{pmatrix} 1 & 1 \\ 0 & 1 \end{pmatrix}.
\end{align}
Together $u$ and $t$ form the modular group with presentation $utu=tut$ and $(tu)^6 = 1$.

The actions of these two generators on the ground state space is similar, but with additional phases.
In particular, the matrix elements in the standard basis $\ket{a^\llll_\mmm}$ are,
\begin{align}
&\begin{aligned}    \braket{a^\llll_\mmm|\Dehn_\llll| b^ \llll_\mmm}& = \braket{a^\llll_\mmm|\sum_x \frac{\theta_x^*d_x}{\tqd} W_\llll(x)| b^ \llll_\mmm} \\
   &=\braket{a^\llll_\mmm|\sum_{e,x} \frac{\theta_x^*d_xN_{xb}^e}{\tqd} | e^ \llll_\mmm} \\
   & = \braket{a^\llll_\mmm|\sum_{e} S_{\bar{b},e} \theta_b^*\theta_e^* | e^ \llll_\mmm} \\
   &= \theta_a^*S_{a,b}^*\theta_b^*.
   \label{eq:DehnL}
\end{aligned}\\
   &\! \! \!  \braket{a^\llll_\mmm|\Dehn_\mmm| b^ \llll_\mmm} = \Theta^*\theta_b\delta_{ab}.\label{eq:DehnM}
\end{align}
To derive the first equation we used the identity Eq.\eqref{eq:SandTheta}: $S_{ab} = \frac{1}{\tqd} \sum_c N_{ab}^c\frac{\theta_a\theta_b}{\theta_c}$.
Here we see that the actions of the $\big( \Dehn_\llll \Dehn_\mmm \Dehn_\llll \big)^\dag$ and $\Dehn_\mmm$ give us the modular $S$- and $T$-matrices respectively.

The Dehn twist operator thus satisfies
\begin{align}
    \Dehn_{\mmm}\Dehn_{\llll}\Dehn_{\mmm} &=  \Dehn_{\llll}\Dehn_{\mmm}\Dehn_{\llll} ,
&   \big( \Dehn_\mmm \Dehn_\llll \big)^6 &= \Theta^{-8} \mathbbm{1}.
\end{align}
We find that the algebra of Dehn twist operators from a projective representation of the modular group.
(This algebra remains valid for a general oriented manifold.)

For an arbitrary modular transformation of the torus, the two points of view give the same overlap matrices although their calculations differ in details.
For example, let's consider the 3-fold rotation of the torus $\R$. Let $\llll$ and $\mmm$ be $120^\circ$ apart with equal length.
The coordinate basis transform under the rotation $\R$ in the following way:
\begin{align}
    \R: \begin{pmatrix}
    \llll,
    \mmm
    \end{pmatrix}  \mapsto \begin{pmatrix}
    -\llll-\mmm,
    \llll
    \end{pmatrix}.
\end{align}
Here we assume that the standard basis transforms in the same way (without phases factors or label permutation): $\R \ket{b_\mmm^\llll} = \ket{b^{-\llll-\mmm}_\llll}$.
Then from the passive point of view, the overlap between the basis is given by
\begin{align}
    \braket{a^{\llll}_\mmm|b^{-\llll-\mmm}_\llll} &= \sum_c \braket{a^{\llll}_\mmm|c^{-\mmm}_\llll}\braket{c^{-\mmm}_\llll|b^{-\llll-\mmm}_\llll} \nonumber \\
    &= \sum_c S_{ac}T^*_{cb}=\left(ST^*\right)_{ab}.
\end{align}
The active point of view, this transition is associated with a element in the modular group,
\begin{align}
    \R \to
    \begin{pmatrix} 0 & -1 \\ 1 & -1 \end{pmatrix}
    = \begin{pmatrix} 0 & -1 \\ 1 & 0 \end{pmatrix} \begin{pmatrix} 1 & 1 \\ 0 & 1 \end{pmatrix}^{-1}
    = st^{-1}.
\end{align}
We again find the overlap is also given by 
\begin{align}
    \braket{a^{\llll}_\mmm|\mathcal{R}|b^{\llll}_\mmm} = \left(ST^*\right)_{ab}.\label{eq:3-foldrotation}
\end{align}
In Sec \ref{sec:3-fold}, we will reexamine the 3-fold rotation in greater details relaxing the phase/permutation assumption made here.


\section{Diagnosing topological order from minimum entangled states}\label{sec:globalsymmetry}

We have so far shown that, given the TQFT of a model on the torus, we can construct the Wilson loop operators and write down the bases of the ground states consistent with the modular data of the TQFT.
Here in this section we tackle the opposite question: how can we detect the topological order given only the ground states, and are there TQFTs cannot be distinguished from the overlap of the ground states alone? 

We begin this section with the labeling of general ground states. Let $\mcC$ be a TQFT model on the torus, $(\llll,\mmm)$ be an arbitrary coordinate basis, $\ket{n_{\mmm}}$ be the ground states with the local minimum bipartite entanglement entropy along the $\mmm$-cut,
and $0\leq n\leq |\mcC|-1$ some arbitrary ordering of the states.
These ground states
are related to the standard basis $\ket{a^\llll_{\mmm}}$ (consistent with $\mcC$) defined in section \ref{sec:defMES},\ref{sec:MT} via
\begin{align}
    \ket{n_{\mmm}} = \gamma_n\ket{{a(n)}^\llll_{\mmm}},\label{eq:labeling}
\end{align}
where $a(n)$ is an anyon in $\mcC$ and $\gamma_n$ are arbitrary phases dependent on $\llll$ and $\mmm$.
We demand that $\ket{0_\mmm}$ corresponds to an Abelian anyon (i.e., have global minimum topological entanglement).
Then we can always choose the first state correspond to the trivial anyon, i.e., $a(0) = 0$, with an appropriate gauge transformation.
The challenge to detecting the topological order from theses states arises because we do not have a priori know the labeling/phases of the ground states in absence of modular data.
Because of this, we cannot obtain all the information about a TQFT model only from the overlap. We will show that the information of a TQFT model that can be extract from the overlap of MESs is the following modular data:
\begin{itemize}
    \item fusion rules $N_{ab}^c$,
    \item triplet spins $\theta_a\theta_b/\theta_c$ for $N_{ab}^c >0$, 
\end{itemize}
where $a,b,c \in \mcC$.
These two constitute a complete set of information in the sense that any information about $\mcC$ from the overlaps (such as $S$ and $T^2$) are combinations of the data listed above.
Conversely, if two models have the same fusion rule and the triplet spins, they cannot be distinguished by the overlap of their MESs alone.

This section is organized as follow: In the section \ref{sec:indistinguishablemodel} and section \ref{sec:algorithm} we present the criterion for when a pair of models that cannot be distinguished by the overlap of MESs. In the Section \ref{sec:algorithm}, We also restate the algorithm form \cite{YiZhangoverlapSmatrix} which give us the maximum data from the overlap we mentioned in last paragraph. Section \ref{sec:MaxInfo} states several equivalent conditions to the indistinguishable models, which gives us a complete set of information that can be obtained from the overlap of ground states. We also discuss the indistinguishable models from the Wilson loop algebra point of view in section \ref{sec:discussion}.

\subsection{Indistinguishable models}\label{sec:indistinguishablemodel}

In this section we show that if two modular categories $\mcC$ and $\tilde{\mcC}$ have the same $S$-matrix and their $T$-matrices differ by a fusion phase,
\begin{align}
\boxed{
    \tilde{S}=S, \quad \tilde{\theta}_a = \frac{S_{r,a}}{S_{0,a}}\theta_a,}
    \label{eq:defindistinguishable}
\end{align}
where $r\in \mcC$ is a self-dual Abelian anyon,
then we cannot tell the difference between these two models from the overlap of MESs. This criteria defines an equivalence class of TQFT models.

Let $\ket{n^{\llll}_{\mmm}}$ be the standard basis of the ground states which are consistent with the Wilson loop algebra of $\mcC$.
We have previously established that for any coordinate basis $(\sss,\uuu)$,
\begin{align}
    \braket{i^{\sss}_{\uuu}|j^{\uuu}_{-\sss}} &\propto S_{i,j},
&   \braket{i^{\sss}_{\uuu}|j^{\sss+\uuu}_{\uuu}} &\propto T_{i,j}.
\end{align}
Then, we can define an alternative set of basis vectors
\begin{align}\begin{aligned}
   {\ket{ \widetilde{n^{\sss}_{\uuu}}}} &= W_{\sss}(r) W_{\uuu}(r)\ket{n^{\sss}_{\uuu}}= \frac{S^*_{r,n}}{S_{0,n}}\ket{rn^\sss_\uuu}\label{eq:newstates}
\end{aligned}\end{align}
whose overlap matrices are consistent with $\tilde{\mcC}$:
\begin{align}
    \braket{\widetilde{i^{\sss}_{\uuu}}|\widetilde{j_{\sss}^{-\uuu}}} &= \braket{i^{\sss}_{\uuu}|W^\dag_{\uuu}(r)W^\dag_{\sss}(r)W_{-\uuu}(r)W_{\sss}(r)|j_{\sss}^{-\uuu}}\nonumber\\
    &\propto  {S}_{i,j}=\tilde{S}_{i,j}, 
\\
     \braket{\widetilde{i^{\sss}_{\uuu}}|\widetilde{j_{\uuu}^{\sss+\uuu}}}
     &= \braket{{i^{\sss}_{\uuu}}|W^\dag_{\uuu}(r)W^\dag_{\sss}(r) W_{\sss+\uuu}(r)W_{\uuu}(r)|{j_{\uuu}^{\sss+\uuu}}}\nonumber \\
     &\propto \braket{{i^{\sss}_{\uuu}}|W_{\uuu}(r)|{j_{\uuu}^{\sss+\uuu}}} = \frac{S_{r,i}}{S_{0,i}}\theta_i \delta_{ij}.
\end{align}
The new states \eqref{eq:newstates} also have minimum bipartite entanglement entropy in the direction $\uuu$. However, these states are not a standard basis for $\mcC$ because they do not satisfy the modular transformation we define in section \ref{sec:MT}. (Notice that a valid gauge transformation of basis \eqref{eq:GTofgroundstates} does not depend on the direction.)
However, without the modular data, specifically the topological spin, we cannot determine whether $\{\ket{ \widetilde{n^{\llll}_{\mmm}}}\}$ are consistent bases of ground state. 

Imagine the situation that we want to determine whether a physical model $P$ is described by a TQFT model $\mcC$ or $\tilde{\mcC}$. If we only have a collection of ground states, the previous argument states that we can always find two sets of basis of $P$ such that one of them gives the $S$- and $T$-matrices for $\mcC$ and another one gives the $S$- and $T$-matrices for $\tilde{\mcC}$. Without further information, we cannot tell whether $P$ belongs to $\mcC$ or $\tilde{\mcC}$. Conversely, if we have two physical models $P_1$ and $P_2$ described by $\mcC$ and $\tilde{\mcC}$ respectively, we can always find a basis for $P_1$ and a basis for $P_2$ such that their overlap are the same for the same modular transformation. So that $P_1$ and $P_2$ are indistinguishable to the overlap. 


The key point of the indistinguishable models is the existence of a self-dual Abelian anyon $r$. The topological spin of $r$ can only be $\pm1,\pm i$ because $\theta_r^{-2} = \tqd S_{r,r} = \pm1$. When $\theta_r = 1$, $\mcC$ and $\tilde{\mcC}$ describe the same TQFT,
related by relabelling the anyons.
For other three situations, $\tilde{\mcC}$ is distinct from $\mcC$, and can be constructed by using the anyon condensation in Appendix~\ref{appx:construction}.

Here we give an example of a pair of indistinguishable models.
Let $\mcC_\mathrm{sem}, \mcC_{\overline{\mathrm{sem}}}$ be the semion and anti-semion models respectively. They have the same fusion rules
\begin{align}
   & 0\times \sigma = \sigma \times 0 = \sigma, &\sigma\times\sigma = 0.
\end{align}
Their $S$-matrices are the same and the $T$-matrices are
\begin{align}
& S_{\mathrm{sem}} =S_{\overline{\mathrm{sem}}} = \frac{1}{\sqrt{2}}\begin{pmatrix}
    1 &1\\
    1&-1
    \end{pmatrix},  \\
   & T_{\mathrm{sem}} = \begin{pmatrix}
    1 &0\\
    0&i
    \end{pmatrix},  \quad T_{\overline{\mathrm{sem}}} = \begin{pmatrix}
    1 &0\\
    0&-i
    \end{pmatrix}.
\end{align}
Let  $\ket{0_\mmm},\ket{1_\mmm},\ket{0_\uuu},\ket{1_\uuu}$ be the ground states of the semion model with minimum bipartite entanglement in the respective directions. The overlap of these states are given by 
\begin{align}
    \braket{i_\mmm|j_\uuu} = \Gamma_\mmm S_{\mathrm{sem}}^{n_1}T_{\mathrm{sem}}^{m_1}\cdots S_{\mathrm{sem}}^{n_k}T_{\mathrm{sem}}^{m_k}\Gamma_\uuu,
\end{align}
where $n_i,m_i$ depend on the modular transformation between the states,  $\Gamma_\mmm,\Gamma_\uuu$ are two diagonal matrix. The non-zero elements of $\Gamma_\mmm,\Gamma_\uuu$ are arbitrary phase in the labeling \eqref{eq:labeling}.

Let $P = \begin{pmatrix}
0&1\\-1&0
\end{pmatrix}$, with the property 
\begin{align}
    PS_{\mathrm{sem}}P^{-1}&=S_{\overline{\mathrm{sem}}}, &
    PT_{\mathrm{sem}}P^{-1}&=T_{\overline{\mathrm{sem}}}.
\end{align}
We can define the alternative states as
\begin{align}
\begin{aligned}
    & \ket{\widetilde{0_\mmm}} =  \ket{1_\mmm} ,&&\ket{\widetilde{1_\mmm}} =  -\ket{0_\mmm},&\\
    & \ket{\widetilde{0_\uuu}} = \ket{1_\uuu} ,&&\ket{\widetilde{1_\uuu}} = -\ket{0_\uuu}.
\end{aligned}
\end{align}
Then, the overlap between the new states are given by the modular matrices of ${\mcC}_{\overline{\mathrm{sem}}}$.
\begin{align}
    \braket{\widetilde{i_\mmm}|\widetilde{j_\uuu}} &= P\Gamma_\mmm S_{\mathrm{sem}}^{n_1}T_{\mathrm{sem}}^{m_1}\cdots S_{\mathrm{sem}}^{n_k}T_{\mathrm{sem}}^{m_k}\Gamma_\uuu P^{-1} \nonumber
    \\
    &=\Gamma_\mmm'S_{\overline{\mathrm{sem}}}^{n_1}{T}_{\overline{\mathrm{sem}}}^{m_1}\cdots S_{\overline{\mathrm{sem}}}^{n_k}{T}_{\overline{\mathrm{sem}}}^{m_k}\Gamma_\uuu'.
\end{align}
Therefore, the overlap of MESs alone cannot distinguish the semion and anti-semion model.
However, they can be distinguished from each other with the aid of other methods, such as the entanglement spectrum~\cite{SemionAntisemionOverlap}.

In this subsection we have established an ``upper bound'' to the information that can be obtained from MES overlap data.
Specifically (in the presence of a self-dual Abelian anyon $r$), it is impossible to get all the topological spins precisely; $\theta_a$ will always have an ambiguity of the form Eq.~\eqref{eq:defindistinguishable}.
Later in Subsection~\ref{sec:MaxInfo}, we will establish that two models with the same fusion rule and the triplet spins are always indistinguishable.
Therefore, for a physical model, we can only obtain at most the fusion rule and the triplet spins from the overlap of MESs.


\subsection{Modular data from MESs}\label{sec:algorithm}
In this section, we present an algorithm to determine the modular data from overlap of MESs, which is a slightly modified version of the algorithm from Ref.~\onlinecite{YiZhangoverlapSmatrix}.
We also show that this algorithm gives all maximum information, i.e., the fusion rules and the triplet spins, we list in the beginning of this section.

Let $\ket{n_\mmm},\ket{n_{\llll}}, \ket{n_{\mmm-\llll}}$ be the MESs of a TQFT model $\mcC$ in the respective directions, where $(\llll,\mmm)$ is a coordinate basis. We demand that the first states $\ket{0_\mmm},\ket{0_\llll},\ket{0_{\mmm-\llll}}$ in three directions are associated with an Abelian anyon, i.e., they have the global minimum bipartite entanglement. 

In practice, the MESs may have numerical errors due to numerical artifacts, finite size effects, etc.  The steps below are still be applicable and will give reasonable approximations to the true modular matrices.

\textbf{Step 1.}
Let $\ket{\widehat{n_\mmm}} = e^{i\phi_1(n)}\ket{n_\mmm}$
such that $\braket{0_\llll|\widehat{n_\mmm}} > 0$ for all $n$. 

We fix the relative phase of $\ket{n_\mmm}$ by using the fact that the first row of $S$ matrix is always positive, so we have the labeling,
\begin{align}
     \ket{\widehat{n_\mmm}} &= \ket{a(n)^\llll_\mmm},
\end{align}
and the $S$-transformation $ \ket{0_\llll} = \sum_{n}S_{0,n}\ket{\widehat{n_\mmm}}$.
Here $a(n)$ is an (arbitrary) assignment of numbers $0,\dots,|\mcC|-1$ to anyon labels (with the requirement that $a(0)$ is the trivial anyon).
The overlaps will give the quantum dimensions $\braket{0_\llll|\widehat{n_\mmm}} = d_n/\tqd$.

In this step we actually choose the gauge such that $\ket{0_\mmm}, \ket{0_\llll}$ correspond to the trivial anyon. Once the gauge has been chosen, the state $\ket{0_{\mmm-\llll}}$ automatically associate to an Abelian anyon $v$. 

\textbf{Step 2.}
Let $
    \ket{\widehat{n_\llll}} = e^{i\phi_2(n)}\ket{n_\llll}
$ and $\ket{\widehat{n_{\mmm-\llll}}} = e^{i\phi_3(n)}\ket{n_{\mmm-\llll}}$ such that the overlaps 
$\braket{\widehat{0_\mmm}|\widehat{n_\llll}}$, $\braket{\widehat{0_\llll}|\widehat{n_{\mmm-\llll}}}$
are positive.

This step fixes the relative phases of $\ket{n_\llll},\ket{n_{\mmm-\llll}}$ by using the fact that the first column of $S$ and $ST$ are always positive. So we have the labeling:
\begin{subequations} \label{eq:labeling2} \begin{align}
       \ket{\widehat{n_\llll}} &=\bigket{P_1(a(n))^{-\mmm}_{\llll}},\\
       \bigket{\widehat{n_{\mmm-\llll}}} &=\ket{vP_2(a(n))^\llll_{\mmm-\llll}}.
\end{align} \end{subequations}
$P_1$ and $P_2$ are two permutations that leave the vacuum anyon invariant because of the degree of freedom in labeling the states in different directions. $v$ is an Abelian anyon we mentioned in step 1.
We will find that the anyon $v$ cannot be detected from the overlap of ground states, and this is why we cannot get all the modular data of the system from the overlap of these three sets of MESs.
In the rest of the section, we omit the anyon assignment $a(\cdot)$, and use the integer $n$ to represent the anyon.

\textbf{Step 3.}
Let $K,Q,R$ be the overlap matrices
\begin{align}
   & K_{ij} = \Braket{\widehat{i_\mmm}|\widehat{j_\llll}}, &&Q_{ij } =  \Braket{\widehat{i_\mmm}|\widehat{j_{\mmm-\llll}}},&& R_{ij} = \Braket{\widehat{i_\llll}|\widehat{j_{\mmm-\llll}}},&
\end{align}
and define the auxiliary matrices
\begin{align}\begin{aligned}
       \tilde{K} &= \frac{d_id_j}{\tqd^2}\frac{K_{ij}}{K_{i0}K_{0j}},\\ \tilde{Q} &= \frac{d_id_j}{\tqd^2}\frac{Q_{ij}}{Q_{i0}Q_{0j}} ,\\ \tilde{R} &=\frac{d_id_j}{\tqd^2}\ \frac{R_{ij}}{R_{i0}R_{0j}}. &
\end{aligned}
\end{align}

The overlap matrices are related to the modular matrices via 
\begin{align}
    \begin{aligned}
       K_{ij}  &\propto  (S P_1)_{ij} , \\
       Q_{ij }  &\propto (T^*S^*T^*N_vP_2)_{ij},&\\
        R_{ij}  &\propto (P_1^{-1} TS^* N_v P_2)_{ij}.
        \label{eq:auxiliarymatrices}
    \end{aligned}
\end{align}
Here $(N_{v})_{ij} = N_{v,j}^{i} $ is the fusion matrix, and $P_1, P_2$ are the matrix representation of the permutation, $ (P_1)_{ij} = \delta_{i,P_1(j)}$, etc.
This form is directly from the labeling \eqref{eq:labeling2}. 
The three auxiliary matrices are given by setting the first column and row to be positive, therefore they can be written as the combination of the $S$-matrix and the permutation. 
\begin{align}
\begin{aligned}
   & \tilde{K}  =  S P_1,&&
  \tilde{Q} =  S^* P_2,&&
  \tilde{R}  = P_1^{-1} S^* P_2.&
\end{aligned}
\end{align}

\textbf{Step 4.}
Solving for the $S$- and permutation matrices:
\begin{align}
    \begin{aligned}
      & S =
      \tilde{Q}^{-1}\tilde{K}\tilde{R}, &&P_1 = \tilde{Q} \tilde{R}^{-1}  ,&& P_2 = \tilde{K} \tilde{R}. &
    \end{aligned}
\end{align}

The $S$-matrix and the permutation $P_1,P_2$ are given by the overlap matrices directly.
(If there were numerical errors in the MESs, the $S$-matrix might not be symmetric or unitary. The symmetricity and unitarity of the $S$-matrix can be used as an indicator of simulation error.)
The fusion rules can be obtained from the $S$-matrix through the Verlinde formula.

\textbf{Step 5.}
Denote $V = P_1 R P_2^{-1}$. 
For any Abelian anyon $a$ such that $V_{\bar{i}a} = V_{ia}$ for all $i$, we have a possible set of solution to the topological spins
\begin{align}
    \theta_i = V_{ia}\tqd /d_i, \label{eq:solutiontotheta}
\end{align}

This expression is straightforward from writing $V = TS^*N_v$ [Eq.~\eqref{eq:auxiliarymatrices}].
The columns of $V$ can be written as $V_{ia} = \theta_{i} S_{i,\bar{a}v}$.
When $a$ is an Abelian anyon, $V_{ia}$ equals to the topological spin times a fusion phase. Together with the restriction for the topological spin $\theta_i = \theta_{\bar{i}}$, we find all possible solutions. And from the expression of the matrix $V$, we know that each pair of possible solutions are differ by a fusion phase.

\textbf{Step 6.} Taking ratios of the topological spins Eq.~\eqref{eq:solutiontotheta} gives the triplet spins $\theta_i \theta_j / \theta_k \propto V_{ia} V_{ja} / V_{ka}$, an expression that always gives the same result whenever $a$ is Abelian.
So the formula for the triplet spins can be expressed simply
\begin{align}
\frac{\theta_{i}\theta_{j}}{\theta_{k}} = 
    \frac{\tqd d_k}{d_id_j}\frac{V_{i0}V_{j0}}{V_{k0}} \text{ for } N_{ij}^k >0.
\end{align}

In this subsection, we give an explicit prescription to compute the $S$-matrix and all the possible $T$-matrices.
We also showed that all possible solutions are related by a fusion phase as Eq.~\eqref{eq:defindistinguishable}.
This establishes that information ``upper bound'' from the overlap can indeed be achieved.


\subsection{Equivalent conditions of the indistinguishable model}\label{sec:MaxInfo}

Due to the existence of the indistinguishable models, we cannot obtain all the modular data of an anyon model from the overlap of MESs. We can get the upper limit of the data from the following equivalence. Let $(\mcC,S,T)$ and $(\tilde{\mcC},\tilde{S},\tilde{T})$ be two fusion categories, the following claims are equivalent.
\begin{itemize}
    \item[(a)] $S = \tilde{S}$, $\tilde{\theta}_a = \theta_a S_{r,a}/S_{0,a} $, where $r$ is a self-dual Abelian anyon.
    \item[(b)] $S = \tilde{S}$, $\tilde{\theta}_a^2 = \theta_a^2,$ and for the anyons in the same class $a\sim b$, $\frac{\theta_a}{\theta_b} =  \frac{\tilde{\theta}_a}{\tilde{\theta}_b}$. 
    
    \item[(c)]$\mcC$ and $\tilde{\mcC}$ have the same fusion rule $N_{ab}^c$, and $f(a,b,c) = \frac{\theta_a\theta_b}{\theta_c} = \frac{\tilde{\theta}_a\tilde{\theta}_b}{\tilde{\theta}_c} = \tilde{f}(a,b,c)$ for $N_{ab}^c>0$.
\end{itemize}
Here in condition (b), $a\sim b$ means for anyon Abelian anyon $r \in \mcC$, $\frac{S_{r,a}}{S_{0,a}} =\frac{S_{r,b}}{S_{0,b}} $.

If $\mcC$ is Abelian, then we have an additional equivalent condition.
\begin{itemize}
    \item[(d)] $S =  \tilde{S}$.
\end{itemize}

We first show the Abelian case.

\textbf{(a) $\Rightarrow$ (d)} is obvious.

\textbf{(d) $\Rightarrow$ (a)}.
Let $\mcC,\tilde{\mcC}$ be two Abelian fusion categories with the same $S$-matrix, then 
\begin{align}
    \tqd S_{a,a} = \frac{\theta_0}{\theta_a\theta_a} =  \frac{\tilde{\theta}_0}{\tilde{\theta}_a\tilde{\theta}_a} = \tilde{S}_{a,a} \;\;\Rightarrow\;\; \theta_a = \pm \tilde{\theta}_a.
\end{align}
Let $\tilde{\theta}_a = e^{i\varphi(a)}\theta_a$, then we have $\varphi(a) = \varphi(\bar{a})$. Since the $S$-matrix for these two model are the same, we have
\begin{align}
	S_{a,b} = \frac{1}{\tqd}\frac{\theta_a\theta_b}{\theta_{a{b}}} = \frac{1}{\tqd}\frac{\tilde{\theta}_a\tilde{\theta}_b} {\tilde{\theta}_{a{b}}} = \frac{1}{\tqd}\frac{e^{i\varphi(a)}e^{i\varphi(b)}}{e^{i\varphi(ab)}}\frac{\theta_a\theta_b} {\theta_{a{b}}}
\end{align}
Therefore, the phase factor obey the fusion rule $e^{i\varphi(ab)}=e^{i\varphi(a)}e^{i\varphi(b)}$, such that $e^{i\varphi(a)} = \pm1$ is a self-dual fusion phase. So if two Abelian models have the same $S$-matrix, then their $T$-matrices only differ by a fusion phase. Therefore, two Abelian models are indistinguishable if and only if they share the same $S$-matrix. 

Then we will show the rest three claims (for general categories) are equivalent.

\textbf{(a) $\Rightarrow$ (b)}. $S = \tilde{S}$ is obvious.
Because $r$ is a self-dual Abelian anyon, we have $S_{r,a}^2 = d_a^2/\tqd^2 = S_{0,a}^2$, and thus
\begin{align}
    \tilde{\theta}_a^2 = \theta_a^2 \frac{S^2_{r,a}}{S_{0,a}}= \theta_a^2.
\end{align}
For the anyon in the same equivalence class $a\sim b$, we have
\begin{align}
    \frac{\tilde{\theta}_a}{\tilde{\theta}_b} = \frac{{\theta}_a}{{\theta}_b} \frac{S_{r,a}}{S_{0,a}}\left(\frac{S_{r,b}}{S_{0,b}}\right)^{-1} = \frac{{\theta}_a}{{\theta}_b}.
\end{align}

\textbf{(b) $\Rightarrow$ (c)}. Since $\mcC$ and $\widetilde{\mcC}$ have the same $S$-matrix, they have the same fusion rule. Because $\theta_a^2=\tilde{\theta}_a^2 $, we can denote $\tilde{\theta}_a = \lambda(a)\theta_a$, where $\lambda(a) \in \{\pm1\}$. For any anyon $a,b\in \tilde{\mcC}$ we have
\begin{align}
    \tilde{S}_{a,b} = \sum_c N_{ab}^c \frac{d_c}{\tqd} \frac{\tilde{\theta}_a\tilde{\theta}_b}{\tilde{\theta}_c} = \sum_c N_{ab}^c \frac{d_c}{\tqd} \frac{{\theta}_a{\theta}_b}{{\theta}_c}\frac{\lambda(a)\lambda(b)}{\lambda(c)} = S_{a,b}.
\end{align}
Because if two anyon $c_1,c_2$ such that $N_{ab}^{c_1},N_{ab}^{c_2} \neq 0 $, then we have $c_1\sim c_2$. So $\lambda(c_1) = \lambda(c_2)$. Therefore we have 
\begin{align}
\begin{aligned}
    S_{a,b} &= \sum_c N_{ab}^c \frac{d_c}{\tqd} \frac{{\theta}_a{\theta}_b}{\theta_c}\frac{\lambda(a)\lambda(b)}{\lambda(c)} \\&= \frac{\lambda(a)\lambda(b)}{\lambda(c)}\sum_c N_{ab}^c \frac{d_c}{\tqd} \frac{{\theta}_a{\theta}_b}{\theta_c}
=\frac{\lambda(a)\lambda(b)}{\lambda(c)}S_{a,b}.
\end{aligned}
\end{align}
Thus 
\begin{align}
    \frac{\lambda(a)\lambda(b)}{\lambda(c)} = 1, \text{ for any } N_{ab}^c \neq 0,
\end{align}
which implies
\begin{align}
    \frac{\tilde{\theta}_a\tilde{\theta}_b}{\tilde{\theta}_c} = \frac{{\theta}_a{\theta}_b}{{\theta}_c},\text{ for any } N_{ab}^c \neq 0.
\end{align}

\textbf{(c) $\Rightarrow$ (a)}. From the fusion rule we can get the quantum dimension for all the anyons. Therefore, $\mcC$ and $\tilde{\mcC}$ have the same $S$-matrix \begin{align}
    \tqd S = \sum_c N_{ab}^c d_c \frac{\theta_a\theta_b}{\theta_c} = \sum_c N_{ab}^c d_c  \frac{\tilde{\theta}_a\tilde{\theta}_b}{\tilde{\theta}_c} = \tqd\widetilde{S}.
\end{align}
For any anyon $a,b,c \in\mcC$ such that $N_{ab}^c>0$ we have 
\begin{align}
    f(a,b,c) = \frac{\theta_a \theta_{b}}{\theta_c} = \tilde{f}(a,b,c) = \frac{\tilde{\theta}_a \tilde{\theta}_{b}}{\tilde{\theta}_c},
\end{align}
so the ratio $\tilde{\theta}_a/\theta_a$ forms a fusion phase. Thus we have $\tilde{\theta}_a = \theta_a S_{r,a}/S_{0,a}$ for a self-dual Abelian anyon $r$.

These equivalences demonstrate that the indeed the fusion rules and triplet spins is in fact the maximum amount of information that one may obtain from the overlap of the MESs solely.

\subsection{Discussions}\label{sec:discussion}

In previous subsections, we demonstrated the existence of indistinguishable models, which cannot be told apart by the overlaps of ground states alone. In this subsection, we approach the indistinguishable models from the Wilson loop algebra point of view. The main point of this subsection is to prove the following two claims:
\begin{itemize}
    \item [1. ] Given (i) the fusion rules $N^{ab}_c$ for a TQFT model $\mcC$, together with (ii) any self-dual solution $\xi(a) = \xi(\bar{a}) \in\mathrm{U}(1)$ to 
    \begin{align}
    \frac{\xi(a)\xi(b)}{\xi(c)} = \frac{\theta_a\theta_b}{\theta_c},  \text{ for } N_{ab}^c>0, \label{eq:solutionXi}
    \end{align}
    one can construct a collection of $|\mcC| \times |\mcC| $ matrices $\widetilde{W}_\sss(a)$, parameterized by anyons ${a}$ and loops ${\sss}$ on the torus, with the property that these matrices agree with the ground state matrix elements of the Wilson loop operators $W_\sss(a)$ up to phase (which depends on $\sss$ and $a$).
    \begin{align}
        \left(\widetilde{W}_\sss(a) \right)_{ij}= \Braket{i^\llll_\mmm|\phi(a,\sss) W_\sss(a)|j^\llll_\mmm}.
    \end{align}
    \item[2. ] Given (i) the fusion rules $N^{ab}_c$, (ii) solution $\xi(a)$ to equation \eqref{eq:solutionXi} and (iii) any collection of $|\mcC| \times |\mcC| $ matrices $\widetilde{W}_\sss(a)$ constructed from previous claim, there exists a collection of bases $\{\ket{a^\llll_\mmm}\}$ of $\mcC$ such that the overlap of them can be calculated from (i), (ii) and (iii). 
\end{itemize}

The first claim states that from the fusion rule $N_{ab}^c$ and a set of phases $\xi(a)$, we can construct a collection of $|\mcC|\times |\mcC|$ matrices which are almost the matrix representation of the Wilson loop operators (differ by a phase depended on anyon species and loop). Here we will first construct these matrices explicitly, and then show that they obey the Wilson loop algebra.

\textbf{Step 1.} First, we know that the self-dual solution $\xi(a) = \xi(\bar{a})$ always exists because $\xi(a) = \theta_a$ is a set of solution. Then we can define two sets of $|\mcC|\times|\mcC|$ matrices.
\begin{align}
    &\left(\widetilde{W}_\llll(a)\right)_{ij} = N_{aj}^i,&
    \left(\widetilde{W}_\mmm(a)\right)_{ij} = \frac{S_{a,i}^*}{S_{0,i}}\delta_{ij}.&
\end{align}
The first matrix is directly from the fusion rules, while the second matrix comes form the following identity:
\begin{align}
    \tqd S_{i,j} = \sum_k N_{ij}^k d_k \frac{\xi(i)\xi(j)}{\xi(k)}.\label{eq:modifiedS}
\end{align}
Here, the quantum dimension $d_k$ for $k\in \mcC$ is easily obtained since it is the 1-dimensional positive representation of the fusion rule. Notice that this two sets of matrices are exactly the matrix representation of the Wilson loop operators $W_\llll(a)$ and $W_\mmm(a)$ in the basis of $\ket{b^\llll_\mmm}$. 

\textbf{Step 2.} Define the ``modified'' Dehn twist operators:
\begin{align}
    \widetilde{\Dehn}_\llll &= \sum_x \frac{\xi(x)^*d_x\widetilde{W}_\llll(a)}{\tqd},\\\widetilde{\Dehn}_\mmm &= \sum_x \frac{\xi(x)^*d_x\widetilde{W}_\mmm(a)}{\tqd}.
\end{align}

\textbf{Step 3.} Similar to the case in section~\ref{sec:basisW}, we can construct the matrix $\widetilde{W}_\sss(a)$ for an arbitrary loop $\sss$ from the basis $\widetilde{W}_\llll(a)$ and $\widetilde{W}_\mmm(a)$ by using the modified Dehn twist operators.
\begin{align}
    \widetilde{W}_{\tau_\mmm(\llll)}(a) = \widetilde{\Dehn}_\mmm  \widetilde{W}_\llll (a) \widetilde{\Dehn}_\mmm^\dag.
\end{align}
 
In particular, if we choose $\xi(a) = \theta_a$, this procedure will produce the matrix representation of the Wilson loop operators in the basis of $\ket{b^\llll_\mmm}$. But we are more interested in the case when $\xi(a) \neq \theta_a$. We notice that all the solution to equation~\eqref{eq:solutionXi}, are given by the topological spin and the fusion phase because
\begin{align}
    \frac{\xi(x)}{\theta_x}\frac{\xi(y)}{\theta_y} = \frac{\xi(z)}{\theta_z}, \text{ if } N_{xy}^z>0.
\end{align}
Thus, we can denote $\xi(x) = \theta_x S_{r,x}/S_{0,x}$ for some self-dual Abelian anyon $r$. Then, if we replace $\theta_x,\theta_y$ with $\xi(x),\xi(y)$ in calculation~\eqref{l+m}, we will get an extra phase $S_{r,b}/S_{0,b}$ in the finial result. And that leads to 
\begin{align}
   \frac{S_{r,a}}{S_{0,a}}{W}_{\llll+\mmm}(a) &= \sum_{x,y} \frac{d_xd_y\xi(x)^*\xi(\bar{y})}{\tqd^2} {W}_\mmm(x){W}_\llll(a){W}_\mmm(y).
\end{align}
Therefore, by repeating the modified Dehn twist equation above it is easy to check these matrices satisfy following algebra:
\begin{align}
    \widetilde{W}_\llll(a)\widetilde{W}_\llll(b) &= \sum_c N_{ab}^c\widetilde{W}_\llll(c),\label{eq:matrixAlgebra1}\\
  \widetilde{W}_{\tau_\mmm(\llll)}(a) &= \widetilde{\Dehn}_\mmm \widetilde{W}_\llll(a)\widetilde{\Dehn}_\mmm ^\dag\label{eq:matrixAlgebra2}\\
     \widetilde{W}^\dag_\llll(a)& = \widetilde{W}_\llll(\bar{a}) =  \widetilde{W}_{\bar{\llll}}(a),
     \label{eq:matrixAlgebra3}
\end{align}
Therefore, we find the matrices we constructed are differ from the standard matrix representation of the Wilson loop operators by a phase,
\begin{align}
    \widetilde{W}_{p\llll+q\mmm} (x) = \left( \frac{S_{r,x}}{S_{0,x}} \right)^{p+q-1}\hat{W}_{p\llll+q\mmm}(x),
\end{align}
where $\hat{W}_{p\llll+q\mmm}(x)$ is the matrix representation of ${W}_{p\llll+q\mmm}(x)$ in the basis of $\ket{b^\llll_\mmm}$. So far, we construct a collection of matrices and show that they equal to the matrix representation of the Wilson loop operators up to a phase. Additionally, these matrices satisfy a similar algebra to the Wilson loop operators.

Let $\ket{a^\sss_\uuu}$ be the standard basis we defined in section \ref{sec:defMES}. Then the basis
\begin{align}
       {\ket{ \widetilde{a^{\sss}_{\uuu}}}} &= W_{\sss}(r) W_{\uuu}(r)\ket{a^{\sss}_{\uuu}}= \frac{S^*_{r,a}}{S_{0,a}}\ket{ra^\sss_\uuu}
\end{align}
is the basis we want in the second claim. The overlap of these states are generated by 
\begin{align}
    \braket{\widetilde{i^{\sss}_{\uuu}}|\widetilde{j_{\sss}^{-\uuu}}} &=  {S}_{i,j} \propto \left(\widetilde{\Dehn}^*_\llll\widetilde{\Dehn}^*_\mmm\widetilde{\Dehn}^*_\llll\right)_{ij}, 
\\
     \braket{\widetilde{i^{\sss}_{\uuu}}|\widetilde{j_{\uuu}^{\sss+\uuu}}}
     &= \frac{S_{r,i}}{S_{0,i}}\theta_i \delta_{ij} \propto \left(\widetilde{\Dehn}_\mmm\right)_{ij}.
\end{align}
These states are exactly the basis we defined in section~\ref{sec:indistinguishablemodel}. In fact, the Wilson loop point of view is equivalent to our former argument. As the discussion in section \ref{sec:gaugetransformation}, the phase factors attached to the Wilson loop operators will induce the rearrangements of the ground states. Thus, to construct a set of matrices $\widetilde{W}$ is equivalent to have all the ground states without knowing the label. If two system have the same fusion rule and triplet spins, we can construct the same sets of matrices $\widetilde{W}$ which leads to the same overlap in certain basis. This consistent with our conclusion from section~\ref{sec:indistinguishablemodel} and \ref{sec:MaxInfo} that the overlaps between the ground states cannot tell them apart. 

We can also show that if a collection of matrices $\widetilde{W}$ equal to the matrix representation of the Wilson loop operators up to a phase and satisfy restriction \eqref{eq:matrixAlgebra2}, then the coefficients $\xi(x)$ in equation \eqref{eq:matrixAlgebra2} is a solution to equation \eqref{eq:solutionXi} 
. This statement is the Wilson loop version of section \ref{sec:algorithm}.

At the end of this section, we categorize the TQFT models with rank no larger than 5 into the indistinguishable classes up to isomorphism (TABLE \ref{tab:table3},\ref{tab:table4}). This classification is based on \cite{ClassificationOfAnyon}\cite{classificationofAnyonrank5}. We noticed that if we can further determine the $\Theta$, then the $S$- and $T$-matrices are uniquely determined by the overlap (up to an anyon relabeling).


\begin{table*}[tbh]
\caption{\label{tab:table3} This table  classifies the TQFT model with rank less than 5. Each cell represent a set of indistinguishable models. In this table, we omit the indistinguishable pairs that corresponding to the anyon relabeling. }

\begin{tabular}{|c|c|c|c|c|}
\hline
\textbf{Rank}&\textbf{Fusion rule}&\textbf{Quantum dimensions}&\begin{tabular}{c}\textbf{Central charge}\\(mod 8)\end{tabular}&{\textbf{Topological spins}} \\
\hline
1& \text{trivial}  &(1)  &0& $(1)$ trivial model \\
\hline 
\multirow{4}{*}{2} &\multirow{2}{*}{$\mathbb{Z}_2$} &\multirow{2}{*}{(1,1)} &1& $(1,i)$ semion model \\
&&&$-1$& $(1,-i)$ anti-semion model  \\
\cline{2-5}
&\multirow{2}{*}{\parbox{12ex}{$\mathrm{SU}(2)_3/\mathbb{Z}_2$\\(Fibonacci)}}&\multirow{2}{*}{\parbox{12ex}{$(1,\frac{1+\sqrt{5}}{2})$}} &${14}/{5}$&$(1,\zeta_5^2)$ Fibonacci model \\
\cline{4-5}
& &&$-{14}/{5}$&$(1,\zeta_5^{-2})$ anti-Fibonacci model \\
\hline
\multirow{12}{*}{3 } &\multirow{8}{*}{\parbox{12ex}{\hfill\\$\mathrm{SU}(2)_2$\\(Ising)}} &\multirow{8}{*}{\parbox{12ex}{$(1,1,\sqrt{2})$}} &1/2 &$ (1,-1,\zeta_{16}^1) $ Ising model \\
& &&9/2&$ (1,-1, \zeta_{16}^9)$ \\
\cline{4-5}
& &&3/2&$ (1,-1, \zeta_{16}^3) $ \\

& &&11/2&$ (1,-1, \zeta_{16}^{11})$ \\
\cline{4-5}
& &&5/2&$(1,-1,\zeta_{16}^5)$ \\

& &&13/2&$(1,-1,\zeta_{16}^{13})$ \\
\cline{4-5}
& &&7/2&$ (1,-1, \zeta_{16}^7)$ \\

& &&15/2&$(1,-1, \zeta_{16}^{15})$ \\
\cline{2-5}
&\multirow{2}{*}{$\mathbb{Z}_3$}&\multirow{2}{*}{(1,1,1)} &2& $ (1,\zeta_3^1,\zeta_3^1)$ \\
\cline{4-5}
&&&$-2$& $ (1,\zeta_3^2, \zeta_3^2)$  \\
\cline{2-5}
&\multirow{2}{*}{$\mathrm{SU}(2)_5/\mathbb{Z}_2$ } &$(1,d,d^2-1),$ & -8/7&$\ (1,\zeta_7^5, \zeta_7^1)$ \\
\cline{4-5}
&&where $d = 2\cos\left(\frac{\pi}{7}\right)$&8/7& $ (1,\zeta_7^{-5}, \zeta_7^{-1})$ \\
\hline
\multirow{18}{*}{4} &\multirow{6}{*}{$\mathbb{Z}_2 \times \mathbb{Z}_2$} &\multirow{6}{*}{(1,1,1,1)} &2& $ (1, i, i, -1)$ sem. $\times$ sem.  \\
&&& $-2$&$ (1, -i, -i, -1)$ anti-sem. $\times$ anti-sem. \\
&& &0&$ (1, i, -i, 1)$ double semion\\
\cline{4-5}
&& &0&$ (1, 1, 1, -1)$ toric code model\\
&&&4& $ (1, -1, -1, -1)$ 3-fermion model\\
\cline{2-5}
&\multirow{4}{*}{\parbox{12ex}{$\mathrm{SU}(2)_3$\\($\mathrm{Fib.}\times\mathbb{Z}_2$)}}&\multirow{4}{*}{$(1,1,\frac{1+\sqrt{5}}{2},\frac{1+\sqrt{5}}{2})$} &19/5& $ (1, i, \zeta^2_5, i\zeta_5^2)$ Fib. $\times $ sem. \\
&&&9/5& $ (1, -i, \zeta^2_5, -i\zeta_5^2)$ Fib. $\times $ anti-sem. \\
\cline{4-5}
&&&$-9/5$&$ (1, i, \zeta^{-2}_5, i\zeta_5^{-2})$ anti-Fib. $\times $ sem. \\
&& &$-19/5$&$ (1, -i,  \zeta^{-2}_5, -i \zeta^{-2}_5)$ anti-Fib.$\times $anti-sem. \\
\cline{2-5}
&\multirow{3}{*}{$\text{Fib.} \times \text{Fib.}$} &\multirow{3}{*}{$(1,\frac{1+\sqrt{5}}{2},\frac{1+\sqrt{5}}{2},\frac{3+\sqrt{5}}{2})$}  &$-12/5$& $ (1, \zeta^2_5,  \zeta^2_5, \zeta^4_5)$  Fib.$\times$ Fib. \\
\cline{4-5}
&& &12/5&$ (1, \zeta^{-2}_5,  \zeta^{-2}_5, \zeta^{-4}_5)$  anti-Fib.$\times$ anti-Fib.  \\
\cline{4-5}
&&&0& $  (1, \zeta^{2}_5,  \zeta^{-2}_5, 1)$ double Fibonacci\\
\cline{2-5}
& \multirow{4}{*}{$\mathbb{Z}_4$ }& \multirow{4}{*}{(1,1,1,1)}
& 1&$ (1,-1, \zeta_8^1,\zeta_8^1)$\\
&&&-1& $ (1,-1,\zeta_8^{-1}, \zeta_8^{-1})$\\
\cline{4-5}
&&&3& $(1,-1, \zeta_8^{3},\zeta_8^3)$ \\
&&&$-3$& $ (1,-1,\zeta_8^{-3}, \zeta_8^{-3})$ \\
\cline{2-5}
&\multirow{2}{*}{$\mathrm{SU}(2)_7/\mathbb{Z}_2$}&$(1,d,d^2-1,d+1)$,
& 10/3&$(1,\zeta_9^2, \zeta_9^6,\zeta_9^{-6})$ \\
\cline{4-5}
&&where $d = 2\cos\left(\frac{\pi}{9}\right)$& $-10/3$&$ (1,\zeta_9^{-2}, \zeta_9^{-6},\zeta_9^{6})$ \\
\hline
\end{tabular}
\end{table*}

\begin{table*}[tbh]
\caption{\label{tab:table4} This table  classifies the TQFT model with rank 5. All these models are distinguishable from each other. }

\begin{tabular}{|c|c|c|c|c|}
\hline
\textbf{Rank}&\textbf{Fusion rule}&\textbf{Quantum dimensions}&\begin{tabular}{c}\textbf{Central charge}\\(mod 8)\end{tabular}&{\textbf{Topological spins}} \\
\hline
\multirow{14}{*}{5 }& \multirow{2}{*}{$\mathbb{Z}_5$ }& \multirow{2}{*}{$(1,1,1,1,1)$ }&0& $(1,\zeta_5^1,\zeta_5^{4},\zeta_5^{4},\zeta_5^1)$
\\
\cline{4-5}
& &&4&$(1,\zeta_5^2,\zeta_5^{3}.\zeta_5^{3},\zeta_5^2)$ 
\\
\cline{2-5}
&\multirow{9}{*}{$\mathrm{SU(2)_4}$} &\multirow{9}{*}{$(1,\sqrt{3},2,\sqrt{3},1)$} &2& $(1, \zeta_8^1, \zeta_3^1, \zeta_8^{5},1)$\\
\cline{4-5}
&&&2& $(1, \zeta_8^2, \zeta_3^1, \zeta_8^{6},1)$\\
\cline{4-5}
&&&2& $(1, \zeta_8^3, \zeta_3^1, \zeta_8^{7},1)$\\
\cline{4-5}
&&&2& $(1, -1, \zeta_3^1, 1,1)$\\
\cline{4-5}
&&&$-2$& $(1, \zeta_8^1, \zeta_3^2, \zeta_8^{5},1)$\\
\cline{4-5}
&&&$-2$& $(1, \zeta_8^2, \zeta_3^2, \zeta_8^{6},1)$\\
\cline{4-5}
&&&$-2$& $(1, \zeta_8^3, \zeta_3^2, \zeta_8^{7},1)$\\
\cline{4-5}
&&&$-2$& $(1, -1, \zeta_3^2, 1,1)$\\
\cline{2-5}
& \multirow{2}{*}{$\mathrm{SU(2)_9}/\mathbb{Z}_2 $ }&{$\left(1,4d^2-1,\frac{1}{\sqrt{2-2d}},8d^3-4d,2d\right)$ }&16/11 &$(1,\zeta^2_{11},\zeta^6_{11},\zeta^1_{11},\zeta^9_{11})$ \\
\cline{4-5}
&&where $d=\cos\frac{\pi}{11}$&$-16/11$&$(1,\zeta^9_{11},\zeta^5_{11},\zeta^{10}_{11},\zeta^2_{11})$\\
\cline{2-5}
&\multirow{2}{*}{$\mathrm{SU(3)_4}/\mathbb{Z}_3 $ }&{$(1,2a+2b,4a+2b,1+2a,2a+2b)$ }&$-18/7$&$(1,\zeta^1_{7},\zeta^4_{7},\zeta^6_{7},\zeta^1_{7})$\\
\cline{4-5}
&&where $a = \cos\frac{\pi}{7},\ b = \sin\frac{\pi}{14} $&18/7&$(1,\zeta^6_{7},\zeta^3_{7},\zeta^1_{7},\zeta^6_{7})$\\
\hline

\end{tabular}
\end{table*}

\section{Rotational symmetry}\label{sec:RotationalSymmetry}

The global symmetry enriches the structure of the TQFT model. The possible symmetry enriched states are classified by the symmetry actions including anyon permutations, the symmetry fractionalization class and the defectification class \cite{SymmetryEnrichedReview,spatialsymmetries}.
In this paper we only consider the action of unitary rotational symmetry $\R$ on the anyons.

The rotation $\R$ act on the anyon change it's species which is denoted as $a \mapsto \R(a)$. And the rotation of the movement operators of anyon will not only switch the directions $\llll \mapsto \R(\llll)$ but also induce phases attached to them. 
Because the $S$- and $T$-matrices remains fixed, 
\begin{align}
    S_{\R(a),\R(b)} &= S_{a,b},&
    \theta_{\R(a)} &= \theta_{a},
\end{align}
the possible transformations on the basis of the Wilson loop operators are given by:
\begin{align}
\begin{aligned}
    \R W_{\llll}(a) \R^\dag &= \frac{S_{r,\R(a)}}{S_{0,\R(a)}} W_{\R(\llll)}\big(\R(a)\big),\\
    \R W_{\mmm}(a) \R^\dag &= \frac{S_{s,\R(a)}}{S_{0,\R(a)}} W_{\R(\mmm)}\big(\R(a)\big),
\end{aligned}
\end{align}
where $r,s$ are two Abelian anyon.
The transformation of an arbitrary Wilson loop operator $W_{\sss}(a)$ can be obtained from the transformation of this basis.
Let $\ket{a^\llll_\mmm}$ be a basis of the ground states, there are three effects of rotation acting on this basis: the modular transformation $(\llll,\mmm) \mapsto (\R(\llll),\R(\mmm))$, the permutation of anyon $a \mapsto \R(a)$ and the rearrangement caused by the phase attached to the Wilson loop operator. Therefore, from on set of basis $\ket{a^\llll_\mmm}$ and the rotation $\R$, we automatically get another set of basis 
\begin{align}
   \R \ket{a^\llll_\mmm} = W_{\R(\llll)}(s)W_{\R(\sss)}(r) \ket{\R(a)^{\R(\llll)}_{\R(\mmm)}},
\end{align}
where $W_{\R(\llll)}(r)$ and $W_{\R(\mmm)}(s)$ are the rearrangement of the basis caused by the phase of the Wilson loop operator.

\begin{figure}[h]
    \centering
    \includegraphics[width = 0.5\textwidth]{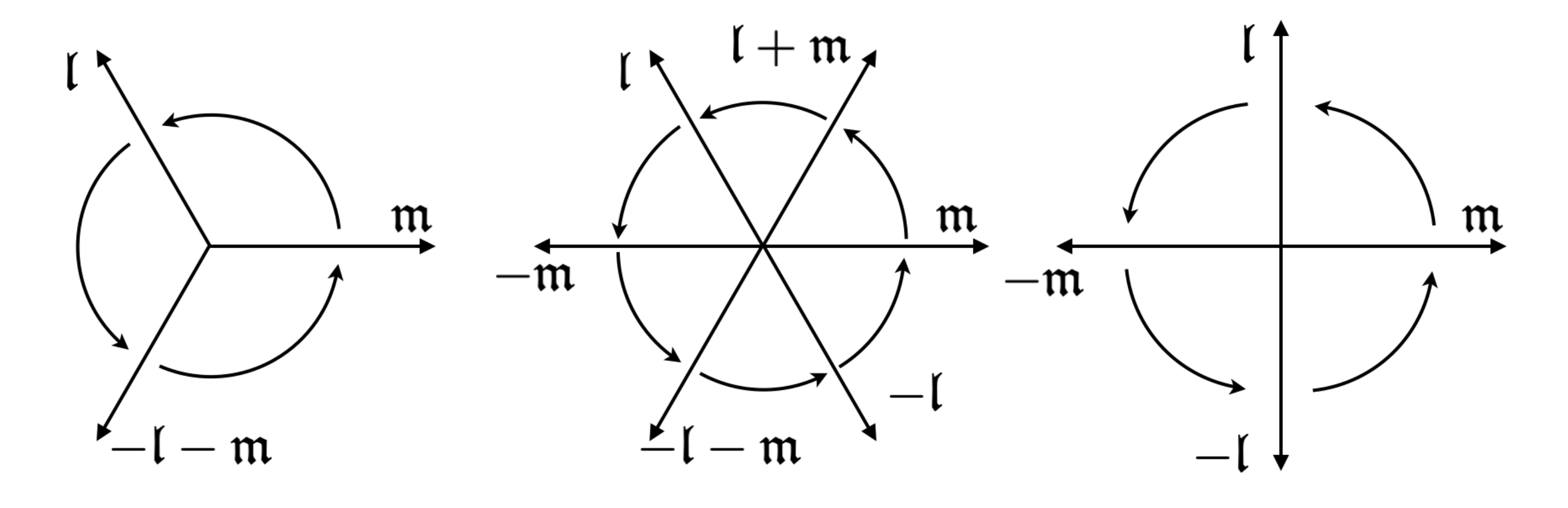}
    \caption{In this section, we consider the 3-fold, 6-fold and 4-fold rotation. Topologically, we can also treat the torus as the lattice on complex plane $\mathbb{C}$ with two basis vectors $1$ and $\tau$. Here $\tau$ is a complex number which is called the modular parameter of the torus. For the 3- and 6-fold rotation we choose the modular parameter of the torus to be $\tau = -\frac{1}{2}+\frac{\sqrt{3}}{2}i$, while for the 4-fold rotation, the modular parameter is $\tau = i$.}
    \label{fig:rotations}
\end{figure}

In this section, we will consider the system bear the the 3-, 6- or 4-fold rotational symmetry (FIG. \ref{fig:rotations}). 

The system may also possess time-reversal symmetry or a mirror reflection symmetry \cite{SETMirrorandTimereversal}.
Such symmetries induce strong restrictions for the model, i.e., $S_{\R(a),\R(b)} = S_{a,b}^* $ and $\theta_{\R(a)}=\theta_a^*$. In fact, for the TQFT models with rank less than 6, only the trivial TQFT, double semion, double Fibonacci, and toric code may coexist with time-reversal or mirror symmetries (models with 0 chiral central charge)\footnote{The $\mathbb{Z}_5$ model with $c=0$ is compatible with the anti-unitary symmetry with order 4, but not one with order 2.}.
These these models are already distinguishable, so we need not consider these symmetries in this paper.

\subsection{3-fold rotation}\label{sec:3-fold}
Given a set of basis of MESs along a single cut and 3-fold rotation, we automatically get three sets of bases along 3 different directions(Fig.~\ref{fig:rotations}). As we discussed in section \ref{sec:discussion}, this guarantees that all the information about a TQFT model mentioned in last section can be obtained from the overlap.
Moreover, the permutation of anyon induced by the rotation is also detectable from the overlap of MESs. The non-trivial permutation, if exists, can be used to determine the TQFT models which are indistinguishable without symmetry. 

Here we consider the torus in the coordinate basis $(\llll,\mmm)$ with modular parameter $\tau = -\frac{1}{2}+i\frac{\sqrt{3}}{2} $. The transformation of the standard basis under the 3-fold rotation $\R$ is given by 
\begin{align}
    \R \ket{a^\llll_\mmm} = \frac{S_{r,\R(a)}}{S_{0,a}} \Ket{s\R(a)^{-\llll-\mmm}_\llll}.
\end{align}
Thus the matrix representation of the 3-fold rotation is 
\begin{align}
    \braket{a^\llll_\mmm|\R|b^\llll_\mmm} = \tqd S^*_{r,s}S_{ra,s\R(b)}\theta_{s\R(b)}^*. \label{eq:3-fold}
\end{align}
If there were no phases attached to the Wilson loop operators nor permutations of anyon during the rotation, this formula reduces to 
$\braket{a^\llll_\mmm|\R|b^\llll_\mmm} = S_{ab}\theta_b^*$, as in Eq.~\eqref{eq:3-foldrotation}.

In this section, we provide the algorithm to calculate the modular data and the permutation of anyon from the MESs with the 3-fold rotation $\R$. Let $\ket{n_\mmm}$ be the ground states with the minimum bipartite entanglement entropy alone $\mmm$-cut, and the state $\ket{0_\mmm}$ is chosen to have the global minimum bipartite entanglement entropy.

\textbf{Step 1.} Calculate the matrix representation of the rotation,
\begin{align}
    R_{ij} = \frac{\braket{i_\mmm|\R|j_\mmm}}{\braket{0_\mmm|\R|0_\mmm}}.
\end{align}

\textbf{Step 2.} Define the auxiliary matrix, 
\begin{align}
    \tilde{R}_{ij} = \frac{d_{i}d_{j}}{\tqd} \frac{R_{ij}}{R_{i0}R_{0j}}.
\end{align}

Here the topological dimension are given by the magnitude of the first row of matrix $R$. Here the matrix $\tilde{R}$ is obtained by set the first row and column of $R$ to be positive, so $\tilde{R}$ is the $S$-matrix with an anyon permutation,
\begin{align}
    \tilde{R}_{ij} = S_{i,\R(j)}.
\end{align}

\textbf{Step 3.} The $S$-matrix is given by 
\begin{align}
    S_{i,j} = (\tilde{R}^{-3})_{ij}.
\end{align}

Here we use the fact that $\R^3(a) = a$ and $S^4 = \mathbbm{1}$. The fusion rules be extract from the $S$-matrix by the Verlinde formula.

\textbf{Step 4.} The permutation of anyon is given by 
\begin{align}
    \delta_{i,\R(j)} = (S^* \tilde{R})_{ij} .
\end{align}

\textbf{Step 5.} Define another auxiliary matrix
\begin{align}
    Q_{ij} = \frac{d_{i}R_{ij}}{R_{i0}R^*_{j0}} .
\end{align}
For each Abelian anyon $i$, we have a set of possible solution to the topological spin,
\begin{align}
    \theta_i = Q_{ij}^*,
\end{align}
provided $Q_{ij} = Q_{i\bar{j}}$ for all anyon $j\in \mcC$.

\textbf{Step 6.} This addition step gives the spin triplets.
\begin{align}
    \frac{\theta_{i}\theta_{j}}{\theta_{k}} &=\frac{Q_{0k}}{Q_{0i}Q_{0j}} \text{ for } N_{ij}^k>0.
\end{align}

Here, we obtain all the information described in Sec.~\ref{sec:globalsymmetry} from the overlap matrix, in addition to the permutation matrix $\delta_{a,\R(b)}$.
The permutation information may be used to distinguish models which previously indistinguishable from just the overlaps.
For example, the toric code $\mcC_{\text{toric}} = \{0,e,m,f\}$ and 3-fermion model $\mcC_{\text{3-fermion}} = \{0,e,m,f\}$ have the same $S$-matrix and their $T$-matrices are differ from a phase 
\begin{align}
    & S_{\text{toric}} = S_{\text{3-fermion}} = \frac{1}{2} \begin{pmatrix}
    1&1&1&1\\
    1&1&-1&-1\\
    1&-1&1&-1\\
    1&-1&-1&1
    \end{pmatrix},\\
      & T_{\text{toric}} = \mathrm{diag}(1,1,1,-1),\\
      & T_{\text{3-fermion}} = \mathrm{diag}(1,-1,-1,-1).
\end{align}
Without symmetry, we cannot distinguish them with the overlap matrices.
However, notice that a 3-fold symmetry with non-trivial permutation of anyon
\begin{align}
    \R: e\to m\to f \to e
\end{align}
can only exists in the 3-fermion model. Therefore, if we observe a non-trivial permutation of anyon under the 3-fold rotation, we can determine the topological order is described by 3-fermion model.

\subsection{6-fold rotation}
Notice that if $\R_{6}$ is a 6-fold rotation, then $\R_{6}^2$ is a 3-fold rotation.
So, by using the algorithm in the previous subsection, we can obtain all the information we mentioned in section \ref{sec:MaxInfo}, i.e, the $S$-matrix and the $T$-matrix up to a fusion phase.

In this section, we calculate the permutation of anyons from the MESs with the 6-fold rotation $\R_{6}$.
We again consider the torus in the coordinate basis $(\llll,\mmm)$ with modular parameter $\tau = -\frac{1}{2}+i\frac{\sqrt{3}}{2} $. Similar to 3-fold rotation, the transformation of the basis under the 6-fold rotation is given by 
\begin{align}
    \R_{6} \ket{a^\llll_\mmm} = \frac{S_{r,\R_{6}(a)}}{S_{0,a}} \ket{s\R_{6}(a)^{-\mmm}_{\llll+\mmm}}.
\end{align}
And the matrix representation of the 6-fold rotation is 
\begin{align}
    \braket{a^\llll_\mmm|\R_{6}|b^\llll_\mmm} = \tqd \theta_a S_{r,s}^* S_{ra,s\R_{6}(b)}.
    \label{eq:6-fold}
\end{align}
We notice that the expression \eqref{eq:3-fold} and \eqref{eq:6-fold} are similar, which indicates the 3-fold and 6-fold rotation give the same information.


Starting from a set of MESs $\ket{n_\mmm}$ along the $\mmm$-cut with $\ket{0_\mmm}$ chosen to have the global minimum bipartite entanglement entropy, the steps are as follows.

\textbf{Step 1.} Calculate the matrix representation of the rotation,
\begin{align}
    (R_{6})_{ij} = \frac{\braket{i_\mmm|\R_{6}|j_\mmm}}{\braket{0_\mmm|\R_{6}|0_\mmm}}.\label{eq:6-folddefinition}
\end{align}

\textbf{Step 2.} Define the auxiliary matrix, 
\begin{align}
    (\tilde{R}_{6})_{ij} = \frac{d_{i}d_{j}}{\tqd} \frac{(R_{6})_{ij}}{(R_{6})_{i0}(R_{6})_{0j}}
\end{align}
such that
\begin{align}
    (\tilde{R}_{6})_{ij} = S_{i,\R_{6}(j)}.
\end{align}

\textbf{Step 3.} The permutation of the anyons can be obtained from the $S$ matrix and the auxiliary matrix,
\begin{align}
    \delta_{i,\R_{6}(i)} = \big(S^* \tilde{R}_{6}\big)_{ij}. \label{eq:6-foldresult}
\end{align}

\subsection{4-fold rotation}
Let $\R_4$ be a 4-fold rotation on the torus with the modular parameter $\tau = i $. We have 
\begin{align}
    \R_4 \ket{a^\llll_\mmm} = \frac{S_{r,\R_4(a)}}{S_{0,a}} \ket{s\R_4(a)^{-\mmm}_{\llll}}.
\end{align}
And the matrix representation of the 4-fold rotation is 
\begin{align}
    \braket{a^\llll_\mmm|\R_4|b^\llll_\mmm} = \tqd  S_{r,s}^* S_{ra,s\R_4(b)}.
    \label{eq:4-fold}
\end{align}
This matrix representation does not contain the topological spins, so that we cannot get any information about the $T$-matrix (that is not already contained in $S$) from the 4-fold rotation of a single set of MESs.
Another way to interpret it is that the 4-fold rotation only involve two directions, according to the discussion \ref{sec:discussion}, to obtain the maximum information from the overlap, we need at least 3 directions. In fact, if we only have the ground states along one cut with the 4-fold rotation, we cannot even get the $S$-matrix generally.

However, if we obtain the $S$-matrix by using the algorithm \ref{sec:algorithm}, we can obtain the permutation of anyon under the 4-fold rotation.
This would be identical to the steps Eqs.~\eqref{eq:6-folddefinition}--\eqref{eq:6-foldresult}, but with $\R_6$ replaced by $\R_4$.

The 4-fold rotation can distinguish the double semion model from the semion square and anti-semion square model. We notice that the semion square and anti-semion square model can have a non-trivial 2-fold anyon permutation while double semion model cannot.

\begin{acknowledgments}
We are grateful to David Aasen and Michael Levin for stimulating conversations.  We are especially indebted to Michael Levin for his thoughtful comments on the manuscript and the construction in Appendix~\ref{sec:Mfusion}.  ZL and RM are supported by the National Science Foundation Grant No.~DMR-1848336.
\end{acknowledgments}

\clearpage
\appendix

\section{Brief review of symbols and diagrams in TQFT} \label{TQFT}
The fusion rules for anyons are described by
\begin{align}
    a\times b = \bigoplus_{c\in\mcC} N_{ab}^c c,
\end{align}
where $N_{ab}^c$ are non-negative integers, which called the fusion symbols.
The anyons, together with the fusion rules forms a commutative, associative algebra:
\begin{align}
    &N_{ab}^c = N_{ba}^c,& & \sum_c N_{ab}^cN_{cd}^e = \sum_c N_{ac}^eN_{bd}^c. &
\end{align}
Fusion and splitting of anyons can be represented by following diagrams.
\begin{align}
    &{\xy
    (-4,-8)="a1";(-4,8)="a2";(4,-8)="b1";(4,8)="b2";
    (0,0) = "c1"; (0,8) = "c2";"c1";"c2"**[PineGreen]\dir{-};{\ar@[PineGreen]@{>}(0,4);(0,4.01)};
    "a1";"c1"**[red]\dir{-};"b1";"c1"**[blue]\dir{-};{\ar@[red]@{>}(-2,-4);(-1.99,-3.98)};{\ar@[blue]@{>}(2,-4);(1.99,-3.98)};
    (-5,-7)*{\scr a};(5,-7)*{\scr b};(2,8)*{\scr c};(2,0)*{\scr \mu};
    \endxy}&{\xy
    (-4,8)="a1";(-4,8)="a2";(4,8)="b1";(4,8)="b2";
    (0,0) = "c1"; (0,-8) = "c2";"c1";"c2"**[PineGreen]\dir{-};{\ar@[PineGreen]@{>}(0,-4);(0,-3.99)};
    "a1";"c1"**[red]\dir{-};"b1";"c1"**[blue]\dir{-};{\ar@[red]@{>}(-2,4);(-2.01,4.02)};{\ar@[blue]@{>}(2,4);(2.01,4.02)};
    (-5,7)*{\scr a};(5,7)*{\scr b};(2,-8)*{\scr c};(2,0)*{\scr \mu};
    \endxy}
\end{align}
The quantum dimension $d_a$ of anyon $a$ is define to be the largest eigenvalue of the fusion matrix $N_a $ whose elements are $(N_{a})_{bc} = N_{ab}^c$. Diagrammatically, the quantum dimension is denote as
\begin{align}
     d_a =  
	{\xy (-5,0)="a";  "a";"a"+(0,1),**\dir{}, "a",{\ellipse(5){-}}; {\ar@{>} (-0.05,0.85);(-0.05,0.9)}; (2,0)*{ a} \endxy} .
\end{align}
The quantum dimension $d_a$ forms an 1-dimensional real representation of the fusion algebra, namely
\begin{align}
    d_ad_b = \sum_cN_{ab}^cd_c.
\end{align}
This formula comes from the following fusion diagram directly.
\begin{align}
\begin{aligned}
    {\xy
    (-2,-8)="a1";(-2,8)="a2";(2,-8)="b1";(2,8)="b2";
    "a1";"a2"**[red]\dir{-};"b1";"b2"**[blue]\dir{-};{\ar@[red]@{>}(-2,0);(-2,0.01)};{\ar@[blue]@{>}(2,0);(2,0.01)};
    (-4,-7)*{\scr a};(4,-7)*{\scr b};
    \endxy}\
   & = \
    \sum_{c,\mu} \sqrt{\frac{d_c}{d_ad_b}} 
    {\xy
    (-2,-8)="a1";(-2,8)="a2";(2,-8)="b1";(2,8)="b2";(0,-4)="c1";(0,4)="c2";
    "a1";"c1"**[red]\crv{(-2,-4)};"b1";"c1"**[blue]\crv{(2,-4)};
    "c1";"c2"**[PineGreen]\dir{-};
     "a2";"c2"**[red]\crv{(-2,4)};"b2";"c2"**[blue]\crv{(2,4)};{\ar@[PineGreen]@{>}(0,1);(0,1.01)};
    (-4,-7)*{\scr a};(4,-7)*{\scr b};(2,0)*{\scr c};(1,3)*{\scr \mu};(1,-3)*{\scr \mu};
    \endxy}, \\ 
    {\xy 
    (0,-8)="a1";(0,8)="a2";(0,-3)="a3";(0,3)="a4";
    "a3";"a4"**[PineGreen]\crv{(-3,0)};
    "a3";"a4"**[Purple]\crv{(3,0)};
    "a2";"a4"**[blue]\dir{-};
    "a1";"a3"**[red]\dir{-};(-2,-7)*{\scr a};(-2,7)*{\scr b};(-3,0)*{\scr c};(3,0)*{\scr e};(2,4)*{\scr \mu};(2,-4)*{\scr \nu};
    \endxy} \
&= \  
    \delta_{ab}\delta_{\mu\nu}\sqrt{\frac{d_cd_e}{d_a}} {\xy 
    (0,-8)="a1";(0,8)="a2";
    "a1";"a2"**[red]\dir{-};{\ar@[red]@{>}(-0,0);(0,0.01)};(-2,-7)*{\scr a};
    \endxy} \ ,
\end{aligned}
\end{align}
where $\mu,\nu$ indicate the fusion channels.
The total quantum dimension of $\mcC$ is
\begin{align}
    \tqd \defeq \sqrt{ \sum_{a\in \mcC} {d_a^2}}.
\end{align}
Two important braiding statistics of the TQFT model, self-statistics $\theta_a$ (topological spin), and mutual-statistics $S_{a,b}$ are defined as follow:
\begin{align}
    \theta_a = \theta_{\bar{a}} \defeq  \frac{1}{d_a} \,
		{\xy (0,0)*{\phdot}="o"; (-3,3)="tl"; (3,3)="tr"; (-3,-3)="bl"; (3,-3)="br"; (3,5)*{a}; 
			"o";"tl"**\dir{-}; "bl",{\ellipse^{}}; "tr"**\dir{-}?(1)*\dir{>}; "br",{\ellipse_{}}; "o"**\dir{-}; \endxy} \,,
\end{align}
\begin{align}
    S_{a,b} = \frac{1}{\tqd} \,
	{\xy (-9,0)*{}="Bl"; (-4,0)="a"; (2,0)="b"; (1,0.7)="aa",*+!L{a}; {\ar@{>} "aa"+(0,-.01);"aa"}; (7,0.7)="bb",*+!L{ b}; {\ar@{>} "bb"+(0,-.01);"bb"};
		"a";"b",**\dir{}, "a",{\ellipse(5):a(154),=:a(-20){-}}; "b";"a",**\dir{}, "b",{\ellipse(5):a(154),=:a(-20){-}}; \endxy} = \frac{1}{\tqd} \sum_{c\in\mcC}N_{ab}^cd_c \frac{\theta_a\theta_b}{\theta_c} .\label{eq:SandTheta}
\end{align}
We also use the $R$-symbol to describe the braiding,
\begin{align}
   & \sum_{\mu}(R^{ab}_c)_{\mu\nu}{\xy
    (-4,8)="a1";(0,4)="a2";(4,8)="b1";
    (0,0) = "c1"; (0,-8) = "c2";"c1";"c2"**[PineGreen]\dir{-};
    "a1";"c1"**[red]\dir{-};
    "b1";"c1"**[blue]\dir{-};
    (-5,7)*{\scr a};(5,7)*{\scr b};(2,-8)*{\scr c};(2,0)*{\scr \mu};
    \endxy} = {\xy
    (-4,8)="a1";(0,4)="a2";(4,8)="b1";
    (0,0) = "c1"; (0,-8) = "c2";"c1";"c2"**[PineGreen]\dir{-};
    "c1";"a2"*{\phdot}**[red]\crv{(2,2)};"a1"**[red]\dir{-};
    "c1";"a2"**[blue]\crv{(-2,2)};
    "a2";"b1"**[blue]\dir{-};
    (-5,7)*{\scr a};(5,7)*{\scr b};(2,-8)*{\scr c};(2,0)*{\scr \nu};
    \endxy}, \\
    & 
    \sum_{\mu}(R^{ba}_c)^\dag_{\mu\nu}{\xy
    (-4,8)="a1";(0,4)="a2";(4,8)="b1";
    (0,0) = "c1"; (0,-8) = "c2";"c1";"c2"**[PineGreen]\dir{-};
    "a1";"c1"**[red]\dir{-};
    "b1";"c1"**[blue]\dir{-};
    (-5,7)*{\scr a};(5,7)*{\scr b};(2,-8)*{\scr c};(2,0)*{\scr \mu};
    \endxy} = {\xy
    (-4,8)="a1";(0,4)="a2";(4,8)="b1";
    (0,0) = "c1"; (0,-8) = "c2";"c1";"c2"**[PineGreen]\dir{-};
    "c1";"a2"**[red]\crv{(2,2)};"a1"**[red]\dir{-};
    "c1";"a2"*{\phdot}**[blue]\crv{(-2,2)};"b1"**[blue]\dir{-};
    (-5,7)*{\scr a};(5,7)*{\scr b};(2,-8)*{\scr c};(2,0)*{\scr \nu};
    \endxy}.
\end{align}
The ribbon relation relates the $R$-symbol and the topological spin:
\begin{align}
  \sum_\nu  (R^{ab}_c)_{\mu\nu} (R^{ba}_c)_{\nu\lambda} = \frac{\theta_c}{\theta_a\theta_b}\delta_{\mu\lambda}.
\end{align}
We let $\Theta$ denote a phase in the modular tensor category theory that relate to the chiral central charge $c_-$.
\begin{align}
    \Theta &= \sum_{a\in \mcC} d_a^2\theta_a = e^{\frac{2\pi i }{8}c_-}.
\end{align}
We also list some useful modular equations here.
\begin{align}
\begin{aligned}
    STSTST &= \Theta S^2,& S^4 &= \mathbbm{1}.
\end{aligned}
\end{align}
The Verlinde formula relate the fusion rule with the $S$ matrix,
\begin{align}
    N_{ab}^c = \sum_{x\in\mcC} \frac{S_{a,x}S_{b,x}S^*_{c,x}}{S_{0,x}}.
\end{align}

\section{Fusion of the movement operators}\label{sec:Mfusion}
This section serves as a supplementary of section \ref{defW}. We will prove the fusion rules for the movement operators and derive the Wilson loop algebra \eqref{eq:fusionM1}--\eqref{eq:antiW1}.
First we list some of the assumptions used to construct our Wilson loop operators.
\begin{itemize}
    \item[(1)] In between every step, the anyons are separated from one another by a distance that is much larger than the correlation length of the ground state.
    \item[(2)] For any anyon and any pair of points $x$ and $x'$, there exist a local unitary movement operator $\M_{x,x'}(a)$ which moves the anyon $a$ from $x$ to $x'$. The ``local" means that the operator $\M_{x,x'}(a)$ is supported in the neighborhood of the interval containing $x$ and $x'$.
    \item[(3)] The movement operators for the trivial anyon $0$ along arbitrary path is the identity $\mathbbm{1}_0$.
    \item[(4)] The composition of the movement operators $\M_f(a)\M_g(a)$ is the movement operator of the composition path $\M_{f+g}(a)$. Especially, the movement operator of the reverse path is the complex conjugate of the movement operator, $\M_{\bar{f}}(a) = \M^\dag_f(a)$.
    \item[(5)] For 3-tuple $(a,b,c)$ such that $N_{ab}^c > 0$ and two well separated position $x_1$ and $x_2$, there exist a local splitting operator $\Phi^{ab}_c$ which take an anyon $c$ at position $x$ as input and output $a$ and $b$ at position $x_1$ and $x_2$ respectively. Here $x$ lies in the neighborhood of the interval containing $x_1$ and $x_2$.
    \item[(6)] It is always possible to close a puncture to get a superposition of single anyon remaining on the expanded manifold%
        ~\footnote{This assumption is based on the conjecture that all vertex basis gauge transformations that leave the $F$- and $R$- symbol invariant up to a gauge transformation are natural isomorphisms.}%
    .
\end{itemize}

Let's first prove the fusion rule of the movement operators,
\begin{align}
    \alpha_\llll(b)\widetilde{\M}_{\llll}(b)
    &\alpha_\llll(a)\widetilde{\M}_{\llll}(a)\nonumber\\
    &\overset{?}{=} \sum_{c,\mu}\sqrt{\frac{d_c}{d_ad_b}}\Phi_{c,\mu}^{ab}\alpha_\llll(c)\widetilde{\M}_{{\llll}}(c)
    \Phi_{ab}^{c,\mu},
    \label{eq:fusionM1}
\end{align}
with the splitting and fusion operators with the normalization condition 
\begin{align}
\sum_{c,\mu}\sqrt{\frac{d_c}{d_ad_b}}\Phi^{ab}_{c,\mu}\Phi_{ab}^{c,\mu} = \mathbbm{1}_{ab},
     \label{eq:normalization1}\\
       \sqrt{\frac{d_c}{d_ad_b}}\Phi_{ab}^{c,\mu}\Phi^{ab}_{c,\nu} = \delta_{\mu\nu}\mathbbm{1}_{c}.
     \label{eq:normalization2}
\end{align}

Let $V_{ab},V_c$ denote the vector space spanned by all the topologically degenerate states with anyons $a,b$ and $c$ at position $x_a,x_b$ and $x_c$ respectively. Then the fusion operator $\Phi_{ab}^{c,\mu}$ is the homomorphism from $V_{ab}$ to $V_c$, while the splitting operator $\Phi^{ab}_{c,\mu}$ is the homomorphism from $V_{c}$ to $V_{ab}$.
 
\begin{figure}[b]
     \centering
     \includegraphics[width=\linewidth]{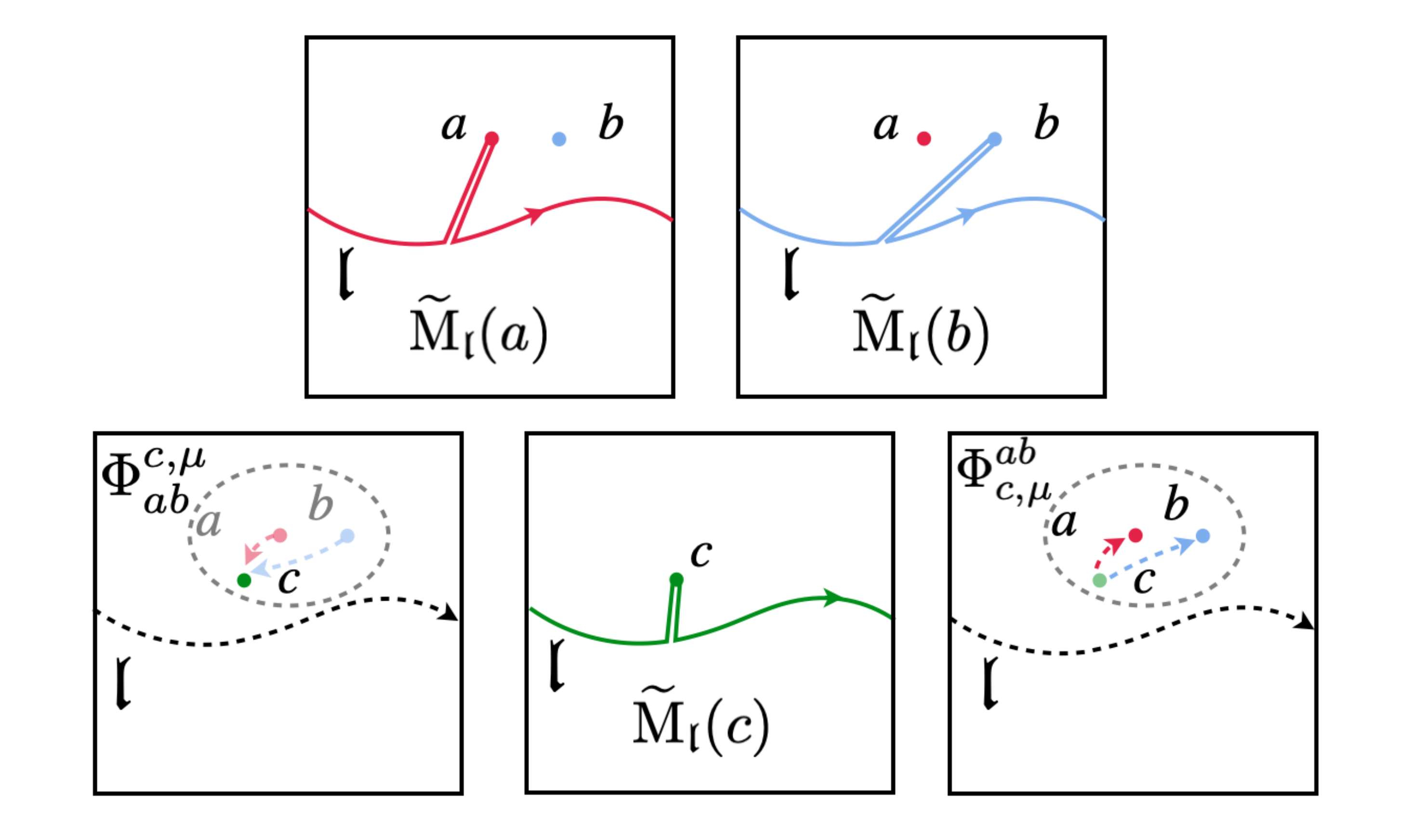}
     \caption{%
     The two rows illustrates two different processes on left/right-hand-side of Eq.~\eqref{eq:fusionM1}.
     On the first row, we move anyon $a$ and $b$ along the path $\llll$ separately.
     On the second row, we first fuse $a$ and $b$ to anyon $c$, and move $c$ along the path $\llll$, and then split $c$ back to $a$ and $b$.
     The main objective of this subsection is show that the two processes are in fact equivalent.}
     \label{fig:MovementEquation}
 \end{figure}

 Diagrammatically, we want to show the two processes in figure \ref{fig:MovementEquation}  are equivalent.


To prove equation \eqref{eq:fusionM1}, we isolate the neighborhood of the torus near $\llll$, which is topologically an annulus.
We first cut an annular neighborhood of $\llll$, we then close one of the punctures to get a disk with an anyon $d$ at the center.
We wish to show that moving $a$ and $b$ around $d$ is equivalent to the steps: fusion $a \times b \to c$, moving $c$ around $d$, and then splitting $c \to a \times b$.
The key construction in our argument is that in the disk geometry, movement of anyon $a$ and $b$ around $d$ is compensated by the movement of $d$ around $a$ and $b$.
Therefore, back to the torus geometry, moving two anyon $a$ and $b$ separately equivalent to moving the fused anyon $c$.
Our method can be easily generalized to higher genus manifolds as long as our assumptions still hold.

\begin{figure}[b]
    \centering
    \includegraphics[width= \linewidth]{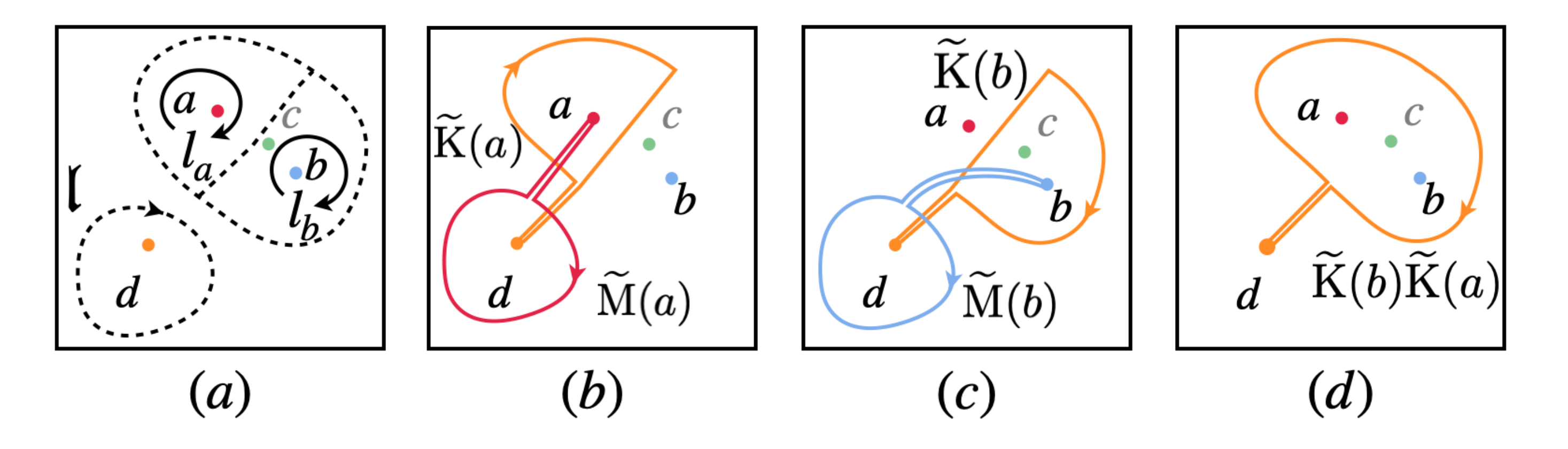}
    \caption{Definition of the Movement operators. $a$ and $b$ are two well separated anyons on the plane. $d$ is the charge carried by the puncture. $\widetilde{\M}(a)$ moves anyon $a$ around the anyon $d$ along the path $\llll$ clockwise, $\widetilde{\K}(a)$ moves anyon $d$ around the anyon $a$ along the path $l_a$ clockwise. $l_a+l_b$ is defined to enclose the anyon $c$.}
    \label{fig:MandD}
\end{figure}

Figure~\ref{fig:MandD} shows the geometry on the disk, with anyon $d$ located at where the puncture used to be.
Here, $\widetilde{\M}(a)$ moves anyon $a$ from $x_a$ to the beginning of $\llll$, then around the anyon $d$ along the path $\llll$ clockwise, and finally returns $a$ to its original position at $x_a$.
Operators $\widetilde{\M}(b)$ and $\widetilde{\M}(c)$ are defined in the similar way. Let $\widetilde{\K}(a)$ moves anyon $d$ around the anyon $a$ along the path $l_a$  clockwise and then returns $d$ to its original position. Operator $\widetilde{\K}(b)$ moves anyon $d$ around the anyon $a$ along the path $l_b$ and $\widetilde{\K}(c)$ moves anyon $d$ around the anyon $c$ along the path $l_a+l_b$.
We choose the paths so that $l_a+l_b$ enclose the anyon $c$. 

The movement operator $\widetilde{\M}(a)$ is equivalent to the operator $\widetilde{\K}(a)$ up to a phase since 
\begin{align}
  \widetilde{\M}(a)  \widetilde{\K}(a)^\dag &=\kappa_\llll(a) \kappa_{l_a}(d) {\xy 
   (-2,-8)="d1"; (-2,-6)="d2";(-2,2)="d3";(-2,4)="d4";(-2,8)="d5";
   (2,-8)="a1"; (2,-4)="a2";(2,-2)="a3";(2,6)="a4";(2,8)="a5";
   "d1";"d2"**[orange]\dir{-};"a2"**[orange]\crv{(-2,-4)};"a3"*{\phdot}**[orange]\crv{(4,-4)&(6,-3)&(4,-2)};"d3"*{\phdot}**[orange]\crv{(-2,-2)};"d5"**[orange]\dir{-};
   "a1";"a2"*{\phdot}**[red]\dir{-};"a3"**[red]\dir{-};"d3"**[red]\crv{(2,2)};"d4"*{\phdot}**[red]\crv{(-4,2)&(-6,3)&(-4,4)};"a4"**[red]\crv{(2,4)};"a5"**[red]\dir{-}; 
   "a1"+(1,0)*{\scr a};"d1"+(-1,0)*{\scr d}
   \endxy}
   =
   \kappa_\llll(a) \kappa_{l_a}(d){\xy 
   (-2,-8)="d1";(-2,8)="d5";
   (2,-8)="a1";(2,8)="a5";
   "d1";"d5"**[orange]\dir{-};
   "a1";"a5"**[red]\dir{-}; 
   "a1"+(1,0)*{\scr a};"d1"+(-1,0)*{\scr d}
   \endxy}\nonumber \\
   & = \kappa_{\llll}(a) \kappa_{l_a}(d) \mathbbm{1},
\end{align}
where $\kappa_\llll(a)$ is the geometric phase attached to $\widetilde{\M}(a)$ while $\kappa_{l_a}(d)$ is the geometric phase attached to $\widetilde{\K}(a)^\dag$.
Crucially, the topological contributions of $\widetilde{\M}$ and $\widetilde{\K}^\dag$ cancels.  Hence the operators are equal up to a phase: $\widetilde{\M}(a) = \kappa_{\llll}(a) \kappa_{l_a}(d) \widetilde{\K}(a)$ (and similarly for $b$ and $c$.)

\begin{figure}[h]
    \centering
    \includegraphics[width=0.45\textwidth]{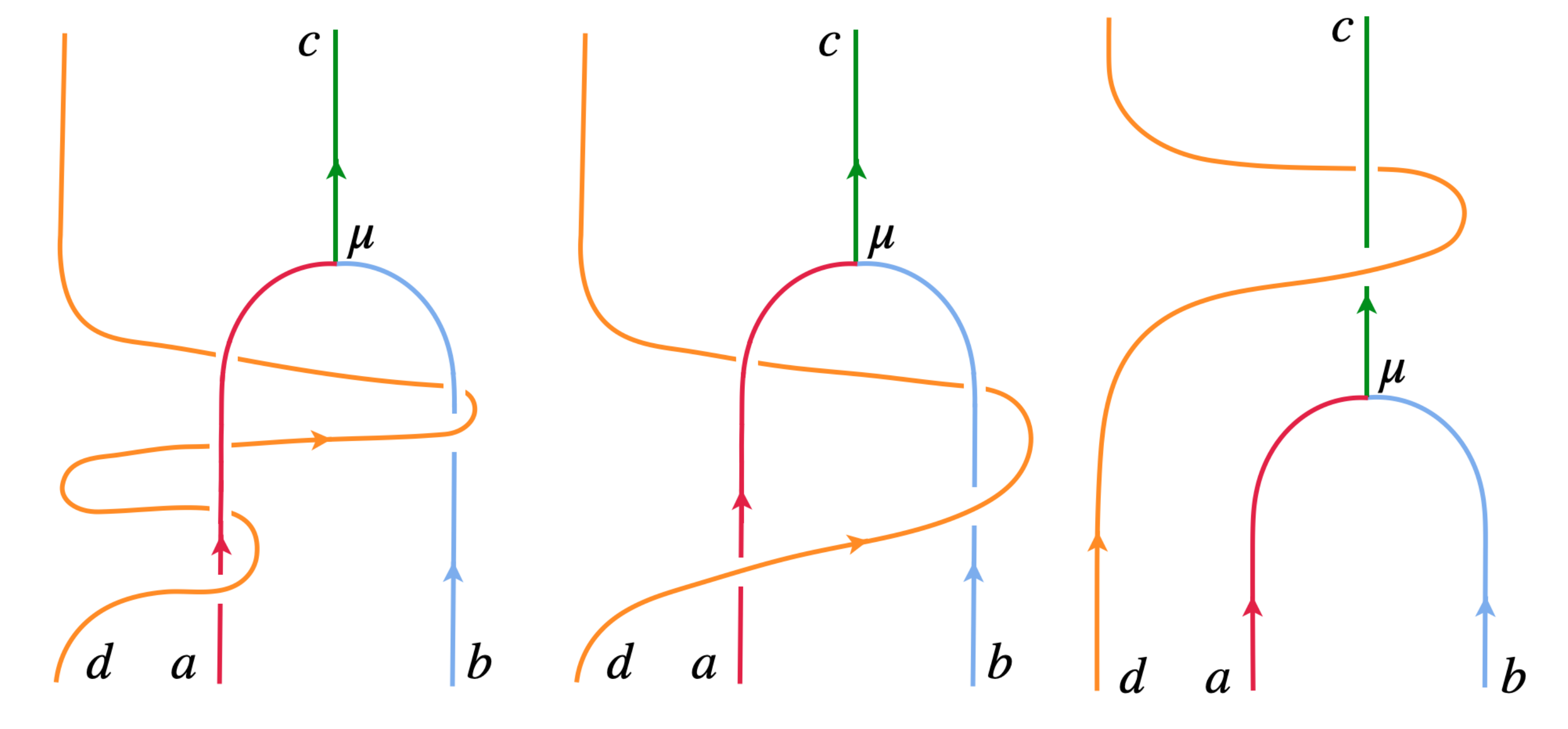}
    \caption{The effects of movement operators $\widetilde{\K}(b)\widetilde{\K}(a)$ and $\widetilde{\K}(c)$ can be divided into two pieces, the geometric part depends on the homotopy class of loop while the topological part depends on the planer diagram. This digram shows that $\Phi_{ab}^c\widetilde{\K}(b)\widetilde{\K}(a) = \widetilde{\K}(c)\Phi_{ab}^c$ because, a) $\widetilde{\K}(b)\widetilde{\K}(a)$ and $\widetilde{\K}(c)$ move the anyon $d$ along the same path so the geometric contribution is the same, b) the topological contribution is the same since the planer diagram is isotopic invariant under such deformation. }
    \label{fig:movefusionoperator}
\end{figure}

Therefore, the movement operators commute with the fusion and splitting operators.
\begin{align}
 \Phi_{ab}^{c,\mu}\widetilde{\M}(b)\widetilde{\M}(a) &= \kappa_\llll(a)\kappa_\llll(b)\kappa_{l_a}(d)\kappa_{l_b}(d)\Phi_{ab}^{c,\mu}\widetilde{\K}(b)\widetilde{\K}(a)\nonumber
\\ 
&= \kappa_\llll(a)\kappa_\llll(b)\kappa_{l_a}(d)\kappa_{l_b}(d)\widetilde{\K}(c)\Phi_{ab}^{c,\mu} \nonumber
\\  
&= \frac{\kappa_\llll(a)\kappa_\llll(b)\kappa_{l_a}(d)\kappa_{l_b}(d)}{\kappa_\llll(c)\kappa_{l_a+l_b}(d)}\widetilde{\M}(c)\Phi_{ab}^{c,\mu} \nonumber
\\  
&= \frac{\kappa_\llll(a)\kappa_\llll(b)}{\kappa_\llll(c)}\widetilde{\M}(c)\Phi_{ab}^{c,\mu} ,
   \label{eq:commute}
\end{align}
where the second step is from figure \ref{fig:movefusionoperator}. Here $\kappa_{l_a}\kappa_{l_b} = \kappa_{l_a+l_b}$ because the geometric phase only depends on the loop and species of anyon. 

Then from the normalization condition \eqref{eq:normalization1}, we have
\begin{align}
   \widetilde{\M}(b)\widetilde{\M}(a) &= \sum_{c,\mu} \sqrt{\frac{d_c}{d_ad_b}}\Phi_{c,\mu}^{ab} \Phi_{ab}^{c,\mu}\widetilde{\M}(b)\widetilde{\M}(a) \nonumber \\
   &=\sum_{c,\mu} \sqrt{\frac{d_c}{d_ad_b}}\Phi_{c,\mu}^{ab} \frac{\kappa_\llll(a)\kappa_\llll(b)}{\kappa_\llll(c)}\widetilde{\M}(c)\Phi_{ab}^{c,\mu}. 
\end{align}
Let $\alpha_\llll(a) = \kappa_\llll(a)^*$, then we get the fusion rule for the movement operator \eqref{eq:fusionM1}.
Then for the Wilson loop operators we have
\begin{align}   &W_\llll(a)W_\llll(b) \nonumber \\ &= \
\alpha_\llll(a)\alpha_\llll(b) \Phi^0_{a\bar{a}} \widetilde{\M}_{\llll}(a)\Phi_0^{a\bar{a}}\Phi^0_{b\bar{b}} \widetilde{\M}_{\llll}(b)\Phi_0^{b\bar{b}} \nonumber\\
    &= \
    \Phi^0_{a\bar{a}}\Phi^0_{b\bar{b}} \sum_{c,\mu}\sqrt{\frac{d_c}{d_ad_b}}\Phi_{c,\mu}^{ab}\alpha_\llll(c)\widetilde{\M}_{{\llll}}(c)
    \Phi_{ab}^{c,\mu}\Phi_0^{b\bar{b}}\Phi_0^{a\bar{a}}\nonumber\\
    &=
    \sum_{c,e,\mu,\nu}{\frac{d_c}{d_ad_b}}\Phi^0_{a\bar{a}}\Phi^0_{b\bar{b}} \Phi_{c,\mu}^{ab}\Phi_{\bar{e},\nu}^{\bar{a}\bar{b}}\alpha_\llll(c)\widetilde{\M}_{{\llll}}(c)
    \Phi_{ab}^{c,\mu}\Phi^{\bar{e},\nu}_{\bar{a}\bar{b}}\Phi_0^{b\bar{b}}\Phi_0^{a\bar{a}}\nonumber\\
    &=\sum_cN_{ab}^c\Phi^{0}_{c\bar{c}}\alpha_\llll(c)\widetilde{\M}_{{\llll}}(c)\Phi_{0}^{c\bar{c}}\nonumber\\
    &=
    \sum_cN_{ab}^cW_\llll(c).
\end{align}
In the third step, we insert the identity \begin{align}
\sum_{e,\nu}\sqrt{\frac{d_e}{d_ad_b}}\Phi^{\bar{a}\bar{b}}_{\bar{e},\nu}\Phi_{\bar{a}\bar{b}}^{\bar{e},\nu} = \mathbbm{1}_{\bar{a}\bar{b}},
\end{align}
and then use the fact 
\begin{align}
     \Phi^0_{a\bar{a}}\Phi^0_{b\bar{b}}\Phi^{{a}{b}}_{c,\mu}\Phi^{\bar{a}\bar{b}}_{\bar{e},\nu} = \sqrt{\frac{d_ad_b}{d_c}}A_{\mu,\nu}\Phi^{0}_{c\bar{c}}\delta_{c,e},
     \label{eq:SplittingFusion}
\end{align}
where $A_{\mu,\nu}$ is a phase dependent on $a,b,c$.

For equations~\eqref{eq:conjureW1} and \eqref{eq:antiW1}, since the Wilson loop operators are independent from the location where the particle-antiparticle pair is created, we only need to prove
\begin{align}
   \alpha^*_{\llll}(a) \widetilde{\M}^\dagger_\llll(a)=\alpha_{\bar{\llll}}(a)\widetilde{\M}_{\bar{\llll}}(a),\\
    \alpha^*_{{\llll}}(a)\widetilde{\M}_{{\llll}}^\dag(a) =\alpha_{{\llll}}(\bar{a}) \widetilde{\M}_{\llll}(\bar{a}).
\end{align}
The first equation can be easily obtained since 
\begin{align}
\begin{aligned}
    \alpha_{\llll}(a)\widetilde{\M}_{\llll}(a)\alpha_{\bar{\llll}}(a)\widetilde{\M}_{\bar{\llll}}(a) &= \kappa_{\llll}(a)^*\kappa_{\bar{\llll}}(a)^*\widetilde{\M}_{\llll}(a) \widetilde{\M}_{\bar{\llll}}(a)\\
    &=\mathbbm{1}_{a}
\end{aligned}
\end{align}
To show the second equation, we use result \eqref{eq:commute} that the fusion operators commute with the movement operators
\begin{align}
    \alpha_{\llll}(a)\alpha_{\llll}(\bar{a})\Phi_{a\bar{a}}^{0}\widetilde{\M}_{\llll}(a)\widetilde{\M}_{\llll}(\bar{a}) = \alpha_{\llll}(0)\widetilde{\M}_{\llll}(0)\Phi_{a\bar{a}}^0
\end{align}
Because the movement operators for the trivial anyon $0$ are just the identity, we have
\begin{align}
    \alpha_{\llll}(a)\alpha_{\llll}(\bar{a})\widetilde{\M}_{\llll}(a)\widetilde{\M}_{\llll}(\bar{a}) = \mathbbm{1}_{a\bar{a}}
\end{align}

\section{Deriving the threading diagram}\label{sec:threading}

\begin{figure}[b]
    \centering
    \includegraphics[width=\linewidth]{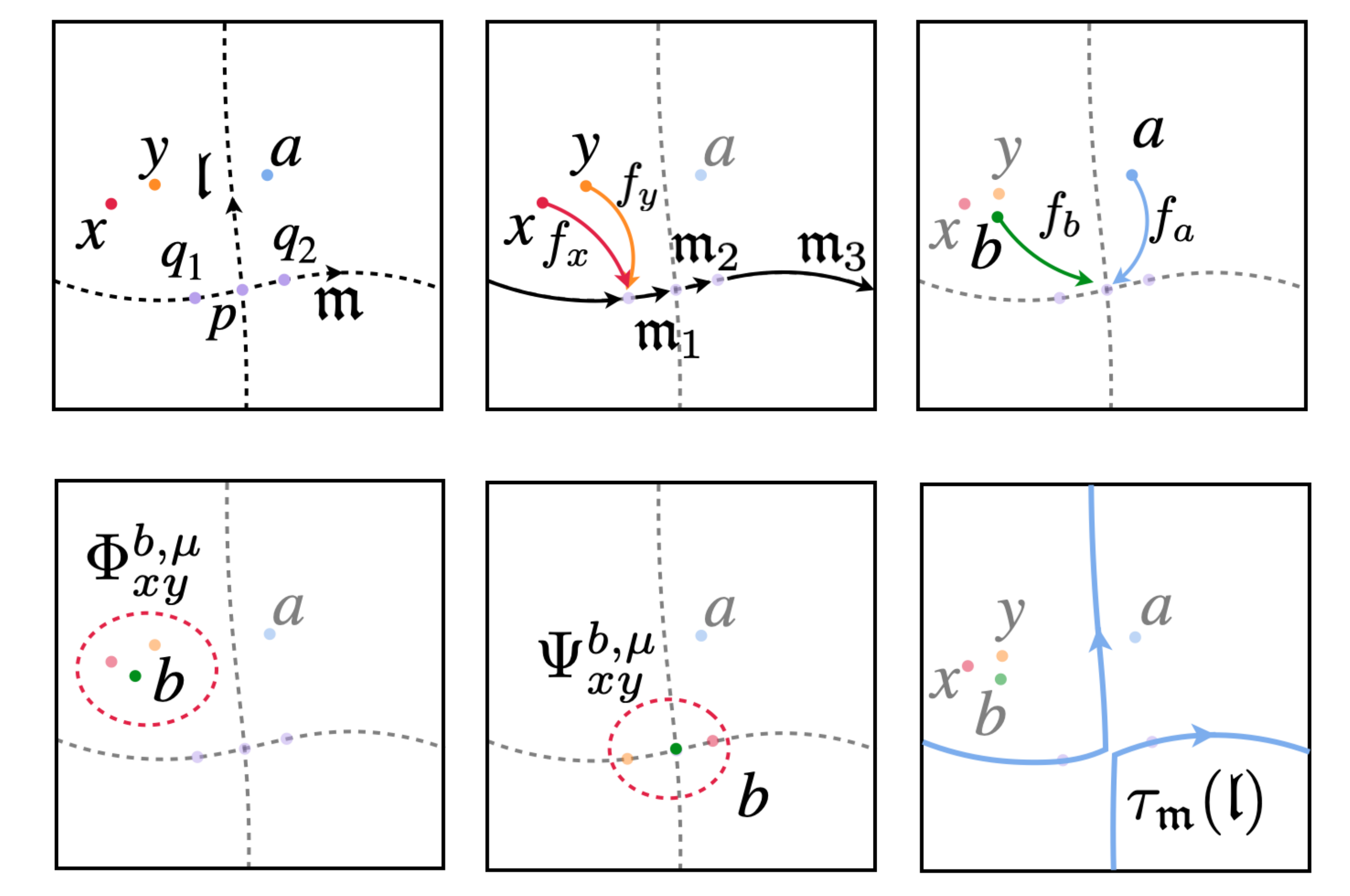}
    \caption{%
    Initially $x$ and $y$ are anyons locate at point $X$ and $Y$ respectively.
    Let $p$ be the intersection of the loops $\llll$ and $\mmm$.
    The two points $q_1$ and $q_2$ are on the path $\mmm$ in the neighborhood of point $p$.
    $b$ is an anyon at point $B$ that satisfy $N_{xy}^b>0$. $f_b$ is the path from the point $B$ to $p$.
    Path $f_x$ connects the point $X$ with the point $q_1$, path $f_y$ links the point $Y$ with the point $q_1$. Loop $\mmm$ is divided into 3 paths $\mmm_1,\mmm_2$ and $\mmm_3$.
    $a$ is an anyon at point $A$. Path $f_a$ is from point $A$ to point $p$. $\Phi_{xy}^{b,\mu}$ and $\Psi_{xy}^{b,\mu}$ are two fusion operators in different locations, so they may different from a arbitrary phase factor. $\tau_\mmm(\llll)= \llll+\mmm$ is the loop after the Dehn twist about loop $\mmm$.}
    \label{fig:paths}
\end{figure}

In this section, we will derive the threading diagram from the microscopic definition of the Wilson loop operators.
Here, the Wilson loop operators $W_{\mmm}(x)$, $W_{\mmm}(y)$ and $W_{\llll}(a)$ are defined based on the paths in Figure \ref{fig:paths}
\begin{align}
    W_{\mmm}(x) &= \Phi_{x\bar{x}}^0\M_{f_x+\mmm+\bar{f}_x}(x)\Phi^{x\bar{x}}_0,
\\  W_{\mmm}(y) &= \Phi_{y\bar{y}}^0\M_{f_y+\mmm+\bar{f}_y}(y)\Phi^{y\bar{y}}_0.
\end{align}
Here we implicitly include the phase factors $\alpha_{\llll}$ within the movement operators, so they will not be explicitly written out.

The geometry is shown in Figure~\ref{fig:paths}.
Let $\Psi_{xy}^{b,\mu}$ be the fusion operators which take anyon $x$ at point $q_1$ and anyon $y$ at point $q_2$ as input and output anyon $b$ at point $p$.
Let $\Phi_{xy}^{b,\mu}$ be the fusion operators that take anyon $x$ at point $X$ and anyon $y$ at point $Y$ as input and output anyon $b$ at point $B$.

\begin{figure*}
    \centering
    \includegraphics[width = \textwidth]{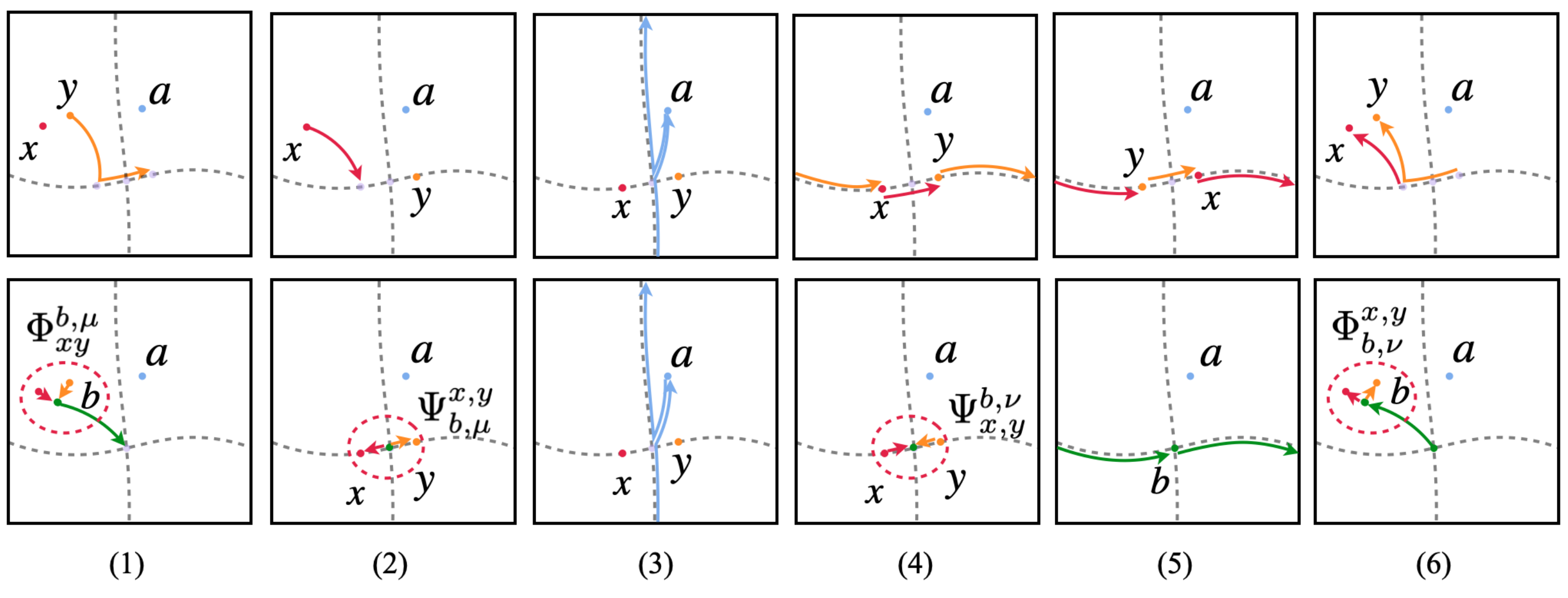}
    \caption{Both rows describe the same process $W_{\llll}(x)W_{\mmm}(a)W_{\llll}(y)$. In the first row, we move the anyon $x$ and $y$ separately, while in the second row, we first fuse $x$ and $y$, and then move the outcome anyon $b$ alone the loop $\llll$. Here we omit the antiparticle $\bar{x},\bar{y},\bar{a}$ and $\bar{b}$. }
    \label{fig:2processes}
\end{figure*}

The RHS of equation \eqref{eq:DehnTwist} can be described by the two equivalent processes in Figure \ref{fig:2processes}. In the first row, we move the anyon $x$ and $y$ separately, while in the second row, we first fuse $x$ and $y$, and then move the outcome anyon $b$ along the loop $\mmm$. The first two diagrams in both rows describe the equivalent process because 
\begin{align}
   &\M_{f_x}(x)\M_{f_y+\mmm_1+\mmm_2}(y) =\sum_{b,\mu}\sqrt{\frac{d_b}{d_xd_y}}\kappa_b\Psi_{b,\mu}^{xy} \M_{f_b}(b)\Phi_{xy}^{b,\mu},
   \label{eq:paths1}
\end{align}
where $\kappa_b$ are arbitrary phases (which also depends on $x$ and $y$). 
From section \ref{defW}, we have the following identity~\eqref{eq:fusionM},
\begin{align}
   &\M_{\mmm}(x)\M_{\mmm}(y) =\sum_{c,\xi} \sqrt{\frac{d_c}{d_xd_y}} \Psi^{xy}_{c,\xi}\M_{\mmm}(c)\Psi_{xy}^{c,\xi}.\label{eq:paths2}
\end{align}
And by reversing all paths in \eqref{eq:paths1} we have,
\begin{align}
   &\M_{\bar{f}_x}(x)\M_{\bar{f}_y+\bar{\mmm}_1+\bar{\mmm}_2}(y) =\sum_{d,\nu}\sqrt{\frac{d_d}{d_xd_y}}\kappa_d^* \Phi^{xy}_{d,\nu}\M_{\bar{f}_b}(d)\Psi_{xy}^{d,\nu}.\label{eq:paths3}
\end{align}
\begin{widetext}
Combining these equations we have,
\begin{align}
\begin{aligned}
   W_{\mmm}(x)W_{\llll}(a)W_{\mmm}(y) &=      \Phi_{x\bar{x}}^0\M_{f_x+\mmm+\bar{f}_x}(x)\Phi^{x\bar{x}}_0
    W_{\llll}(a)
 \Phi_{y\bar{y}}^0\M_{f_y+\mmm+\bar{f}_y}(y)\Phi^{y\bar{y}}_0 \\
 & = \Phi_{x\bar{x}}^0\Phi_{y\bar{y}}^0 \M_{\bar{f}_x}(x)\M_{\bar{f}_y+\bar{\mmm}_1+\bar{\mmm}_2}(y)\M_{\mmm}(x)\M_{\mmm}(y)W_{\llll}(a) \M_{f_x}(x)\M_{f_y+\mmm_1+\mmm_2}(y)
 \Phi^{y\bar{y}}_0\Phi^{x\bar{x}}_0 \\
 &=  
 \sum_{b,d,\mu,\mu',\nu,\nu'}{ \sqrt{\frac{d_b}{d_xd_y}}}{ \sqrt{\frac{d_d}{d_xd_y}}}B_{\nu,\nu'}
 \Phi_{d\bar{d}}^{0}\Phi_{\bar{x}\bar{y}}^{\bar{d},\nu'}\kappa_d^*\M_{\bar{f}_b}(d) \M_{\mmm}(d)\Psi_{xy}^{d,\nu}
W_\llll(a)
\Psi_{b,\mu}^{xy} \kappa_b\M_{f_b}(b)
B_{\mu,\mu'}^*\Phi_{\bar{b},\mu'}^{\bar{x}\bar{y}}\Phi_{0}^{b\bar{b}} \\
& = \sum_{b,\mu}{ \sqrt{\frac{d_b}{d_xd_y}}} \Phi_{b\bar{b}}^0\M_{\llll+\bar{f}_b}(b)\Psi_{xy}^{b,\mu} W_{\llll}(a) \Psi_{b,\mu}^{xy} \M_{f_b}(b)\Phi^{b\bar{b}}_0.
    \label{eq:threading}
\end{aligned}
\end{align}
\end{widetext}
Here the third line is obtained from identities \eqref{eq:paths1}, \eqref{eq:paths2}, \eqref{eq:paths3} and the pivoting identity
\begin{align}
   \Phi_{x\bar{x}}^{0} \Phi_{y\bar{y}}^{0} \Phi_{d,\nu}^{xy} = \sum_{\nu'} B_{\nu,\nu'} \Phi_{d\bar{d}}^{0}\Phi_{\bar{x}\bar{y}}^{\bar{d},\nu'},
\end{align}
where $B_\nu$ is a unitary matrix (dependent on $x,y,d$).
The contraction
\begin{align}
    \Phi^{\bar{d},\nu'}_{\bar{x}\bar{y}}\Phi_{\bar{b},\mu'}^{\bar{x}\bar{y}} = \delta_{\mu',\nu'}\delta_{b,d} { \sqrt{\frac{d_xd_y}{d_b}}}\mathbbm{1}_{\bar{b}}.
\end{align}
enforces $b = d$ and $\mu = \nu$ which allows simplification in the last line.
(Recall that during the entire process, the anyons $\bar{x}$, $\bar{y}$ and $\bar{b}$ do not move once they are created.
Also, during this process, we carefully choose the path for anyon $x$ and $y$ along the loop $\mmm$, so that they do not collide during the movement.)

Now we get the term $\Psi_{xy}^{b,\mu} W_{\llll}(a) \Psi_{b,\mu}^{xy}$, which is the interlocked loops in figure \ref{fig:threading}, and also represented by the diagrams (2)--(4) in the second row of figure~\ref{fig:2processes}.
Since the splitting and fusion operator are defined at the same point, there are no geometric phases caused by the movement operators (red line and orange line in the calculation \eqref{l+m}).

Then, by plugging the result from equation \eqref{l+m}, i.e.,
\begin{align}
    \sum_{x,y,\mu}\frac{1}{\tqd^2}  \sqrt{d_xd_yd_b}\theta_x^*\theta_y\Psi_{xy}^{b,\mu} W_{\llll}(a) \Psi_{b,\mu}^{xy} = \M_{\llll}(a)\delta_{ab} ,
\end{align}
into the interlocked equation \eqref{eq:threading}, we will have
\begin{align}
\begin{aligned}
     &\sum_{x,y}\frac{d_xd_y\theta_x^*\theta_y}{\tqd^2}W_{\mmm}(x)W_{\llll}(a)W_{\mmm}(y)\\ 
     &= \sum_b\Phi_{b\bar{b}}^0\M_{\mmm+\bar{f}_b}(b)\M_{\llll}(a)\delta_{ab} \M_{f_b}(b)\Phi^{b\bar{b}}_0 \\
     &=W_{\llll+\mmm}(a) 
   \end{aligned}
\end{align}
Therefore, we have the Dehn twist equation 
\begin{align}
    	\Dehn_{\mmm} W_{\llll}(a)\Dehn_{\mmm }^\dagger =  W_{(\mmm\times\llll)\mmm+\llll}(a)=W_{\tau_{\mmm }(\llll)}(a).
\end{align}

\section{The eigenstates of Wilson loop operators are MESs}
The $n$\textsuperscript{th} Renyi bipartite entanglement entropy for a ground states $\ket{\Phi} = \sum_i c_i \ket{i^\llll_\mmm}$ along the cut $\mmm$ are given by \cite{YiZhang2012GroundStateEntanglement,Dong2008EntanglementEntropy}:
\begin{align}
    S_n(\ket{\Phi})  = \alpha_n L - \gamma_n',
\end{align}
where $\ket{i^\llll_\mmm}$ are the standard basis of the ground states, (and they are eigenstates of $W_\mmm(a)$), $\alpha_n$ is the non-universal boundary law term, $L$ is the length of the boundary and $\gamma_n'$ is the topological entanglement entropy  
\begin{align}
   \gamma_n' = 2\ln(\tqd) + \frac{1}{n-1} \ln\left(\sum_i |c_i|^{2n}d_i^{2(1-n)}\right).
\end{align} When $n=1$ we get the von Neumman entropy:
\begin{align}
\gamma_1' = 2\ln(\tqd) + \sum_i |c_i|^2 \ln\left(\frac{|c_i|^2}{d_i^2}\right).
\end{align}
We can show that for all $n$, the eigenstates of $W_\mmm(a)$ are the states that have the local minimum of bipartite entanglement entropy $S_n$. 

Without loss of generality, we show the state $\ket{c^\llll_\mmm}$ has the local minimum of $S_n$. We have
\begin{align}
    S_n(\ket{c^\llll_\mmm}) =
    \alpha_n L -2\ln(\tqd) +2 \ln\left(d_c\right). 
\end{align}
So we only need to show that $\gamma'-2\ln(\tqd) = -2 \ln\left(d_c\right)$ is the local maximum.
Consider the perturbation state $\ket{\Phi} = \frac{1}{\sqrt{1+|\Delta|^2}}(\ket{c^\llll_\mmm}+\Delta\ket{\varphi})$, where $\ket{\varphi}$ is the state orthogonal to $\ket{c^\llll_\mmm}$, $|\Delta|\ll1$. For $n>1$ we have
\begin{align}
\begin{aligned}    
\gamma_n'-2\ln(\tqd) &=  \frac{1}{n-1} \ln \left( \frac{d_c^{2(1-n)}}{(1+|\Delta|^2)^n}+ \sum_{i \neq c} \cdots \right) \\
   & < \frac{1}{n-1} \ln \left( \frac{d_c^{2(1-n)}}{(1+|\Delta|^2)^n}+ \frac{|\Delta|^{2n}d_{max}^{2(1-n)}}{(1+|\Delta|^2)^n}\right)\\
   &=-2\ln(d_c)-\frac{2n}{n-1}|\Delta|^2 <-2\ln(d_c).
\end{aligned}
\end{align}
Here $d_{max}$ is the maximal quantum dimension and we only keep the the first order of $|\Delta|^2$.
For $n<1$ we have
 \begin{align}
\begin{aligned}    
\gamma_n'-2\ln(\tqd) &=  \frac{1}{n-1} \ln \left( \frac{d_c^{2(1-n)}}{1+|\Delta|^2}+ \sum_{i \neq c} \cdots \right) \\
   & < \frac{1}{n-1} \ln \left( \frac{d_c^{2(1-n)}}{(1+|\Delta|^2)^n}+ \frac{|\Delta|^{2n}d_{0}^{2(1-n)}}{|\mcA|^{2n}(1+|\Delta|^2)^n}\right)\\
   &=-2\ln(d_c)+\frac{1}{n-1}\ln(1+|\Delta|^2n) \\
   & \phantom{=} +\frac{1}{n-1}\ln(1+\frac{|\Delta|^2n}{|\mcA|^{2n}{d_c}^{2(1-n)}})
   \\
   &<-2\ln(d_c).
\end{aligned}
\end{align}  
When $n =1$ we have
\begin{align}
    \begin{aligned}
       \gamma_1'-2\ln(\tqd) &= \frac{1}{1+|\Delta|^2} \ln \left(\frac{1}{1+|\Delta|^2}\frac{1}{d_c^2}\right) +\sum_{i\neq c}\cdots \\
       &<\frac{1}{1+|\Delta|^2} \ln \left(\frac{1}{1+|\Delta|^2}\frac{1}{d_c^2}\right) \\
       &\phantom{=}+\frac{|\Delta|^2}{1+|\Delta|^2}\ln(\frac{|\Delta|^2}{1+|\Delta|^2})\\
       &=-2\ln(d_c)+|\Delta|^2(\ln(|\Delta|^2)+\ln(d_c^2)-1)
       \\&<-2\ln(d_c)
    \end{aligned}
\end{align}
Here we only keep the first order of $|\Delta|^2$. Therefore, the eigenstates of $W_\mmm(a)$ have the minimal bipartite entanglement entropy along the $\mmm$-cuts.

Thus for any $n>0$, the eigenstates of $W_\mmm(a)$ have the local minimal bipartite entanglement entropy to the cut parallel to the loop $\mmm$. In particular, the states associated with the Abelian anyon have the global minimum bipartite entanglement entropy.

Let $\ket{\Phi}$ has the minimum entanglement entropy in the $\mmm$ direction. Then for arbitrary Wilson loop operators $W_\mmm(a), W_\mmm(b)$ the connected correlation function is 0, i.e
\begin{align}
    \braket{\Phi|W_\mmm(a)W_\mmm(b)|\Phi} = \braket{\Phi|W_\mmm(a)|\Phi}\braket{\Phi|W_\mmm(b)|\Phi}.
\end{align}
From equation \eqref{eq:fusionW1} we have
\begin{align}
    \braket{\Phi|W_\mmm(a)W_\mmm(b)|\Phi} = \sum_c N_{ab}^c \braket{\Phi|W_\mmm(c)|\Phi}
\end{align}
Let $\beta(a)$ be the expectation value of Wilson loop operator $W_\mmm(a)$, $\beta(a) = \braket{\Phi|W_\mmm(a)|\Phi}$, then $\{\beta(a)\}$ forms a fusion character
\begin{align}
    \beta(a)\beta(b) = \sum_c N_{ab}^c\beta(c)
\end{align}
Since all the fusion characters are given by the eigenstate of Wilson loop operators, $\ket{\Phi}$ is the eigenstate of the Wilson loop operators.

\section{Wilson loop gauge transformations}\label{sec:gaugetransformation}
The Wilson loop operators defined in section \ref{defW} are not unique, there are alternate phase factors that can be attached to the movement/Wilson loop operators that still yields a consistent Wilson loop algebra.

Assume $W_\llll(a)$ is a set of Wilson loop operators which satisfies Eqs.~\eqref{fusionW}, \eqref{conjureW} and \eqref{antiW}.
Let
\begin{align}
    W'_\llll(a) = \lambda_\llll(a) W_\llll(a) .
    \label{eq:W_gauge_transform}
\end{align}
Here $ \lambda_\llll(a) $ is the fusion phase attached to the Wilson loop operators in the $\llll$ direction. Then $W'_\llll(a)$ is also a set of Wilson loop operators that remains compatible with Eqs.~\eqref{fusionW}. We will refer to Eq.~\eqref{eq:W_gauge_transform} as ``gauge transformations''. Recall from section \ref{sec:notations}, the fusion phase take the form $
    \lambda_\llll(a) = \frac{S_{r,a}}{S_{0,a}},
$
where $r \in \mcA$ is an Abelian anyon.

To get a set of compatible Wilson loop operators among different loops on the torus, there are additional restrictions of factor $\lambda_\llll(a)$.  Let $W_\llll(a)$ and $W_\mmm(a)$ be the basis of the Wilson loop operators, and then we attached
\begin{align}
   &\lambda_{\llll}(a) = \frac{S_{r,a}}{S_{0,a}},& \lambda_{\mmm}(a) = \frac{S_{s,a}}{S_{0,a}}&
\end{align}
to these operators respectively. Then we have:
\begin{gather}
   \lambda_\llll^*(a) = \lambda_{{\llll}}(\bar{a})=
    \lambda_{\bar{\llll}}(a) ,\label{eq:conjureP}\\
    \lambda_{\llll+\mmm}(a)=\lambda_\llll(a)\lambda_\mmm(a).\label{eq:DehnP}
\end{gather}
The first one is directly from the the Wilson loop algebra Eqs.~\eqref{antiW} and \eqref{conjureW}. The last one is from the Dehn twist equation~\eqref{eq:DehnTwist}:
\begin{align}
W_{\llll+\mmm}'(a) &= \sum_y \sum_x \frac{d_xd_y \theta_x^*\theta_y}{\tqd^2}W_\mmm'(x)W_\llll'(a)W_\mmm'(y)\nonumber\\
&=\lambda_\llll(a)\sum_u \sum_v \frac{d_ud_v \theta_u^*\theta_vS_{s,a}}{\tqd^2S_{0,a}}W_m(u)W_\llll(a)W_\mmm(v)\nonumber\\
&=\lambda_\llll(a)\lambda_\mmm(a)W_{\llll+\mmm}(a)\nonumber\\
& = \lambda_{\llll+\mmm}(a) W_{\llll+\mmm}(a)
\end{align}
Here we use the identities $
    S_{a,b} = \sum_c N_{ab}^c d_c\frac{\theta_a\theta_b}{\theta_c},$ and 
    \begin{align}
    W_\mmm(a)W_\llll(b) = \frac{\tqd S^*_{a,b}}{{d_ad_b}} W_\llll(b)W_\mmm(a).
\end{align}
The second identity holds if $a$ or $b$ is Abelian.

From section \ref{sec:basisW}, we know that all the Wilson loop operators are generated by two sets of Wilson loop operators that defined along the coordinate basis $(\llll,\mmm)$. Therefore, to get the gauge transformation of all the Wilson loop operators, we only need to determine the phase attached to these two set of Wilson loop operators.
Then, from restriction~\eqref{eq:conjureP} and \eqref{eq:DehnP}, the phase attached to an arbitrary simple loop $\sss = p\llll+q\mmm$ is given by
\begin{align}
    \lambda_\sss(a) = \lambda^p_\llll(a)\lambda^q_\mmm(a) = \frac{S_{r^ps^q,a}}{S_{0,a}} .\label{eq:cohomology}
\end{align}
This equation assigns each loop on the torus with an Abelian anyon $r^ps^q$, therefore it is a function that maps the set of close loops on the manifold to the Abelian group $\mcA$. More generally, the gauge transformations on a manifold $\mathcal{M}$ is isomorphic to the cohomology group $H^1(\mathcal{M}, \mcA)$.

There are $|\mcA|$ different fusion phases in $\mcC$, so that there are $|\mcA|^2$ different ways to assign phases to the Wilson loop operators. Therefore, we have $|\mcA|^2$ different gauge transformations of the Wilson loop operators in total. And the Wilson loop operators generated by these gauge transformations automatically satisfy the Wilson loop algebra: both $W_\llll(a)$ and $W'_\llll(a)$ are valid solutions to equation~\eqref{fusionW}-\eqref{conjureW} and the Dehn twist operator~\eqref{eq:DehnTwist}. So none of them are special, and that is why we call these transformations the gauge transformations of the Wilson loop operators.

\section{Rearrangement of ground states labeling}\label{sec:rearrange}
The basis of the ground states are labeled with the anyons in $\mcC$ by their eigenvalues \eqref{defMm}. The transformation defined in last section \eqref{eq:GaugeOfL} changes the eigenvalues of the Wilson loop operators, so that it will also change the labels of the basis. Furthermore, since the relative phases are fixed by the Wilson loop operators \eqref{defMm}, the gauge transformation also attaches a phase factor to the basis. 
%
%
The explicit form of the gauge transformation can be easily obtained from the definition of the basis.

Let's consider the gauge transformation
\begin{align}
    {W}_\mmm(a)'&=\lambda_\mmm(a){W}_\mmm(a)=\frac{S_{s,a}}{S_{0,a}}{W}_\mmm(a),\label{eq:GaugeOfM}\\
    {W}_\llll(a)'&=\lambda_\llll(a){W}_\llll(a)=\frac{S_{r,a}}{S_{0,a}}{W}_\llll(a) ,\label{eq:GaugeOfL}
\end{align}
where $r,s \in \mathcal{C}$ are Abelian anyons.
As it turns out, it is possible to write
\begin{align}
    {W}_\mmm(a)'&={G}{W}_\mmm(a){G}^\dagger ,
&   {W}_\llll(a)'&={G}{W}_\llll(a){G}^\dagger ,
\end{align}
with
\begin{align}
    {G} = {W}_\llll({s}){W}_\mmm(\bar{r}) .
\end{align} 
Because $W_\llll(a)$ and $W_\mmm(a)$ are the generators of the Wilson loop operators, for an arbitrary direction $\sss$ we have
\begin{align}
    W_\sss(a)' = G  W_\sss(a) G^\dag = \lambda_\sss(a)  W_\sss(a).
\end{align}
Then let $ \ket{b^\sss_\uuu}$ be a set of standard basis for the coordinate $(\sss,\uuu)$, the gauge transformation $G$ generates another set of standard basis:
\begin{align}
	\ket{\widehat{b^\sss_\uuu}}  ={G}\ket{b^\sss_\uuu} = {W}_\llll({s}){W}_\mmm(\bar{r})\ket{b^\sss_\uuu} .\label{eq:GTofgroundstates}
\end{align}
In particular, for the coordinate $(\llll,\mmm)$ we have
\begin{align}
    \ket{\widehat{b^\llll_\mmm}} = \frac{S_{r,b}}{S_{0,b}}\ket{sb^\llll_\mmm}. 
\end{align}
Notice that the gauge transformation of $\{W_\mmm\}$ permutes the label of basis $\ket{b^\llll_\mmm}$ while the gauge transformation of $\{W_\llll\}$ changes the relative phases, which is equivalent to a permutation of basis $\ket{b_\llll^{-\mmm}}$.
Indeed, the two transformations are independent from each other. 

It is straightforward to check that this ground states rearrangement is compatible with the modular transformation we define in section \ref{sec:MT}.
\begin{align}
    \braket{\widehat{a^\llll_\mmm}|\widehat{a^{-\mmm}_\llll}} &= \braket{{a^\llll_\mmm}| G^\dag G|{a^{-\mmm}_\llll}} = \braket{{a^\llll_\mmm}|{a^{-\mmm}_\llll}} = S_{ab},
    \\
    \braket{\widehat{a^\llll_\mmm}|\widehat{a^{\llll+\mmm}_\mmm}} &= \braket{{a^\llll_\mmm}| G^\dag G|{a^{\llll+\mmm}_\mmm}} = \braket{{a^\llll_\mmm}|{a^{\llll+\mmm}_\mmm}}=T_{ab}.
\end{align}

From subsection \ref{sec:gaugetransformation} we know that there are $|\mcA|^2$ gauge transformations in total, therefore, there are $|\mcA|^2$ different ways to rearrange the MESs.
Since the gauge transformation permutes the label of the basis, if the system contains an Abelian anyon besides the vacuum, then there the assignment of anyons to ground states (compatible with the Wilson loop algebra) is not unique.

For example, the semion model contains two Abelian anyon $0$ and $\sigma$, which obey the fusion relation
\begin{align}
   & 0\times \sigma = \sigma \times 0 = \sigma, &\sigma\times\sigma = 0.
\end{align}
Then we have four gauge transformations
\begin{align}
\begin{aligned}
     {G} \ = \ \mathrm{id},  \quad\hat{W}_{\mmm}(\sigma),  \quad\hat{W}_{\llll}(\sigma),  \quad\hat{W}_{\llll}(\sigma)\hat{W}_{\mmm}(\sigma).
\end{aligned}
\end{align}
Under the gauge transformations, the overlap matrices are invariant,
so there are 4 ways to rearrange the basis of the ground states.

\section{Construction of indistinguishable pairs of TQFTs}\label{appx:construction}
Let $\mcC$ be a modular tensor category with a self-dual Abelian anyon $r$. Here we will show how to construct the indistinguishable pair $(\mcC,\widetilde{\mcC})$ from $\mcC$ and $r$.

The topological spin of $r$ can only be $\pm1, \pm i$ because
\begin{align}
    \tqd S_{r,r} = \frac{\theta_0}{\theta_r\theta_r} = \pm 1.
\end{align}
And the anyon in $\mcC$ can be divided into two sets $\mathcal{A} = \{a| S_{a,r}/S_{a,0} =1\}$ and $\mathcal{B} = \{b| S_{b,r}/S_{b,0} =-1\}$. The anyons in $\mathcal{A}$ have the same topological spin in two fusion categories while the anyons in $\mathcal{B}$ have the opposite topological spin. 
\begin{align}
    &\theta_a = \tilde{\theta}_a, \text{ for } a\in \mathcal{A},&& \theta_b = \tilde{\theta}_b\text{ for } b\in \mathcal{B}.&
\end{align}

If $\theta_r =1$, we have 
\begin{align}
 &  \frac{S_{b,r}}{S_{b,0}} = \frac{\theta_{br}}{\theta_b \theta_r} = -1 && \Rightarrow &&\theta_{br} = -\theta_b.&
\end{align}
Then, the model $\widetilde{\mcC}$ are constructed by rename the anyon in $\mcC$. 
\begin{align}
    &a \to a, \text{ for } a\in \mathcal{A},&& b \to br, \text{ for } b\in \mathcal{B}.&
\end{align}
It is easy to check these two model have the same $S$-matrix
\begin{align}
\begin{aligned}
    &\tilde{S}_{i,j} = S_{i,j}, &\text{ if }i,j \in \mathcal{A}\\
    &\tilde{S}_{i,j} = \frac{S_{i,j}S_{r,j}}{S_{0,j}} = S_{i,j}, &\text{ if } i \in \mathcal{B}, j \in \mathcal{A},\\
    &\tilde{S}_{i,j} = \frac{S_{i,j}S_{r,i}S_{r,j}S_{r,r}}{S_{0,i}S_{0,j}} = S_{i,j}, &\text{ if } i,j \in \mathcal{B}.
\end{aligned}
\end{align}

If $\theta_r =-1$, consider the three-fermion model $\mcC_{3f} = \{0,e,m,f\} $.
\begin{align}
    S_{3f} =\frac{1}{2} \begin{pmatrix}
    1 & 1&1&1\\
    1 & 1&-1&-1\\
    1 & -1&1&-1\\
    1 & -1&-1&1
    \end{pmatrix}, T_{3f}=\begin{pmatrix}
    1 & 0&0&0\\
    0 & -1&0&0\\
    0 & 0&-1&0\\
    0 & 0&0&-1
    \end{pmatrix}
\end{align}
It is easy to check that after the anyon condensation of boson $(r,f)$, the anyons in $\mcC\times \mcC_{3f} $ that not confined are
\begin{align}
    &(a,0),(a,f) \text{ for } a\in \mathcal{A}\\
    &(b,e),(b,m) \text{ for } b \in \mathcal{B}.
\end{align}
Define $\tilde{\mcC}=\{[(a,0)],[(b,e)]\}$ as the fusion category after the condensation of $(r,f)$. We have
\begin{align}
   & \theta_{(a,0)} = \theta_{a}, && \theta_{(b,e)} = -\theta_{b}. &
\end{align}
It is easy to check that $\mcC$ and $\tilde{\mcC}$ have the same $S$-matrix.

If $\theta_r =-i$, consider the semion square model $C_\text{2sem}= \{0,e,m,f\} $.
\begin{align}
    S_{\text{2sem}} =\frac{1}{2} \begin{pmatrix}
    1 & 1&1&1\\
    1 & -1&1&-1\\
    1 & 1&-1&-1\\
    1 & -1&-1&1
    \end{pmatrix}, T_{\text{2sem}}=\begin{pmatrix}
    1 & 0&0&0\\
    0 & i&0&0\\
    0 & 0&i&0\\
    0 & 0&0&-1
    \end{pmatrix}
\end{align}
After the anyon condensation of boson $(r,e)$, the anyons in $\mcC\times \mcC_\text{2sem}$ that not confined are
\begin{align}
    &(a,0),(a,m) \text{ for } a\in \mathcal{A}\\
    &(b,e),(b,f) \text{ for } b \in \mathcal{B}.
\end{align}
Define $\tilde{\mcC}=\{[(a,0)],[(b,f)]\}$ as the fusion category after the condensation of $(r,e)$. We have
\begin{align}
   & \theta_{(a,0)} = \theta_{a}, && \theta_{(b,f)} = -\theta_{b}. &
\end{align}
It is easy to check that $\mcC$ and $\tilde{\mcC}$ have the same fusion rule.

If $\theta_r =i$, consider the anti-semion square model $C_{\overline{\text{2sem}}}= \{0,e,m,f\} $.
\begin{align}
    S_{\overline{\text{2sem}}} =\frac{1}{2} \begin{pmatrix}
    1 & 1&1&1\\
    1 & -1&1&-1\\
    1 & 1&-1&-1\\
    1 & -1&-1&1
    \end{pmatrix}, T_{\overline{\text{2sem}}}=\begin{pmatrix}
    1 & 0&0&0\\
    0 & -i&0&0\\
    0 & 0&-i&0\\
    0 & 0&0&-1
    \end{pmatrix}
\end{align}
After the anyon condensation of boson $(r,e)$, the anyons in $\mcC\times \mcC_{\overline{\text{2sem}}}$ that not confined are
\begin{align}
    &(a,0),(a,m) \text{ for } a\in \mathcal{A}\\
    &(b,e),(b,f) \text{ for } b \in \mathcal{B}.
\end{align}
Define $\tilde{\mcC}=\{[(a,0)],[(b,f)]\}$ as the fusion category after the condensation of $(r,e)$. We have
\begin{align}
   & \theta_{(a,0)} = \theta_{a}, && \theta_{(b,f)} = -\theta_{b}. &
\end{align}
It is easy to check that $\mcC$ and $\tilde{\mcC}$ have the same fusion rule.
\clearpage
\bibliography{ref.bib}

\end{document}